\documentclass[prc,twocolumn,preprintnumbers,showpacs,superscriptaddress]{revtex4}    
\usepackage[colorlinks=true,linkcolor=blue,citecolor=blue,urlcolor=darkred]{hyperref}
\usepackage{amsmath,aas_macros}
\usepackage[T1]{fontenc}
\usepackage{amsmath}
\usepackage{natbib}
\usepackage{soul}
\usepackage{xcolor}
\usepackage{lipsum}
\usepackage{float}
\usepackage{graphicx}	
\usepackage{multirow,makecell}
\usepackage{placeins}
\usepackage{bm,amsfonts,hhline,amssymb,microtype,float,latexsym,epsf,mathtools,times,makecell,array}
\usepackage{tabularx}
\usepackage{epstopdf}
\usepackage{gensymb}
\usepackage{pifont}
\usepackage{hyperref}
\usepackage{lineno}
\usepackage{natbib}

\definecolor{ForestGreen}{HTML}{228B22}

\usepackage[normalem]{ulem}
\definecolor{darkred}{rgb}{0.7,.1,.2}

\begin{document}
\title{Topology of the Superconducting Heart of Neutron Stars: Effects of Microphysics and Gravitational-Wave Signatures}

\author{Mayusree Das}
\email{mayusreedas@iisc.ac.in}
\affiliation{Joint Astronomy Programme, Department of Physics, Indian Institute of Science, Bangalore, 560012, India}

\author{Armen Sedrakian}
\email{armen.sedrakian@uwr.edu.pl}
\affiliation{Frankfurt Institute for Advanced Studies, D-60438 Frankfurt am Main, Germany}
\affiliation{Institute of Theoretical Physics, University of Wroclaw, 50-204 Wroclaw, Poland}

\author{Banibrata Mukhopadhyay}
\email{bm@iisc.ac.in}
\affiliation{Department of Physics, Indian Institute of Science, Bangalore, 560012, India}
\affiliation{Joint Astronomy Programme, Department of Physics, Indian Institute of Science, Bangalore, 560012, India}
\begin{abstract}
	We present a general-relativistic study of the distribution of proton superconductivity in strongly magnetized neutron stars (NSs), using the \texttt{XNS} code to solve the coupled Einstein–Maxwell equations. We investigate equilibrium configurations with both toroidal and poloidal magnetic field geometries and incorporate complex many-body effects through microscopically derived proton pairing gaps. The models employ equations of state (EoS) obtained from microscopic many-body theory—including realistic two- and three-body nuclear interactions—as well as from relativistic mean-field approaches.
  We compare superconducting topologies across our collection of EoS and explore the influences of magnetic field geometry in stellar models parameterized by central density. Our models confirm the absence of $S$-wave superconductivity in the inner core and, importantly, reveal that non-superconducting regions exhibit complex three-dimensional geometries: doughnut-shaped for toroidal fields and prolate-shaped for poloidal fields -- spatial structures that are inherently absent in one-dimensional analyses.  We also compute magnetic deformations and ellipticities for several millisecond pulsars (MSPs), estimating their continuous gravitational wave strain. While these MSPs remain undetectable by current detectors, next-generation instruments such as the Einstein Telescope and Cosmic Explorer may detect their signals, opening an observational window into internal superconductivity and internal magnetic field of NSs, as well as the fundamental microphysics of dense matter.
\end{abstract}

\maketitle

\section{Introduction}

Neutron stars (NSs) cool rapidly after birth via neutrino emission. This causes their internal temperature ($T$) to drop below the critical values for  neutron and proton Cooper pairing --  typically in the range  $\sim 10^9-10^{10}$~K --
within days~\cite{Page06, Blaschke2013PhRvC}. Hence, neutron superfluidity and proton superconductivity set in. In NS crusts and cores, nucleonic pairing operates in the weak-coupling
Bardeen–Cooper–Schrieffer (BCS) regime. Here, the pairing gap (of the order of 1~MeV or less) is much smaller than the characteristic Fermi energies of neutrons and protons (of the order of several tens of MeV). This weak-coupling regime, where the coherence length exceeds the interparticle spacing, justifies the application of the BCS framework \cite{bcs, Abrikosov:Fundamentals, Sedrakian2019EPJA}.

Nucleonic superfluidity and superconductivity in NSs manifest through profound observational signatures. Cooper pairing affects stellar cooling through two primary mechanisms: enhanced neutrino emission via pair-breaking and formation channels (potentially explaining the accelerated cooling of Cassiopeia A~\cite{Page2011PhRvL,Shternin2011}), and internal heating from Josephson effect~\cite{Sedrakian_2025} or crustal vortex creep~\cite{Larson1999,Fujiwara2024}. These superconducting phases also play crucial roles in pulsar timing anomalies (glitches)~\cite{Baym1969Natur,Andersson1975,Sedrakian1999,Haskell2013,2014Goglu,Chamel2017, Haskell18, GlitchesZhou2022, Antonopoulou2022RPPh},  in magnetic field evolution~\cite{Pons2009AA, Ascenzi2024MNRAS},  and continuous gravitational wave (CGW) generation \cite{Alford2014ApJ,Piccinni2022,Andersson2021Univ,Suvorov2025}.

  Shortly after the discovery of pulsars, Baym et al.~\cite{BPP1969} proposed that NS cores contain a type-II superconductor threaded by quantized magnetic flux tubes. Although observed pulsar magnetic fields are below the lower critical field $H_{c1}$ for flux-tube formation, it was argued that flux tubes would nevertheless form, since magnetic-flux expulsion occurs on timescales far exceeding the ages of NSs. Simultaneously, microscopic studies of nucleonic pairing in neutron-star matter established from first principles the BCS pairing of neutrons and protons~\cite{Tamagaki1970,Hoffberg1970}. Refs.~\cite{Clark1972,Takatsuka1972,Takatsuka1973} provided the first microscopic treatments of proton superconductivity; developments in subsequent decades are  reviewed in Ref.~\cite{Sedrakian2019}.

The arrangement of magnetic flux into a two-dimensional lattice of quantized flux tubes produces a spatially averaged stress tensor several orders of magnitude larger than the Maxwell stress associated with a uniform field. This implies that proton superconductivity can generate substantial internal magnetic distortions~\cite{Jones1975,SedrakyanHayrapetyan2015}. 
Such magnetically induced deformations have significant implications for CGW emission from rotating NSs~\cite{Cutler2002,Haskell2008,freibenrezz2012,Akgun2018,sold2021main}. Toroidal flux tubes interact strongly with neutron vortices and may influence spin evolution and glitch dynamics of pulsars~\cite{GugerCAlpar2014,GugerCAlpar2016}.

The interplay between proton superconductivity and strong magnetic fields in the range relevant to strongly magnetized NSs, i.e., magnetars, was considered previously in Refs.~\cite{Sinha2015PPN, Sinha2015PhRvC} using spherically symmetrical models with radially dependent magnetic field. These studies suggested that type-II superconductivity typically forms in the outer core, transitions to type-I at higher densities, and vanishes in the inner core. Higher-partial wave pairing -- proton-proton $^3P_2$-$^3F_2$-wave or neutron-proton $^3D_2$-wave -- is possible under extreme density or isospin asymmetry, but likely not feasible in standard NS matter~\cite{Raduta2019MNRAS, Alm1996NuPhA}.

In our previous study \cite{Das2025} (hereafter DSM2025), we investigated the structure of superconducting regions in magnetars using equilibrium solutions of Einstein-Maxwell equations with the \texttt{XNS} solver \cite{sold2021main}, thereby extending previous studies restricted to models with one-dimensional superimposed magnetic field and proton pairing profiles~\cite{Sinha2015PPN, Sinha2015PhRvC}. 
To treat the matter properties, we employed a relativistic mean-field (RMF) equations of state (EoSs) and a temperature-dependent superconducting gap based on bare two-body (2B) proton interactions from Ref.~\cite{Baldo1992}.

The nature of proton superconductivity, a central point of the present work,  is closely linked to the microscopic parameters of the condensate and to the associated critical magnetic fields. These fields delineate non-superconducting regions, type-II superconducting regions (characterized by magnetic flux tubes), and type-I regions, which may form a mixed-phase, characterized by domains of superconducting and normal matter.  While the critical fields for type-II superconductivity~\cite{Sedrakian1995ApJ, Sinha2015PhRvC} and the conditions of the onset of type-I superconductivity~\cite{Sedrakian1997MNRAS, Buckley2004, Sedrakian2005PhRvD, Alford2005} were discussed separately, Ref.~\cite{Haber2017PhRvD} computed a general phase diagram for type-I and type-II phases in coupled neutron–proton superfluids within the Ginzburg–Landau formalism, incorporating both upper and lower critical fields. For simplicity, in the present study, we neglect corrections due to the coupling between the proton superconductor and the neutron superfluid via entrainment, as discussed in Ref.~\cite{Haber2017PhRvD}, and instead use the standard expressions for the critical fields assuming no proton–neutron coupling.

Here, we extend DSM2025 in several directions. First, we incorporate zero-temperature pairing gaps computed with medium effects, including screening of the pairing interaction through polarization of the medium, wave-function renormalization, and three-body (3B) forces~\cite{Baldo1997, Baldo2007}. These
computations employ single-particle spectra which are obtained using Brueckner-Hartree-Fock (BHF) many-body theory~\cite{Brueckner1958,Baldo1997}, which properly accounts for short-range correlations in nuclear matter. Given the uncertainty in effective pairing interaction strength at high densities, we bracket the expected physical range by considering gaps both without 3B forces (upper bound) and with full medium effects (lower bound).

Second, we investigate how the choice of EoS, particularly its stiffness, affects superconductivity in dense matter. We consider two representative models: a microscopic BHF EoS with two- and three-body forces~\cite{Baldo1997}, and the RMF EoS based on DDME2 parametrization~\cite{gppva, Lala2005}. The BHF EoS is grounded in vacuum nuclear physics reproducing phase shifts, deuteron binding energy, and light nuclei properties. It is consistent with subnuclear structure \cite{Taranto2013}, gravitational wave (GW) constraints from GW170817 \cite{Abbott2018}, and NS cooling data~\cite{Fortin2018}. However, its maximum NS mass of $1.94 M_{\odot}$ falls slightly
below the current lower bound of $2.14_{-0.09}^{+0.10} M_{\odot}$
from PSR J0740+6620~\cite{Cromartie2019}. In contrast, DDME2 is a phenomenological RMF model that matches nuclear saturation properties by construction and supports NS masses up to $2.48 M_{\odot}$. While both EoS are consistent at saturation, they differ significantly in their treatment of nuclear interactions, since DDME2 employs meson-exchange couplings, while BHF uses non-relativistic many-body theory. By comparing these two models, we assess how microphysical uncertainties in the EoS affect both the structure and size of superconducting regions in dense matter. This highlights the differences that arise when transitioning from a moderately soft (BHF) to a stiff (DDME2) EoS.

While the NS core may host additional or alternative degrees of freedom at high density, like hyperons or deconfined quarks, we choose to work here with a minimal nucleonic composition which is charge-neutral, $\beta$-stable mixture of neutrons ($n$), protons ($p$), electrons ($e$), and muons ($\mu$), avoiding thus poorly constrained physics of exotic phases.

Third, we investigate how magnetic field geometry, toroidal versus poloidal, affects the topology of  superconducting phases within the star. We also investigate mixed-field configurations, while toroidally dominated mixed fields are thought to be more dynamically stable and may represent the long-term field structure in magnetars~\cite{Braithwaite2009}. 

Our models do not account for magnetic field rearrangement or the magnetic back-reaction of proton superconductivity on the matter. Instead, we focus on identifying the regions where superconductivity sets in. 
Our stellar structure calculations are performed using the \texttt{XNS} code under the single-fluid approximation, neglecting the multifluid nature of neutron-proton mixtures. Although Newtonian two-fluid superconducting models have been developed \cite{AW2008MNRAS, Lander12}, their general relativistic counterparts remain computationally demanding and are not yet implemented.

A key astrophysical implication of our study is the potential for CGW emission from millisecond pulsars (MSPs). These rapidly rotating, weakly magnetized NSs can emit CGWs if they possess even a small magnetic-field-induced deformation measured via ellipticity, $\epsilon$, and a misalignment between magnetic and spin axes. LIGO and Virgo have already placed upper limits on $\epsilon$ for several MSPs based on non-detections~\cite{ligo2022}. Using internal magnetic field configurations from the general relativistic simulations, we calculate $\epsilon$ for representative MSPs. Our results indicate that superconductivity can significantly enhance magnetic deformations compared to normal matter, potentially boosting CGW signals. Future detectors such as the Einstein Telescope and Cosmic Explorer may thus detect these signals, offering indirect evidence for superconductivity and internal magnetic fields, thereby shedding light on the microphysics of dense matter.

This paper is organized as follows. In Sec.~\ref{sec:model}, we outline the general relativistic framework used to model magnetized NSs. Sec.~\ref{sec:gap} introduces the microphysical inputs, including the pairing gap models. The criteria for the onset of superconductivity are discussed in Sec.~\ref{sec:tc}. In Sec.~\ref{sec:supregions}, we present our results on the spatial topology of superconducting regions. Sec.~\ref{sec:msp} focuses on the implications for GW signatures from MSPs. We summarize our key findings,  discuss limitations of our approach, and future research directions in Sec.~\ref{sec:Conclusions}.

\section{Magnetized neutron star modeling}
\label{sec:model}

Equilibrium axisymmetric models of rotating, magnetized NSs in general relativity, as computed by the \texttt{XNS} solver \cite{sold2021main}, require the simultaneous solution of Einstein’s equations for the spacetime metric and the equations of general relativistic magneto-hydrostatic equilibrium for the stellar matter and electromagnetic fields. The metric is assumed to be conformally flat; thus, the line element is given by 
\begin{equation}
ds^2 = -\alpha^2 dt^2 + \psi^4 \left[dr^2 + r^2 d\theta^2 + r^2 \sin^2\theta (d\phi + \beta^\phi dt)^2 \right],
\label{eq:metric}
\end{equation}
where $\alpha$ is the lapse function, $\psi$ is the conformal factor, and $\beta^\phi$ is the shift vector component \cite{pili2014}. The matter and electromagnetic fields are described in the ideal magnetohydrodynamics framework, with the energy-momentum tensor given by
\begin{equation}
T^{\mu\nu} = (e + P + b^2)u^\mu u^\nu - b^\mu b^\nu + \left(P + \frac{1}{2}b^2\right)g^{\mu\nu},
\label{eq:tmunu}
\end{equation}
where $e$ is the energy density, $P$ is the pressure, $u_\mu$ 
is the fluid four-velocity, whose azimuthal part describes rotation, $b^\mu$ is the magnetic field in the fluid (comoving) frame related to the stellar magnetic field $B^\mu$ whose spatial components are $B^r$, $B^\theta$, and $B^\phi$, and $g^{\mu\nu}$ is the metric tensor. The equilibrium configuration defining the stellar structure is determined by the condition of magnetohydrostatic equilibrium,
\begin{equation}
\nabla_\nu T^{\mu\nu} = 0,
\label{eq:ET}
\end{equation}
along with the Maxwell equations for the electromagnetic field,
\begin{equation}
\nabla_\nu F^{\mu\nu} = 4\pi J^\mu, \qquad \nabla_\nu {}^*F^{\mu\nu} = 0,
\label{eq:max}
\end{equation}
where $F^{\mu\nu}$ is the electromagnetic field tensor, ${}^*F^{\mu\nu}$ is its dual, and $J^\mu$ is the electromagnetic four-current, which acts as the source of the field. Closure is provided by a suitable EoS, e.g. $p = p(\rho)$, relating pressure to density $\rho$.

\begin{figure}[t!]
  \centering \includegraphics[width=\columnwidth]{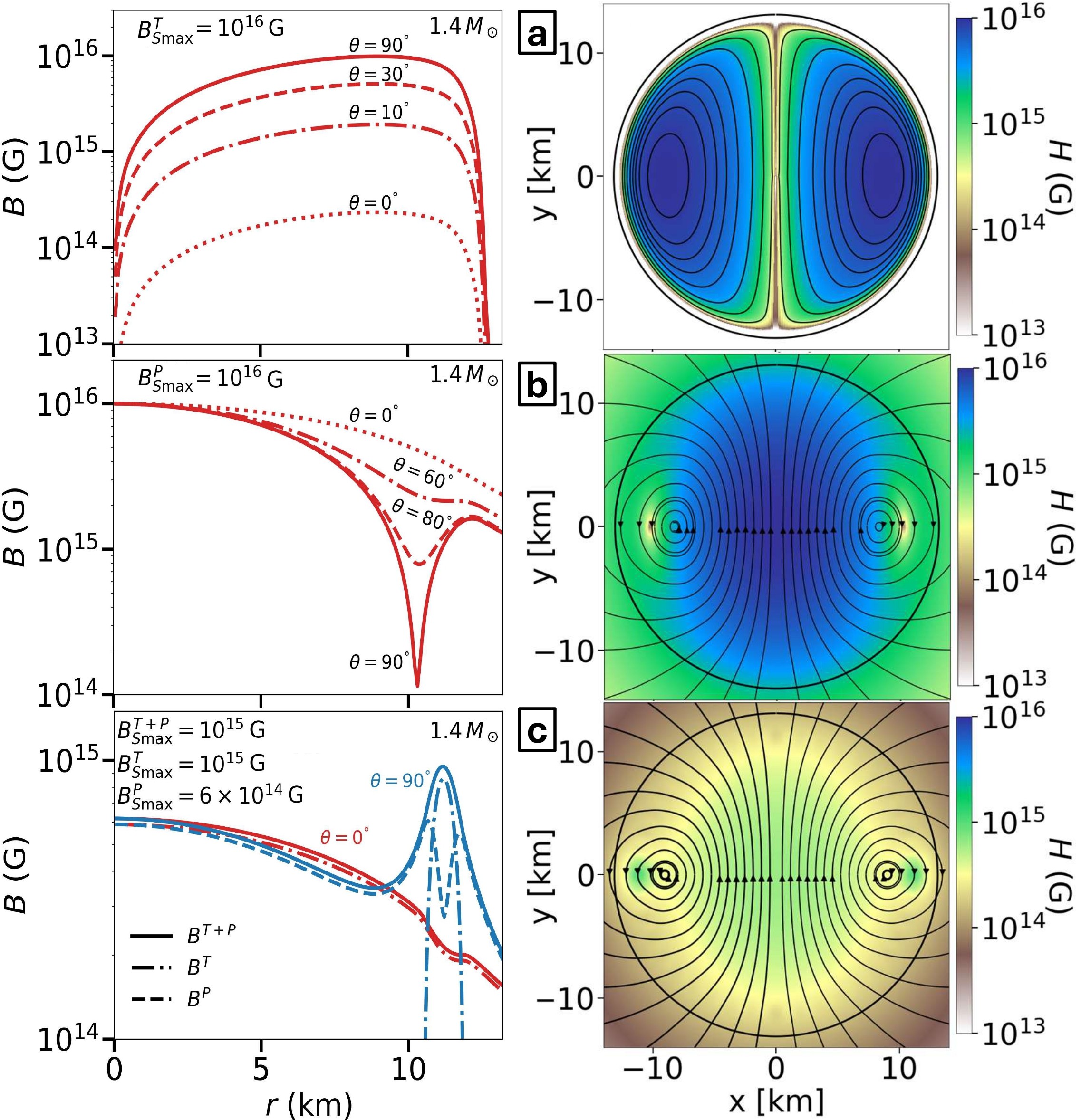} 
  \caption{Left panels: The magnitude of the stellar magnetic field is shown as a function of the radial coordinate for fixed values of the polar angle $\theta$. Each model is specified by the stellar mass and the maximum field strength, as indicated in the corresponding panel. The upper and middle panels present the purely toroidal and purely poloidal configurations, respectively, while the lower panel corresponds to the twisted-torus configuration and displays separately the toroidal and poloidal components, and the total magnetic field, as labeled.
  Right panels: Color maps illustrating the spatial distribution of the stellar magnetic field in the $x$–$y$ plane of Cartesian coordinates, with the magnetic axis aligned along the $y$-direction. The top, middle, and bottom panels correspond to the toroidal, poloidal, and twisted-torus configurations, respectively.}
\label{fig:1}
\end{figure}

The \texttt{XNS} solver takes the central density, maximum magnetic field ($B_{S\text{max}}$), magnetic field configuration (poloidal, toroidal, or mixed), the star’s angular rotation frequency (uniform throughout), and EoS as inputs. It numerically solves the Einstein-Maxwell Eqs.~\eqref{eq:metric}-\eqref{eq:max} self-consistently, yielding the stellar mass ($M$), stellar radius ($R$), and stellar magnetic field profile ($B_S$). The \texttt{XNS} solver assumes a zero-temperature matter, which is appropriate for our study of superconducting NSs, where core temperature remains well below $10^{10}$ K, so that the thermal effects are negligible.

In Sec.~\ref{sec:supregions}, we consider non-rotating stellar models featuring poloidal, toroidal, or twisted-torus magnetic field configurations in order to explore the spatial topology of superconducting regions. This approximation is appropriate for magnetars, which are known to be slowly rotating compact stars (with observed spin frequencies typically of a few Hz) and therefore exhibit negligible rotational deformation.
By contrast, in Sec.~\ref{sec:msp}, where we discuss CGW emission from MSPs, we consider stellar models rotating at frequencies matched to the observed spin frequency of each individual MSP. In this context, we conjecture the presence of a strong toroidal magnetic field beneath the stellar surface, which does not affect the relatively low magnetic-field strength inferred from the spin-down of MSP pulsars.

In the following, we use $H$ to denote the magnetic field intensity (external or applied field) and $B$ for the magnetic induction, which includes the material’s response, see Refs. \cite{Sedrakian1995ApJ,Haber2017PhRvD}. The stellar magnetic field $B_S$, computed from equilibrium models (e.g., \texttt{XNS}), is associated with $H$-field in normal NS matter where magnetization is negligible (e.g., $B_S\approx H$).

We present visualizations of three $1.4\,M_\odot$ NS (magnetar) models, each with a distinct magnetic field configuration: purely toroidal, purely poloidal, and mixed-field. The left panels of Figs.~\ref{fig:1}a and \ref{fig:1}b show the radial profiles of toroidal and poloidal local $H$-field, respectively, for various fixed azimuthal angles $\theta$ in spherical coordinates. The right panels show $x$--$y$ cut orthogonal to the $z$ axis in Cartesian coordinates with the $y$-axis along the magnetic axis, for toroidal (a) and poloidal (b) fields. The maximum fields for toroidal field $B_{S\text{max}}^T$ and poloidal field $B_{S\text{max}}^P$ are $10^{16}$ G. It is seen that the maximum field strength is achieved along the equatorial radial coordinate ($\theta=90^\circ$) for toroidal field and along the pole ($\theta=0^\circ$) for poloidal field. The field strength is minimal along the magnetic axis for the former and near the equator for the latter. Note that for the toroidal field, the field attains its maximum away from the star's center within the outer core, the location of maximum depending on the angle $\theta$. For the poloidal field,  the maximum field value is attained at the star's center.

Magnetic configurations with finite helicity are expected to evolve toward a stable equilibrium containing both poloidal and toroidal components  \cite{Braithwaite2006, Braithwaite2009}. One of such mixed field geometries is the so-called \textit{twisted-torus} structure. It features a highly localized toroidal component confined to a ring-shaped region in the outer core, while the poloidal field extends smoothly into the magnetosphere. An example of such a configuration is shown in Fig.~\ref{fig:1}c, with the left panel displaying radial field profiles (toroidal, poloidal, and total fields) for different $\theta$, and the right panel showing the total field in the $x$--$y$ plane. The maximum total field  for the model is $B_{S\text{max}}^{T+P}=10^{15}$ G.

As the NS models in this work possess strong internal magnetic fields, it is essential to assess their dynamical stability. A commonly adopted criterion to ensure stability requires the ratio of magnetic to gravitational energy $ \lesssim 10^{-3}$ \cite{Braithwaite2009}. The ratios for above mentioned toroidal, poloidal, and twisted-torus models are $4.7\times 10^{-5}$, $2.3\times 10^{-5}$, and $10^{-7}$.

\section{Proton superconducting gap}
\label{sec:gap}
As in laboratory superconductors, where electrons form Cooper pairs
via phonon-mediated attraction, degenerate protons in neutron stars
can also pair through the attractive component of the nuclear force
generated by meson exchange. In both cases, the Cooper-pair size is
much larger than the interparticle spacing, placing the system in the
weak-coupling BCS regime. In the BCS theory, the zero-temperature gap is
$$
 \Delta \approx \mu \exp\!\left[-\frac{1}{V_{\rm eff} N(0)}\right],
$$
where $\mu$ is the proton chemical potential, $V_{\rm eff}$ the
effective proton–proton interaction, and $N(0)$ the density of states
at the Fermi surface, exhibiting the exponential sensitivity of
$\Delta$ to the interaction strength. In nuclear matter, the normal-state quasiparticle spectrum is typically described within
Brueckner–Hartree–Fock theory, which accounts for short-range
correlations by softening the repulsive core of the nucleon
interaction. While pairing is often treated using bare two-body forces, this neglects medium polarization and three-body effects, the latter introducing additional repulsion at high densities and
suppressing the gap, thereby confining proton superconductivity to intermediate-density regions and eliminating it in the innermost core~\cite{Baldo2007}.

 We use $\Delta(0)$ calculated in Ref.~\cite{Baldo2007}, based on 2B Argonne $v_{18}$ potential \cite{V18} along with 3B Urbana UIX force \cite{UIX}, with both the microscopic EoS and pairing treated self-consistently with BHF treatment. We have fitted these results for $\Delta(0)$ corresponding to 2B interaction only, as well as two- and three-body forces (2B+3B)
 interaction with the results, respectively,
 \begin{eqnarray}
   \label{eq:D_2B}
\Delta(0) &=& -7.12831 + \frac{7.12145}{\exp(k_{F_p}^2)} + 7.5122\,k_{F_p} + 7.12145\,k_{F_p}^2 \nonumber \\
&& - 15.9434\,k_{F_p}^3 + 6.45121\,k_{F_p}^4 \nonumber \\
          &=& 0 \quad \text{for} \quad k_{F_p} > 1.3, \\
    \label{eq:D_3B}
\Delta(0) &=& -11.8655 + \frac{11.9175}{\exp(k_{F_p}^2)} + 11.1052\,k_{F_p} + 11.9175\,k_{F_p}^2 \nonumber \\
&& - 33.0354\,k_{F_p}^3 + 17.5324\,k_{F_p}^4 \nonumber \\
&=& 0 \quad \text{for} \quad k_{F_p} > 0.95, \label{eq:gap0}
\end{eqnarray}
where $k_{F_p} = \left( 3\pi^2 n_p\right)^{1/3}$ in fm$^{-1}$ and $n_p$ is the proton
number density. Once the superconducting gap at zero-temperature $\Delta(0)$ is known, the gap at a finite $T$ can be calculated via the fit formula to the BCS relation, defining the
finite $T$ gap given by~\cite{Muhlschlegel1959ZPhy}
\begin{equation}
\frac{\Delta(\tau)}{\Delta(0)} = 
\begin{cases}
\displaystyle 1 - \sqrt{2\gamma \tau} e^{-\pi / (\gamma \tau)}, & 0 \leq \tau \leq 0.5, \\[0pt]
\displaystyle \sqrt{3.016(1-\tau) - 2.4(1-\tau)^2}, & 0.5 < \tau \leq 1,
\end{cases}
\label{eq:gapt}
\end{equation}
where   $\tau=T/T_c$ and the critical temperature $T_c$ is related to $\Delta(0)$ via
\begin{equation}\label{eq:tc}
T_c=\Delta(0)/1.76k_B,
\end{equation} 
where $k_B$ is the Boltzmann constant and $\gamma = 1.781$.  Eqs.  \eqref{eq:gapt} and \eqref{eq:tc} hold in the weak-coupling BCS limit, an excellent approximation since the proton gap is much smaller than the proton chemical potential. To evaluate $\Delta$ in NS models constructed using \texttt{XNS} solver, we express it as a function of density using $n_p = n Y_p$ with the proton fraction $Y_p(n)$ that can be taken from the EoS tables. Although the nucleon–nucleon potential is attractive in the $^1S_0$ channel for both neutrons and protons in the same energy range, the smallness of $Y_p$, and consequently the proton density, implies that $S$-wave pairing occurs at higher baryon densities than for neutrons.

\begin{figure}[bt]
\centering
\includegraphics[width=\columnwidth]{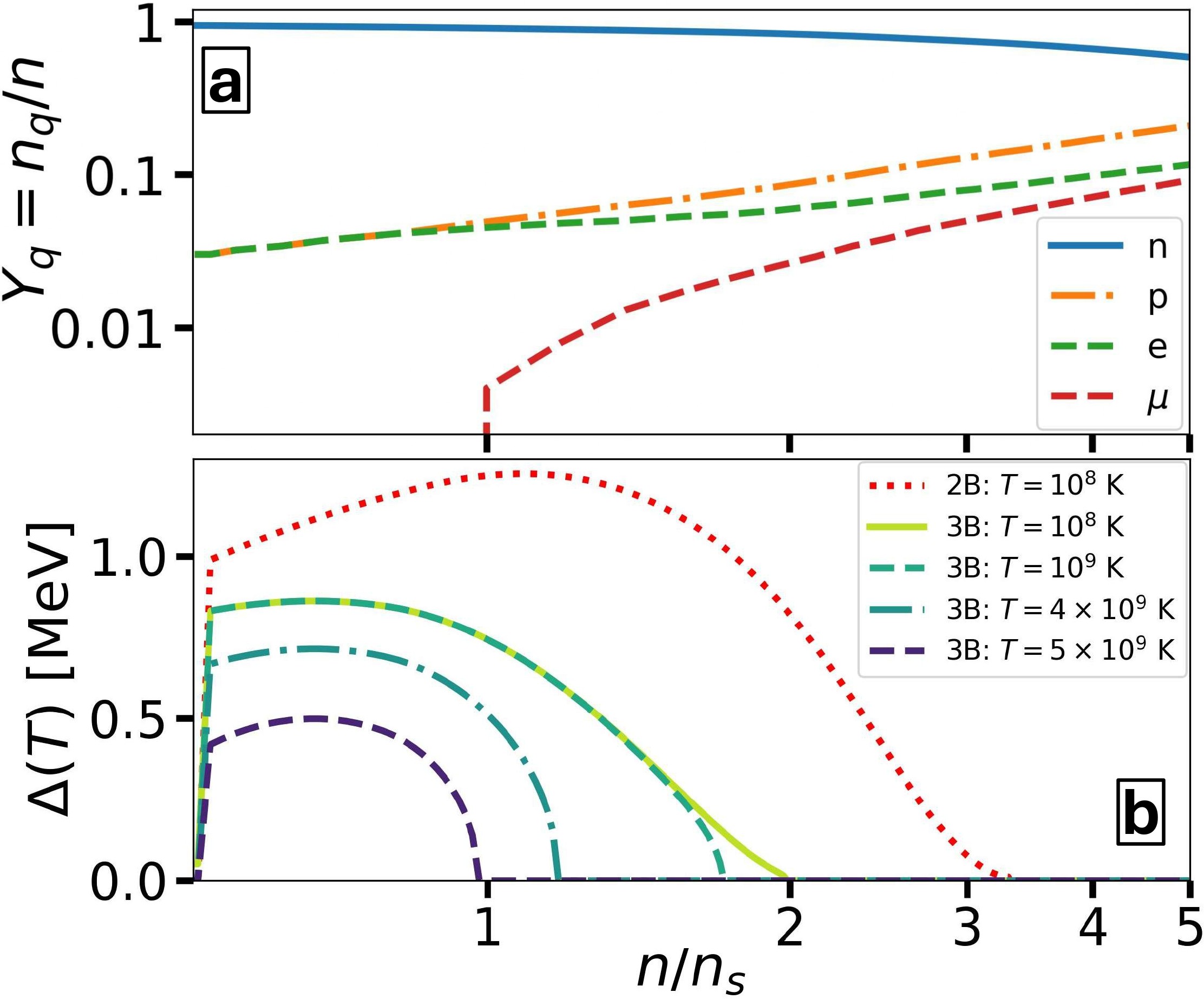}
\caption{(a) The fractions of constituents $Y_{q}=n_q/n$, $q\in n,\, p,\, e,\,  \mu$, as functions of normalized density $n/n_s$~\cite{Sharma2015}. (b) Dependence of proton pairing gap $\Delta$  on normalized density at different temperatures, computed at zero temperature using Eqs.~\eqref{eq:D_2B} or \eqref{eq:D_3B} and extended to finite temperature using Eq.~\eqref{eq:gapt}.}
\label{fig:2}
\end{figure}

Fig. \ref{fig:2}a shows the particle fractions $Y_q=n_q / n$, as computed in Ref.~\cite{Sharma2015}, where $q \in\{n, p, e, \mu\}$, as functions of the normalized baryon density $n / n_s$, with $n$ being the baryon number density and $n_s=0.16$~fm$^{-3}$ being the nuclear saturation density. It is important to note that in the crustal region of NSs, protons are confined within nuclear clusters. As a result, bulk superconductivity does not occur at densities below $n / n_s \lesssim 0.5$, corresponding to the crust-core interface.  Fig. \ref{fig:2}b displays the proton pairing gap $\Delta(T)$ as a function of density for various $T$, calculated using Eqs.~\eqref{eq:gap0} and \eqref{eq:gapt}. Thus, as the star cools and 
the temperature drops below  the critical temperature, the gap emerges and increases in magnitude, marking the onset and growth of superconducting regions within the star. This occurs during the neutrino cooling era, which involves rapid (of the order of hours to days) cooling down to $T \sim 10^9 $~K, followed by a slower decline to $T \sim 10^8$~K over the next $10^4$ years~\cite{Page06}.

\section{Critical temperature and critical magnetic fields of superconductors}
\label{sec:tc}

For a given temperature $T$, superconductivity arises where $T < T_c(\rho)$ and the local magnetic field is below the upper critical field $H_{c2}$. For $S$-wave pairing, the critical temperature follows from the zero-temperature gap via the BCS relation given by Eq. \eqref{eq:tc}.

On mesoscopic scales, superconductivity is governed by two
characteristic lengths: the London penetration depth
$\lambda$ (field decay scale) and the coherence length $\xi$ (Cooper
pair size).
Their ratio defines the {\it Ginzburg-Landau parameter}
$\kappa = \lambda / \xi$~\citep{Abrikosov:Fundamentals}, which determines the superconductivity type.

At high densities, where $\kappa < 1/\sqrt{2}$, the protons form a type-I superconductor. The magnetic field is expected to be expelled if $H < H_{cm}$. However, because the expulsion is ineffective~\citep{Baym1969}, it penetrates through normal domains which are alternating with superconducting domains, in which the field is screened beyond scales of the order of $\lambda$. The domain geometry depends on the proximity to type-II regions, magnetic history, and superconducting-normal interface energy. For $H > H_{cm}$, the normal state is favored.

As density decreases toward the crust–core boundary, $\kappa$ increases and type-II superconductivity becomes favored for $\kappa > 1/\sqrt{2}$. The negative surface energy between phases allows magnetic flux to enter via quantized flux tubes when $H_{c1} \le H \le H_{c2}$.  For $H < H_{c1}$, flux expulsion is expected but  is ineffective due to the high core conductivity~\citep{Baym1969}. Local field rearrangements may lead to the formation of  flux-tube clusters in regions  with $H> H_{c1}$~\cite{Sedrakian1999}. Neutron vortices can form coaxial proton flux tube  clusters due to their magnetization via the entrainment effect~\cite{Sedrakian1995ApJ}. At extremely strong fields, superconductivity is destroyed when the proton Larmor radius becomes smaller than $\xi$, setting the upper critical field $H_{c2}$~\cite{Sinha2015PhRvC}.

The mesoscopic parameters are given by
\begin{equation}
\label{eq:meso}
\xi=\frac{\hbar^2 k_{F_p} }{ \pi m_p^* \Delta }, \quad 
\lambda = \sqrt{\frac{m_p^{*2} c^2}{4 \pi e^2 \rho_{s}}} .
\end{equation}
where $\hbar$ is Planck's constant, $c$ is speed of light, and $e$ is electron's charge. 
The proton effective mass $m_p^*$ in the case of only 2B force is given by the fit formula \cite{Baldo1992, sinha15}
\begin{equation}
\frac{m_p^*}{m_p} = 1.00661 - 0.649838\,k_{F_p} + 0.34416\,k_{F_p}^2 - 0.0441441\,k_{F_p}^3.
\label{A4}
\end{equation}
In the 2B+3B case, where the three-body force is also included, the fit is given by \cite{Baldo2014}
\begin{align}
\frac{m_p^*}{m_p} &= 1 - (a_1 + b_1 Y_p + c_1 Y_p^2)\, n + (a_2 + b_2 Y_p + c_2 Y_p^2)\, n^2 \nonumber \\
&\quad - (a_3 + b_3 Y_p + c_3 Y_p^2)\, n^3,
\end{align}
where $a_1 = 1.56,\ b_1 = 1.31,\ c_1 = -1.89,\ a_2 = 3.17,\ b_2 = 1.26,\ c_2 = -1.56,\ a_3 = 0.79,\ b_3 = 3.78,\ c_3 = -3.81$. In Eq.~\eqref{eq:meso} $\rho_s$ is the density
of superconducting protons, which is related to 
$\rho_p$ by \citep{landau1980statistical}
\begin{equation}
\frac{\rho_s}{\rho_p} = 
\begin{cases}
1 - \sqrt{2\pi \Delta(0)/k_B T} e^{-\Delta(0) / k_B T}, & 0 \leq \tau \leq 0.5, \\[0pt]
2 \left( 1 - \tau \right), & 0.5 < \tau \leq 1.
\end{cases}
\label{eq:ns}
\end{equation}
It is seen that at low $T$, the number of excitations out of the condensate is suppressed
exponentially by the Boltzmann factor, while for $T\to T_c$ the density of superconducting protons vanishes linearly.

\begin{figure}[bt]
\centering
\includegraphics[width=\columnwidth]{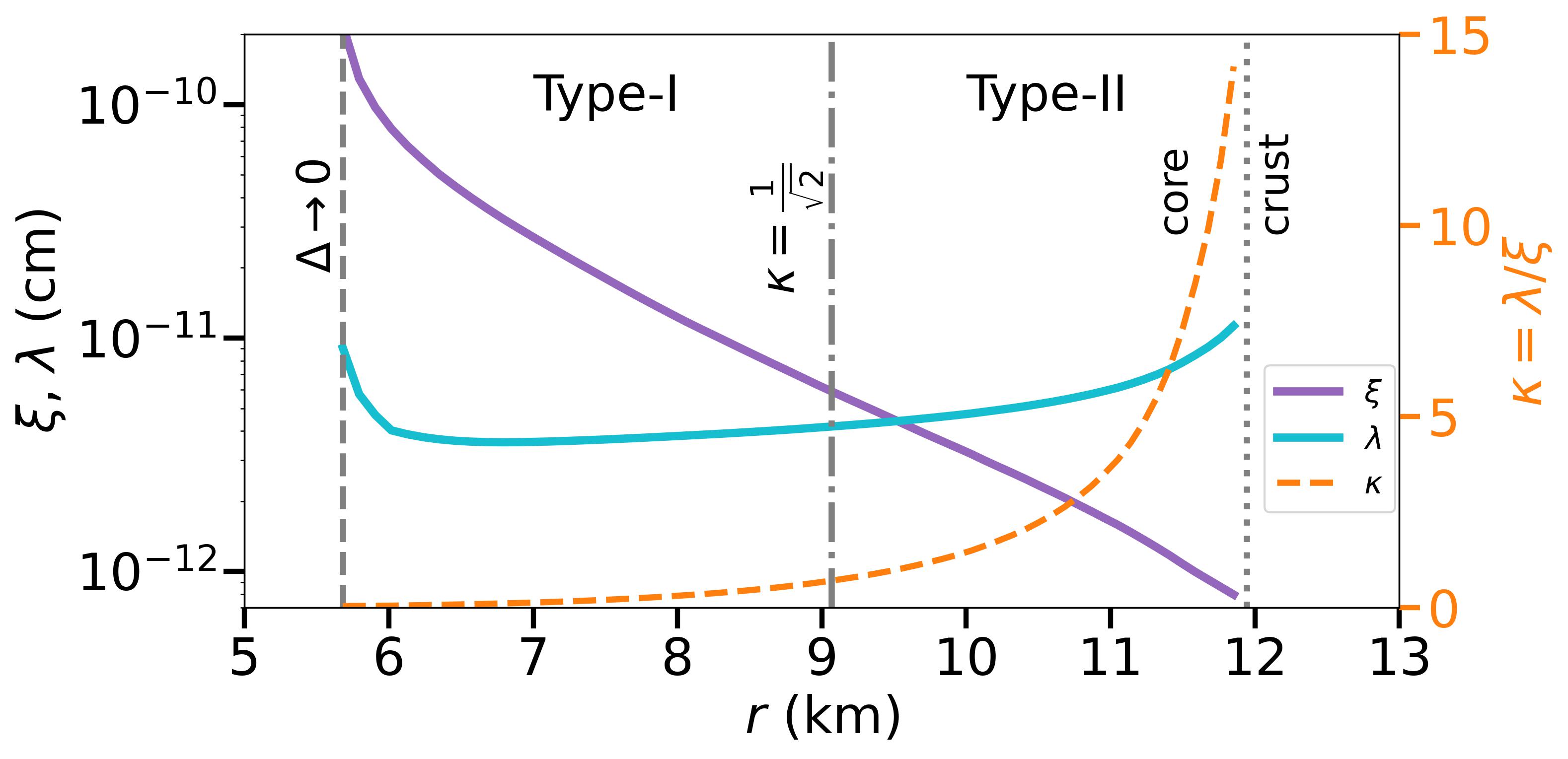}
\caption{$\lambda$, $\xi$ and $\kappa$ for a $1.4M_\odot$ NS as functions of equatorial radial coordinate, showing the possible superconducting region of type-I ($\kappa<1/\sqrt{2}$) and type-II ($\kappa>1/\sqrt{2}$).}
\label{fig:3}
\end{figure}

As $\kappa$ increases with $r$, two key transitions characterize the superconducting structure of the NS core, as illustrated in Fig. \ref{fig:3}: (1) the onset of superconductivity, marked by the normal-to-superconducting phase transition where $\Delta$ becomes non-zero, and (2) the transition from type-I to type-II superconductivity, occurring at the location where $\kappa = 1/\sqrt{2}$, known as Bogomolny limit.

The size of the superconducting region, for any given value of temperature $T\le T_c$, is further determined by the local $H$-field, the critical fields $H_{c1}$ and $H_{c2}$ (for a type-II superconductor), and $H_{cm}$ (for a type-I superconductor). Specifically, the following options arise:
\begin{itemize}
\item type-II superconducting region, $\kappa(\rho) > 1/\sqrt{2}$:
  \begin{itemize}
     \item  $H> H_{c2}$,  normal state is preferred;
    \item  $H_{c1} \leq H \leq H_{c2}$,  superconductor features flux tube arrays with magnetic induction $B\simeq H$ except when $H\to H_{c1}$.
    \item $H < H_{c1}$, the Meissner state is preferred over the mixed state with $B=0$, but an (inhomogenous) lattice of flux tubes
      may arise with $B\simeq H$ as explained above.
    \end{itemize}
\item type-I superconducting region, $\kappa(\rho) < 1/\sqrt{2}$:
\begin{itemize}
    \item  $H > H_{cm}$,  normal state is preferred;
    \item $H < H_{cm}$, the Meissner state with $B=0$ is preferred, which may, under specific conditions, break down into 
      domains of normal and superconducting states with $B\neq 0$.
    \end{itemize}
  \end{itemize}

\subsection{Critical fields: type-II superconductor}

In type-II superconductors, the lower critical field $H_{c 1}$ marks the onset of the mixed state: for $H\ge H_{c1}$ it becomes energetically favorable for a single quantized flux tube (vortex) to enter the superconducting region, which for lower fields is in the Meissner state. 
The upper critical field $H_{c 2}$ corresponds to a second-order transition where the superconducting order parameter vanishes and the system becomes unpaired, i.e., normal-conducting. Below, we will adopt the common definition of the lower critical field, which is given by~\cite{Baym1969,Sedrakian1995ApJ,Gusakov2016a,Rau2020}
  \begin{equation}
    \label{eq:Hc1}
 H_{c 1}=\frac{\Phi_0}{4 \pi \lambda^2} \left[\ln \kappa +C_1 (\kappa)\right], \quad  H_{c 2}=\frac{\Phi_0}{2 \pi \xi^2} ,
\end{equation}
where $C_1 (\kappa) = 0.5 + {(1 + \ln 2)}/{(2\kappa - \sqrt{2} + 2)}$, arises from numerical computations of the exact flux-tube
energy in a type-II superconductor \cite{Hu1972, Brandt2003}, 
$\Phi_0=\pi\hbar c/e$ is the flux quantum.

\subsection{Critical field: type-I superconductor}

In the case of a type-I superconductor, there is a single critical magnetic field $H_{cm}$ that marks a first-order phase transition between the superconducting Meissner state and the normal-conducting state, i.e., it is the field at which the Gibbs free energies of these normal and superconducting phases become equal. Using standard Ginzburg–Landau relations, it can be obtained as~\cite{Abrikosov1957a, Tinkham1996}
  \begin{equation}
    \label{eq:Hcm}
    H_{cm}=  \frac{\kappa\Phi_0}{2^{3/2}\pi \lambda^2 }  =  
    \frac{\Phi_0}{2^{3/2}\pi \xi^2 \kappa} = \frac{H_{c 2}}{\sqrt{2} \kappa}.
  \end{equation}

\section{Stellar models}
\label{sec:supregions}

To investigate the size and topology of superconducting regions in magnetars, we model them using the \texttt{XNS} code by fixing the EoS and the microscopic input that determines superconducting parameters over a relevant range of temperatures. Each stellar model is assumed to be isothermal at a given $T$, and superconductivity is identified by comparing the local $T$ to $T_c$. Additionally, the local magnetic field strength $H$, computed self-consistently by the \texttt{XNS} solver, is compared to the local values of the critical magnetic fields: $H_{c 1}$ and $H_{c 2}$ in regions where type-II superconductivity is predicted ($\kappa >1/\sqrt{2}$), and $H_{c m}$ for type-I superconducting regions ($\kappa <1/\sqrt{2}$). In this study, we systematically vary the proton pairing gap (reflecting different microscopic nuclear interactions), the stiffness of the EoS, and the imposed magnetic field structure, by considering either toroidal, poloidal, or mixed-field configurations.

\subsection{Varying pairing gap of protons}
\label{sec:eosmicro}

\begin{figure}[hbt]
\centering
    \includegraphics[width=\columnwidth]{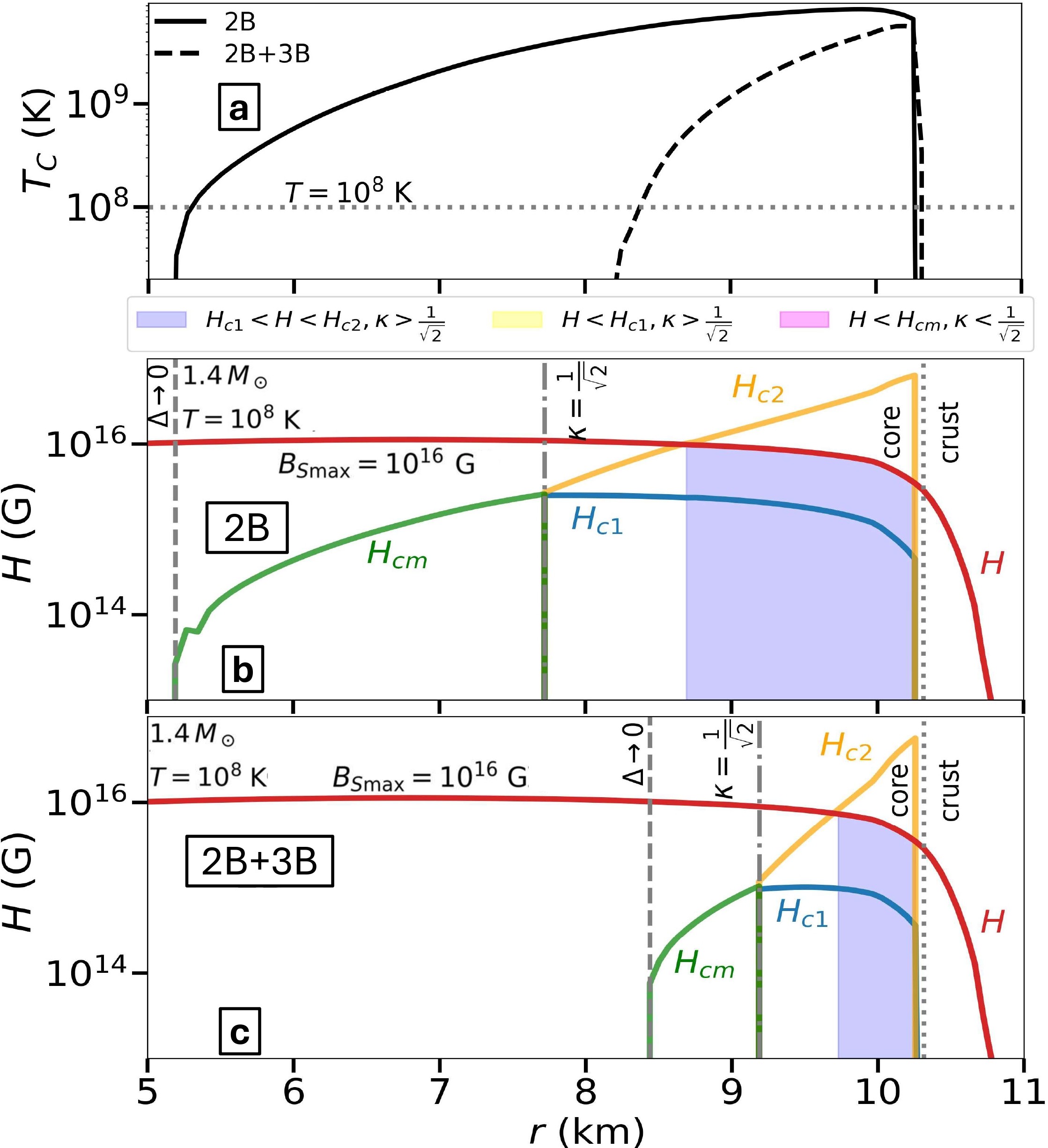}
    \caption{Panel (a) shows the critical temperature $T_c$; panels (b) and (c) show the critical fields $H_{c1}$, $H_{c2}$, and $H_{cm}$ for 2B and 2B+3B force models with the gaps defined via Eqs.~\eqref{eq:D_2B} and \eqref{eq:D_3B} respectively, along with the $H$-field derived from the Einstein-Maxwell
solutions, as a function of equatorial-plane radial coordinate (plane perpendicular to the external field). The NS model has $M = 1.4\,M_\odot$, $R=11$ km, $T = 10^8$ K, and maximal toroidal magnetic field $B_{S\text{max}} = 10^{16}$ G and was constructed using the BBB EoS. Two vertical lines indicate the transition points: one at $\kappa = 1/\sqrt{2}$ marking the boundary between type-I and type-II superconductivity, and another where the superconducting gap vanishes ($\Delta \to 0$), indicating a transition to the normal (unpaired) state. Shaded regions in this and the following similar figures highlight superconducting phases. For type-II superconductors, these include the Meissner state ($H < H_{c1}$) and the flux-tube array state ($H_{c1} < H < H_{c2}$); for type-I superconductors, they correspond to the Meissner or layered-domain state with $H < H_{cm}$ (see text). In the present figure, only flux-tube phases occur for the given magnetic-field ($H$) profile.}
\label{fig:4}
\end{figure}

\begin{figure}
\centering
  \includegraphics[scale=0.45]{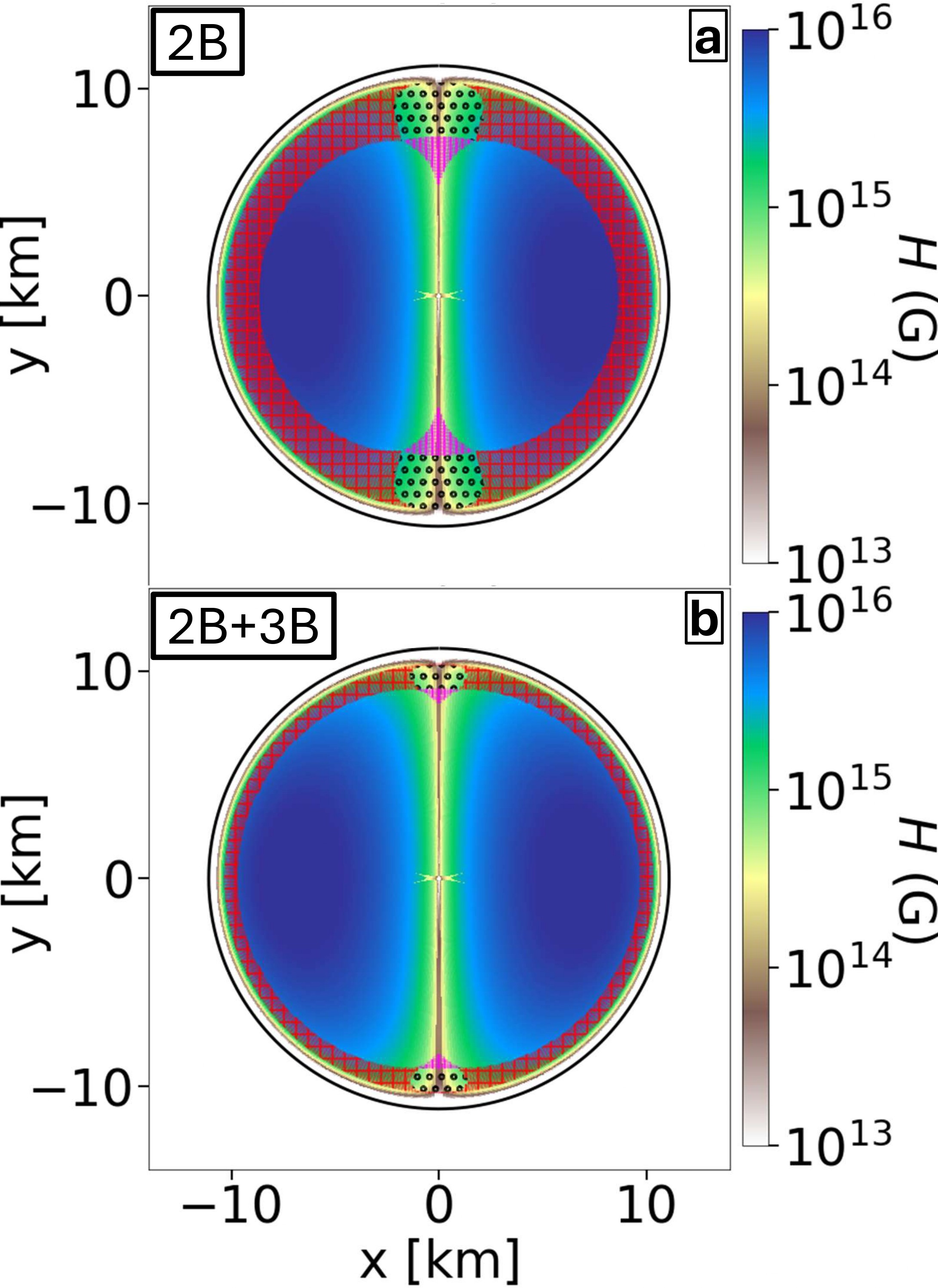}
  \caption{ Color maps of the distribution of the toroidal $H$-field in the $x$–$y$ plane, where the magnetic axis is aligned along the $y$-direction in Cartesian coordinates. Superconducting regions for NS models using the BBB EoS with $M = 1.4\,M_\odot$, $T = 10^8$\,K, $R=11$ km, and toroidal $B_{S\text{max}} = 10^{16}$\,G are shown for 2B and 2B+3B nucleonic force models in panels (a) and (b), respectively. The superconducting phases are identified as follows: red crosses mark the type-II region where $\kappa > 1/\sqrt{2}$ and $H_{c1} < H < H_{c2}$, indicating the presence of a flux tube lattice; black dots represent the Meissner state within a type-II region, characterized by $\kappa > 1/\sqrt{2}$ and $H < H_{c1}$; and magenta vertical hatching denotes the Meissner or layered-domain state in the type-I region where $\kappa < 1/\sqrt{2}$ and $H < H_{cm}$. Note that, if Meissner expulsion of the flux is fast, the magnetic induction vanishes in type-II regions when $H\le H_{c1}$. The same applies to type-I regions when $H\le H_{cm}$. However, the slow field-expulsion timescales \cite{Baym1969} imply that flux tubes in type-II regions and normal domains in type-I regions may still exist. Unshaded regions indicate nonsuperconducting zones where $H > H_{cm}$ in the type-I case or $H > H_{c2}$ in the type-II case.}
\label{fig:5}
\end{figure}

Here, we explore the width of superconducting regions based on upper and lower bounds of the gap, corresponding to models neglecting and including 3B forces, respectively. These represent theoretical limits within which the actual gap is expected to lie. To model the stellar structure, we adopt the Baldo, Bombaci, and Burgio (BBB) EoS~\cite{Baldo1997} for cold nucleonic matter from the {\sc CompOSE} repository \cite{CompOSE2022EPJA,Dexheimer2022}. Since the available tables of the results of Ref.~\cite{Baldo1997} do not correspond to cold $\beta$-equilibrated matter, the proton fraction $Y_p$ is adopted from Ref.~\cite{Sharma2015}, which employs a microscopic many-body framework of a similar type but for $\beta$-equilibrated matter. This EoS is derived using BHF formalism with consistent many-body forces, corresponding to the same microscopic treatment used to compute the gap. The EoS reproduces empirical nuclear structure properties at (sub)nuclear densities \cite{Taranto2013} and is consistent with constraints from GW170817 \cite{Abbott2018} and NS cooling observations \cite{Fortin2018}. Although theoretically well-motivated and explicitly incorporating many-body correlations, this EoS yields a maximum mass of $\sim1.94\,M_\odot$. While this lies marginally below the observational lower bound of $2.14^{+0.10}_{-0.09}\,M_\odot$ inferred from PSR J0740+6620 \cite{Cromartie2019}, it remains useful for exploring the sensitivity of NS properties to the underlying nuclear microphysics, when dealing with the density range relevant to the present study.

As is well known, the stellar structure of NSs can be decoupled from their thermal evolution, and finite-temperature effects can be neglected for cold NSs. Thus, we use zero-temperature EoS to compute the stellar structure, which is appropriate for studying superconductivity in cold, $T \ll T_{F,n/p}$ regime, where $T_{F,n/p}$ are the Fermi temperatures of neutrons and protons. Once the equilibrium stellar configuration is computed via the \texttt{XNS} solver
with zero-temperature EoS, the temperature- and density-dependent superconducting gap can be mapped onto the stellar structure for any given temperature below $T_c$. Note 
that it is guaranteed that the regions where $T \leq T_c$ are superconducting, as long as the local magnetic field is not strong enough to suppress superconductivity.

Fig.~\ref{fig:4}a shows $T_c$, calculated from Eq.~\eqref{eq:tc} using $\Delta(0)$ obtained using 2B and 2B+3B interactions via Eqs. \eqref{eq:D_2B} and \eqref{eq:gap0}, as a function of the equatorial radial coordinate ($\theta=90^{\circ}$) for NS models with $M = 1.4\,M_\odot$, $R=11$ km, $T = 10^8$ K, corresponding to a central density of $10^{15}$ g cm$^{-3}$. Note the steep drop in $T_c$ at higher densities towards the density threshold beyond which proton superconductivity vanishes, while $T_c$ is truncated at the crust–core boundary, where protons are bound in nuclei. For the 2B+3B model, the  region in which $T < T_c$ extends over a narrower range ($8.2 \lesssim r \lesssim 10.2~\mathrm{km}$) compared to the 2B model ($5.2 \lesssim r \lesssim 10.2~\mathrm{km}$). This reduction is expected, since the three-body force introduces additional repulsion at high densities, suppressing proton pairing ($\Delta \rightarrow 0$) and thereby limiting the size of the potential superconducting region. The size of the superconducting region can be further modified by the magnetic field, as discussed below.

In the following step, we explore the possible presence of type-I and type-II superconductivity within the $T < T_c$ regions. We compare the critical fields with the stellar toroidal magnetic field obtained from Einstein-Maxwell solutions for the same NS with $B_{S\max} = 10^{16}\,\mathrm{G}$. Figs~\ref{fig:4}b and \ref{fig:4}c illustrate the critical fields $H_{c1}$, $H_{c2}$, and $H_{cm}$ [computed using Eqs.~\eqref{eq:Hc1} and \eqref{eq:Hcm}] as functions of the equatorial radial coordinate for the 2B and 2B+3B models, respectively. In each panel, the shaded region corresponds to the superconducting regime. In this case, we find only the type-II flux tube state, which extends within the range $8.7\lesssim r\lesssim 10.2$~km and $9.7\lesssim r\lesssim 10.3$~km for 2B and 2B+3B models, respectively. However, type-I superconducting regions arise for other values of the field $H$ and angle $\theta$, as discussed below.

We now extend the analysis from one-dimensional (1D) radial profiles to the two-dimensional (2D) axisymmetric representation. Fig.~\ref{fig:5} displays the 2D structure of the superconducting regions corresponding to the same NS model used in Fig.~\ref{fig:4}. The domains of type-I, type-II, and non-superconducting matter are distinguished by different hatching patterns, as detailed in the figure caption.
The main trends observed in the 1D profiles are also present in the 2D results; for example, the suppression of superconductivity due to the inclusion of three-body forces remains evident. However, by covering the full range of polar angles $\theta$, the 2D calculations reveal the true topology of superconducting regions within the same NS model -- information that is not accessible in the 1D profiles limited to the equatorial plane. While this example considers a purely toroidal magnetic-field configuration, we will see below that the shape and size of superconducting domains are sensitive to  magnetic-field geometry.

In Sec.~\ref{sec:tor}, we compare the results based on the microscopic BBB EoS described above with those obtained using the RMF EoS based on DDME2 parametrization~\cite{gppva,Lala2005}, along with the effect of chosen field geometry.
The DDME2 model, available through the {\sc CompOSE} repository, provides not only the pressure as a function of energy density but also additional thermodynamic quantities, including the proton fraction $Y_p$. It is stiffer than the BBB EoS used above and is based on a phenomenological framework calibrated to reproduce properties of finite nuclei, as well as nuclear and neutron matter. The model predicts a maximum neutron-star mass of $\sim 2.48\,M_\odot$, comfortably exceeding the mass of the heaviest precisely measured pulsar, PSR J0740+6620~\cite{Cromartie2019}.

Our comparison, presented below, illustrates how the predicted size of superconducting regions varies across EoS with different stiffness, reflecting uncertainties in the underlying nuclear microphysics. Throughout this comparison and in the remainder of this work, we adopt the same proton-pairing model based on 2B+3B forces.
This is justified by the widely used ``decoupling approximation", which assumes that the calculation of the pairing gap can be performed independently of the EoS, provided the single-particle spectrum entering the gap equation is properly extracted from the EoS. In the case of BHF calculations, this corresponds to the non-relativistic spectrum, with momentum-dependent real part of the self-energy, whereas in the RMF case this corresponds to a relativistic spectrum with the effective Dirac mass determined by the mean value of the $\sigma$-field.

\subsection{Varying magnetic field structure}
\label{sec:tor}

\begin{figure}[t!]
\centering
    \includegraphics[width=\columnwidth]{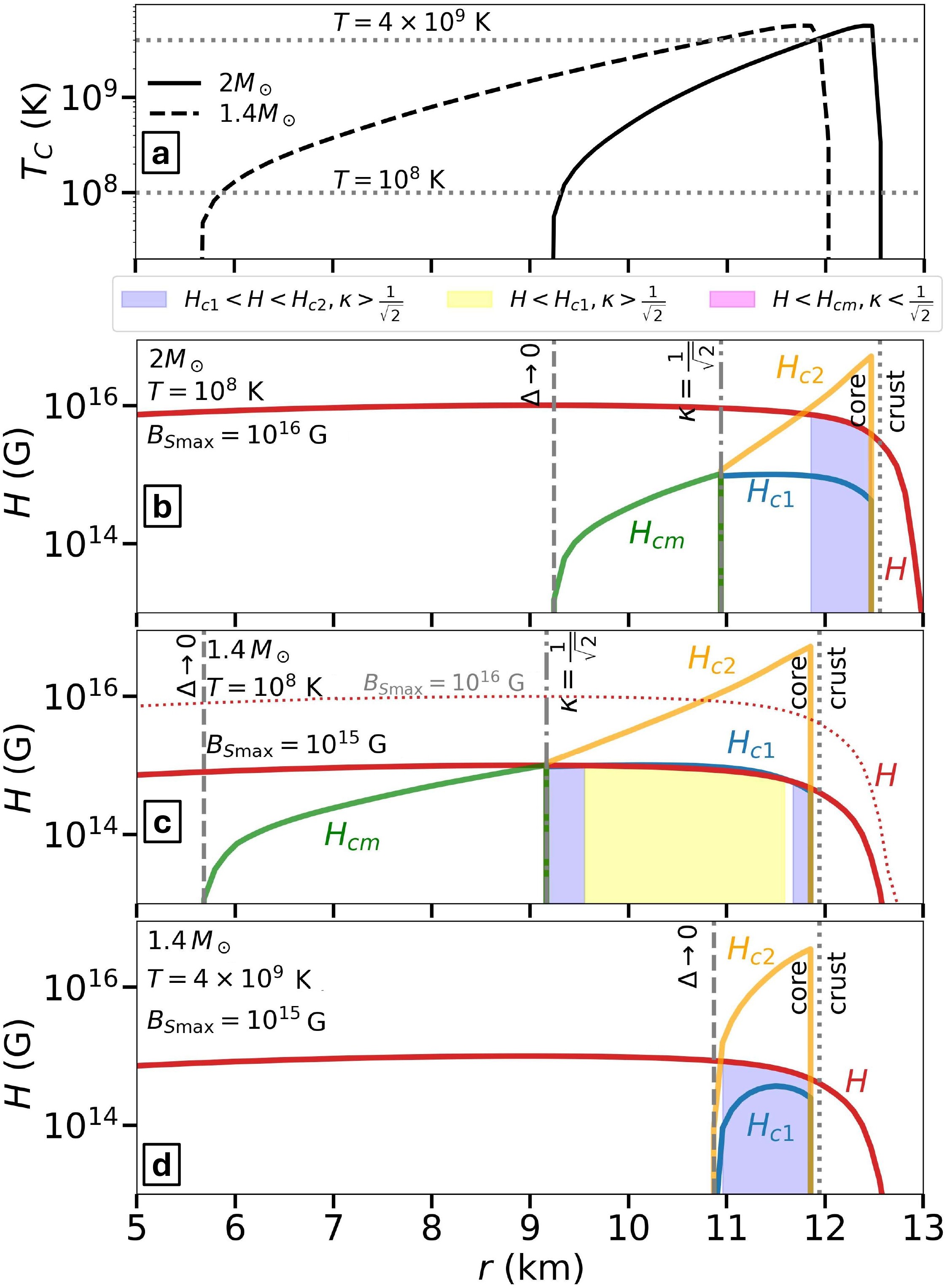}
    \caption{
     The dependence of (a) the critical temperature, and (b, c, d) the critical magnetic fields: $H_{c1}$, $H_{c2}$, and $H_{cm}$, along with the local magnetic field $H$, on the equatorial radial coordinate for NS models with masses $1.4\,M_\odot$ and $2\,M_\odot$ for the DDME2 EoS, with varying temperature, and toroidal magnetic field strengths. The chosen $M$, $T$, and maximum toroidal field strength $B_{S\text{max}}$ are indicated. The shaded regions denote superconducting zones, as detailed in Fig.~\ref{fig:4}.}
\label{fig:6}
\end{figure}

Next, we investigate the influence of various magnetic field configurations -- toroidal, poloidal, and twisted-torus -- on the distribution and size of superconducting regions within the magnetars with DDME2 EoS. For completeness, a concise summary of the magnetic stability criteria and additional relevant references are provided in the Appendix.
We start with models of magnetars with an internal {\it toroidal magnetic field}.  For the 1D NS configuration with $M=1.4\,M_\odot$ and $T = 10^8$ K -- corresponding to a central density of $5.5 \times 10^{14}$~g cm$^{-3}$, as shown in Fig.~\ref{fig:6}a -- the $T<T_c$ region indicating possible superconducting regime extends over $5.5 \lesssim r \lesssim 12$~km, provided the local magnetic field does not exceed the local value of the critical field. This layer is considerably thicker than in NSs modeled with the BBB EoS, because the stiffer DDME2 EoS reaches the same $M$ at a lower central density.  As a result, the superconducting phase extends further inside the core.

\begin{figure}[hbt]
\centering
    \includegraphics[width=\columnwidth]{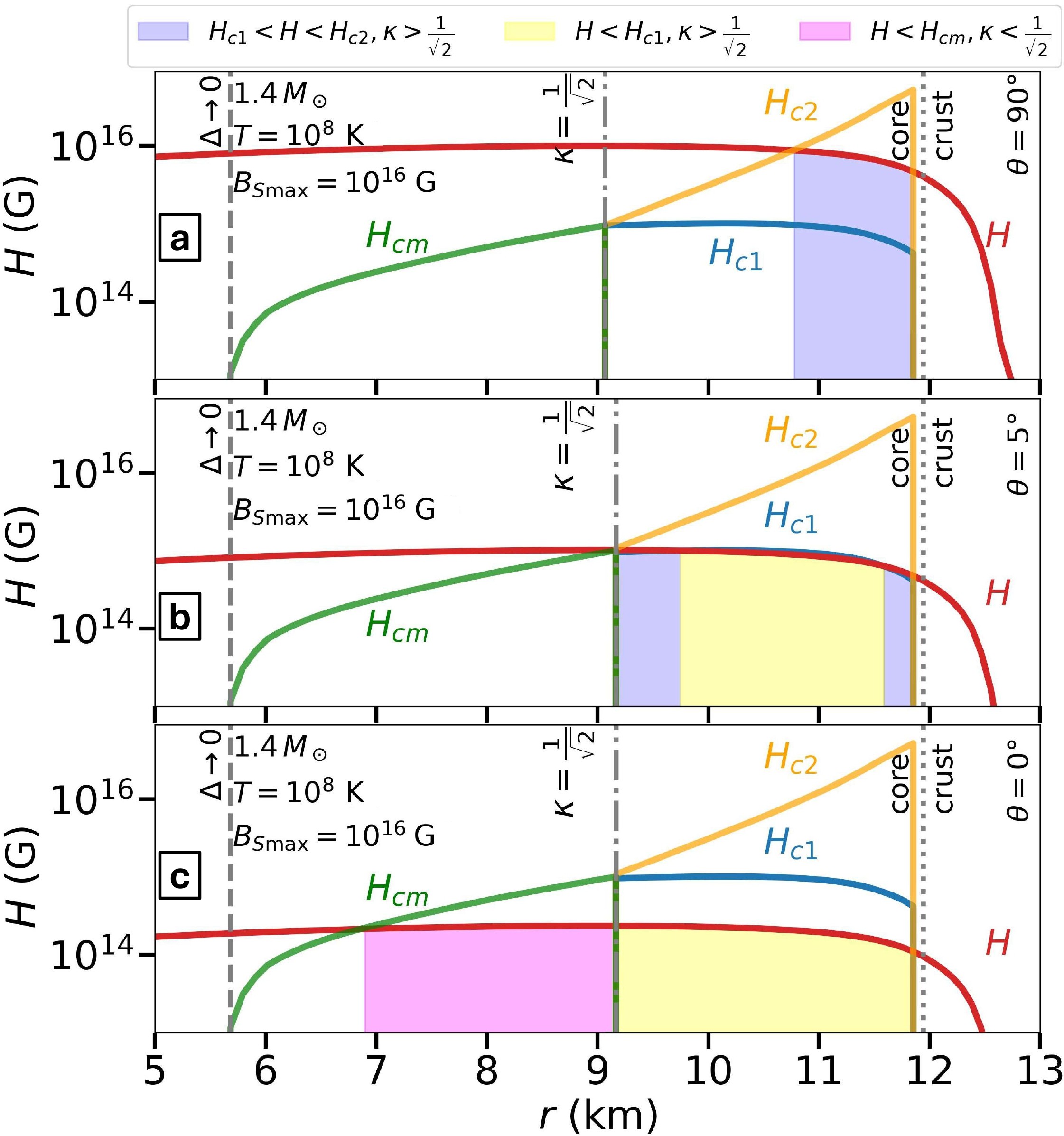}
    \caption{
    The radial profiles of various magnetic fields and
    ranges occupied by various superconducting regions 
      for $1.4\,M_\odot$ NS constructed using the DDME2 EoS, for a toroidal field with $B_{S\text{max}} = 10^{16}$ G and $T = 10^8$ K, and for three polar angles $\theta =$ 
       $90^\circ$~(a), $5^\circ$~(b), and $0^\circ$~(c), with the limiting values corresponding to the equator and the pole, respectively. The
      conventions are the same as in Fig.~\ref{fig:4}.
    }
\label{fig:7}
\end{figure}

\begin{figure}
\centering
  \includegraphics[scale=0.82]{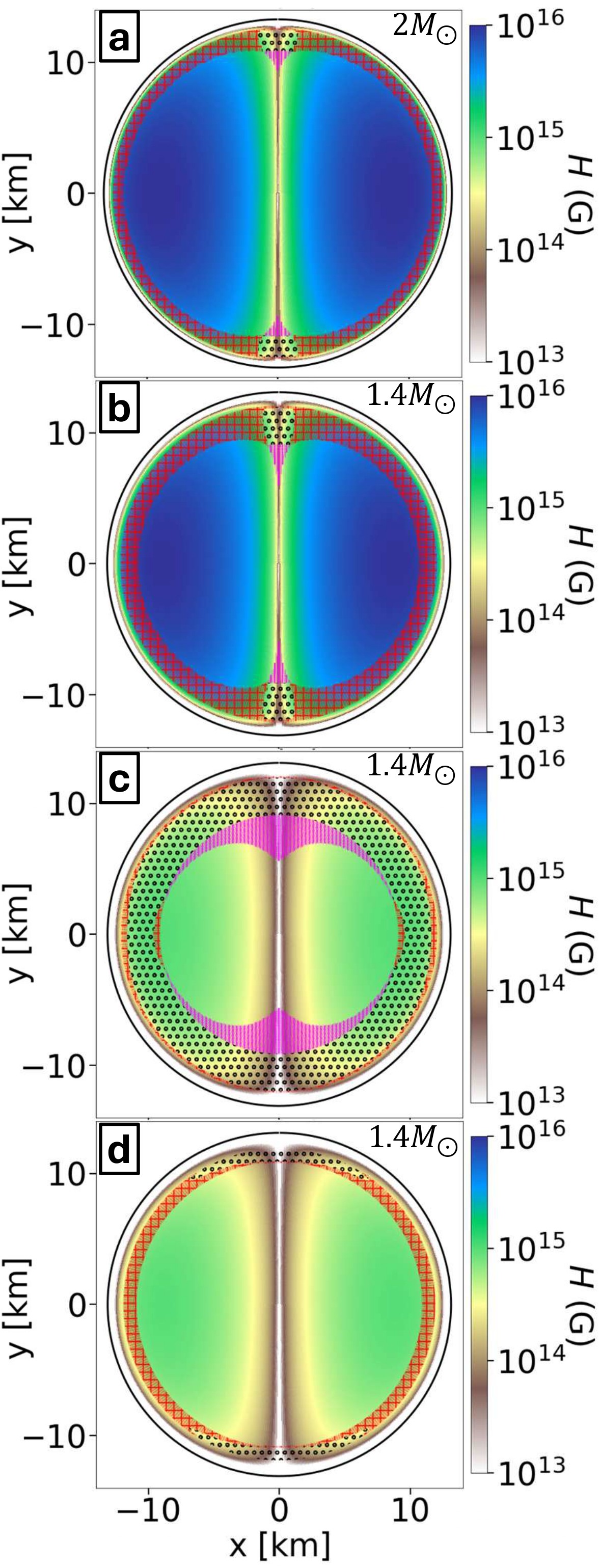}
    \caption{Color maps of the distribution of the toroidal $H$-field in the $x$–$y$ plane. Panels correspond to NS models constructed using the DDME2 EoS for the following parameter sets: (a) $M=2 M_{\odot}, B_{S \max }=10^{16} \mathrm{G}, T=10^8 \mathrm{ K}$; (b) $M=1.4 M_{\odot}, B_{S \max }=10^{16} \mathrm{G}, T=10^8 \mathrm{ K}$; (c) $M=1.4 M_{\odot}, B_{S \max }=10^{15} \mathrm{G}, T=10^8 \mathrm{ K}$;
(d) $M=1.4 M_{\odot}, B_{S \max }=10^{15} \mathrm{G}, T=4 \times 10^9 \mathrm{ K}$.
The hatching conventions are the same as in Fig.~\ref{fig:5}.
}
\label{fig:8}
\end{figure}

Fig.~\ref{fig:6}a shows $T_c(r)$ function for two stellar models with $M=2\,M_\odot$, $R=13.2$ km and $1.4\,M_\odot$, $R=13$ km corresponding to central densities of $7.7 \times 10^{14}$ and $5.5 \times 10^{14}$ g cm$^{-3}$, respectively. For $T=10^8$ K, the $2\,M_\odot$ model exhibits a narrower size of the superconducting region satisfying $T<T_c$ ($9.2 \lesssim r \lesssim 12.5$ km) compared to that of the $1.4\,M_\odot$ model ($5.5 \lesssim r \lesssim 12$ km). This is expected, as the low-density region that supports the paired state is more extended in the $1.4\,M_\odot$ case.
The horizontal dotted lines correspond to constant core temperatures of $T = 10^8$~K and 
$T = 4 \times 10^9$~K, illustrating how the size of the superconducting region expands or shrinks as the temperature decreases or increases, respectively.

The panels b, c, d of Fig.~\ref{fig:6} display the critical fields $H_{c1}$, $H_{c2}$, and $H_{cm}$ as functions of the equatorial radial coordinate. Each panel corresponds to a different combination of stellar mass ($1.4\,M_{\odot}$ or $2\,M_{\odot}$), temperature ($T = 10^8$~{K} or $T = 4 \times 10^9$ ~K), and maximum stellar magnetic field ($B_{S\mathrm{max}} = 10^{15}$~G or $10^{16}$~G). Comparing corresponding panels reveals how variations in these parameters control the actual superconducting phases and their spatial size, i.e., distribution of type-I and type-II superconducting regions. In each plot, the shaded areas indicate the regions where either type-I or type-II superconductivity occurs in analogy to the conventions already introduced in Fig.~\ref{fig:4}.

\begin{figure}[ht!]
\centering
    \includegraphics[scale=0.6]{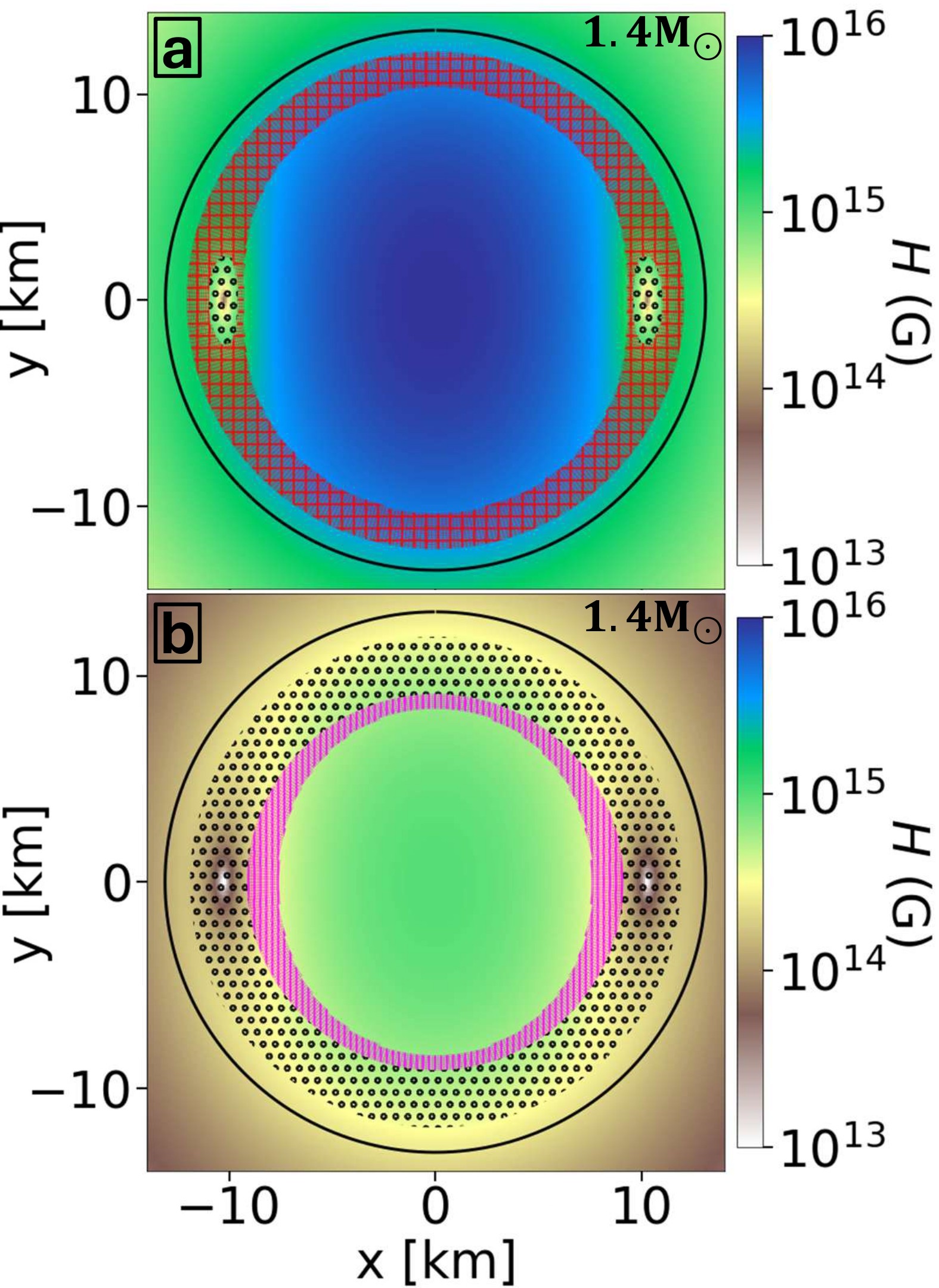}
\caption{Color maps of the poloidal $H$-field in the $x$–$y$ plane, with the magnetic axis aligned along the $y$-direction of the Cartesian coordinate system, are shown for a star with $M = 1.4\,M_\odot$ and $T = 10^8$~K, constructed using the DDME2 equation of state. Panel (a) corresponds to $B_{S\text{max}} = 10^{16}$~G, while panel (b) corresponds to $B_{S\text{max}} = 10^{15}$~G. The various superconducting regions are identified as in Fig.~\ref{fig:5}.} 
\label{fig:10}
\end{figure}

\begin{figure}[t!]
\centering
    \includegraphics[scale=0.48]{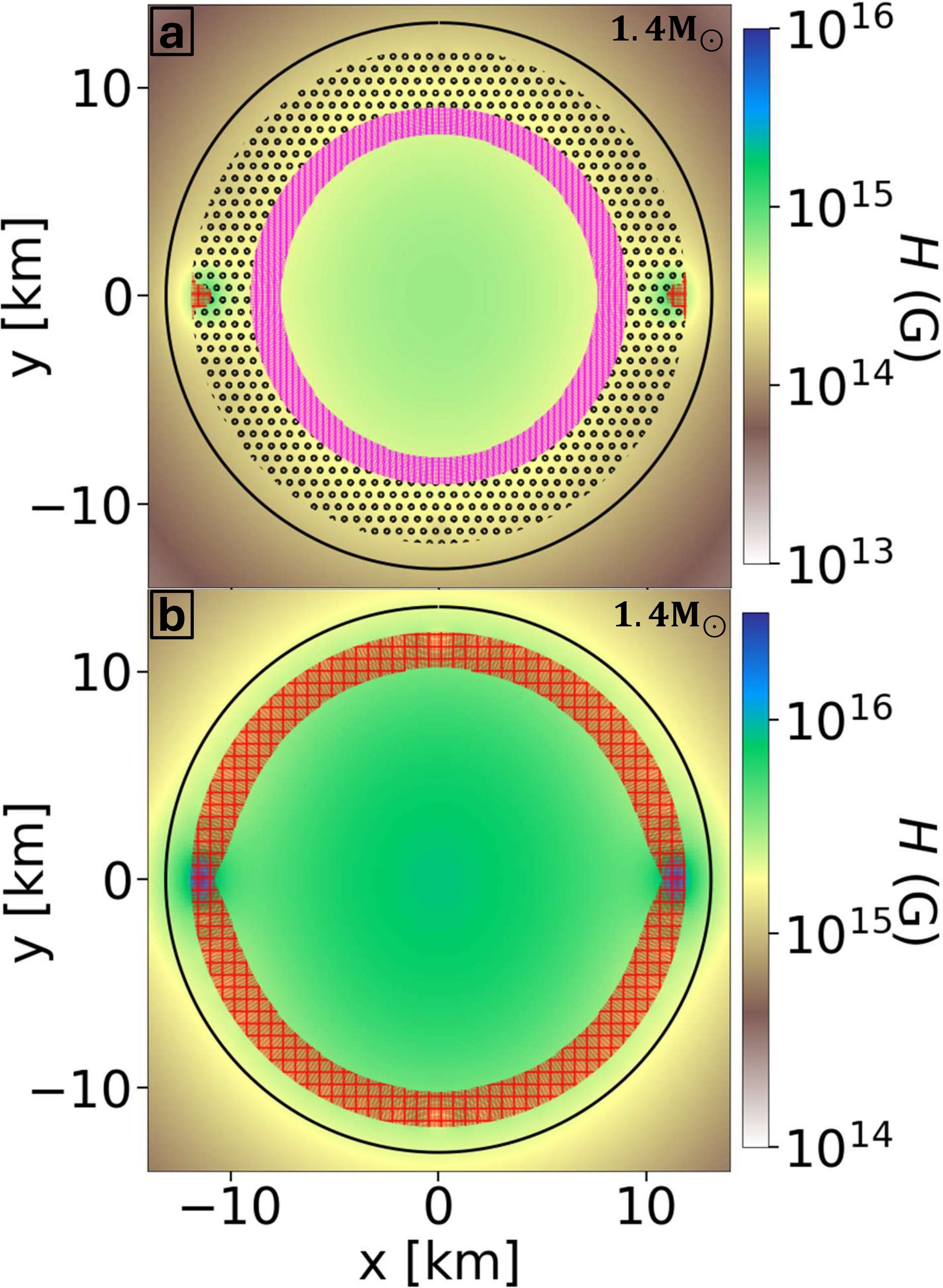}
\caption{The distribution of the poloidal $H$-field in the $x$–$y$ plane, where the magnetic axis is aligned along the $y$-direction in Cartesian coordinates. The figure corresponds to NSs with $M = 1.4\,M_\odot$, $T = 10^8$ K, modeled using DDME2 EoS for twisted-torus configurations with $B_{S\text{max}}^{T+P} = (a)\, 10^{15}$~G, and (b) $2\times 10^{16}$~G. The various superconducting regions are identified following the same scheme as in Fig. \ref{fig:5}.} 
\label{fig:tt2}
\end{figure}

We observe in Fig.~\ref{fig:6} several notable trends in the structure of the superconducting
regions:
\begin{itemize}
  
\item  As $M$ increases, the size of the superconducting region becomes thinner (see Figs \ref{fig:6}b and \ref{fig:6}c).
  
\item In NSs with very strong $H$-fields, where $H> H_{c2}$, i.e., the corresponding
  curve  fails to intersect any of the critical fields, superconductivity is absent (not shown here; however, the non-superconducting regions are visible in Figs.~\ref{fig:5}, \ref{fig:8}, \ref{fig:10} and \ref{fig:tt2}). If $H$ crosses only $H_{c2}$, a superconducting phase with flux tubes appears adjacent to the crust-core boundary.
  
\item Conversely, if $H$ intersects only $H_{c1}$, there are two flux tube regions sandwiching a type-II region where flux tubes are energetically unfavorable - a structure characteristic of toroidal fields. When $H$ lies below $H_{c1}$, the Meissner phase becomes energetically favorable.
  
\item Increasing $T$ (Fig.~\ref{fig:6}d) reduces the spatial size of the superconducting phases, as the pairing gap $\Delta(T)$ decreases with $T$. However, the ordering of the type-I and type-II regions remains unchanged; varying $T$ simply stretches or shrinks the overall superconducting volume.
  
\item  The behavior of the type-I superconducting region is relatively simple. When the local magnetic field is sufficiently weak, a Meissner or layered-domain phase begins to form at the $r$ where $\kappa = 1/\sqrt{2}$ and extends inward up to $r$ where $\Delta\rightarrow0$.
  \end{itemize}

  For clarity and illustration, Fig.~\ref{fig:7} shows how the size and states of the superconducting region change from the equator ($\theta = 90^\circ$) to the pole ($\theta = 0^\circ$) for an NS with $M = 1.4\,M_\odot$, $T = 10^8$~K, and $B_{S\text{max}} = 10^{16}$~G. This pattern results from combining the radial profiles of the microscopic parameters with the axially symmetric toroidal magnetic field. The figure indicates that the flux tube phase is suppressed near the poles, whereas at the equator, a robust flux tube region emerges adjacent to the crust-core boundary.
 
  We now turn to the 2D axisymmetric models and the distribution of type-I, type-II, and non-superconducting regions in space. Fig.~\ref{fig:8} shows these phases
  for models with $M = 1.4\,M_\odot$ and $2\,M_\odot$, $T = 10^8\,\mathrm{K}$ and $4 \times 10^9\,\mathrm{K}$, and $B_{S\text{max}} = 10^{15}$ and $10^{16}\,\mathrm{G}$. The qualitative features seen in the 1D results (Figs.~\ref{fig:6} and \ref{fig:7}) persist along the equator and polar directions. In particular, in the polar regions, where the toroidal $H$-field is significantly weaker, the superconducting phase topology changes significantly. For example, in the $2\,M_\odot$ model (a), a narrow type-II flux tube layer forms near the equator’s crust-core boundary, expanding toward the poles where a type-II Meissner region becomes dominant; a type-I zone arises at higher densities. In contrast, the $1.4\,M_\odot$ configuration (b) features more extended superconducting regions with similar phase topology. This behavior arises from the expansion of the superconducting density range in lower mass stars. A comparison of two $1.4\,M_\odot$ models with differing field strengths, i.e., (b) vs. (c), reveals that stronger fields carve out a torus-like non-superconducting zone around the region where $H$ is maximal. The type-II topology also changes: the  Meissner region is extended throughout the outer core at lower $H$, but it exists only around the polar regions at higher $H$. Likewise, comparing two $1.4\,M_\odot$ models at different temperatures, i.e., panel (c) vs. (d), confirms that higher $T$ suppresses superconductivity, narrowing the allowed region. Furthermore, the disappearance of the type-I region in panel (d) can be attributed to the fact that the decrease in $\rho_s$ [according to Eq.~\eqref{eq:ns}] with increasing temperature implies larger $\lambda$, and therefore $\kappa$. 
  Additionally, variations in the superconducting geometry are evident -- most notably in Fig.~\ref{fig:8}c -- where an outer prolate-shaped, inner torus-shaped (following magnetic geometry) type-I region extends prominently toward the poles. A consistent trend across all configurations is the absence of superconductivity in the central core. This results from the extinction of the $S$-wave superconducting phase at densities exceeding $\rho > 4.3 \times 10^{14}$g\,cm$^{-3}$. For NS models employing the DDME2 EoS, superconductivity persists down to the center only in very low-mass stars; in canonical or high-mass cases, the core becomes nonsuperconducting due to the vanishing of attractive proton-proton interaction in the $S$-wave channel at high densities.

\begin{figure}[t!]
\centering
    \includegraphics[width=\columnwidth]{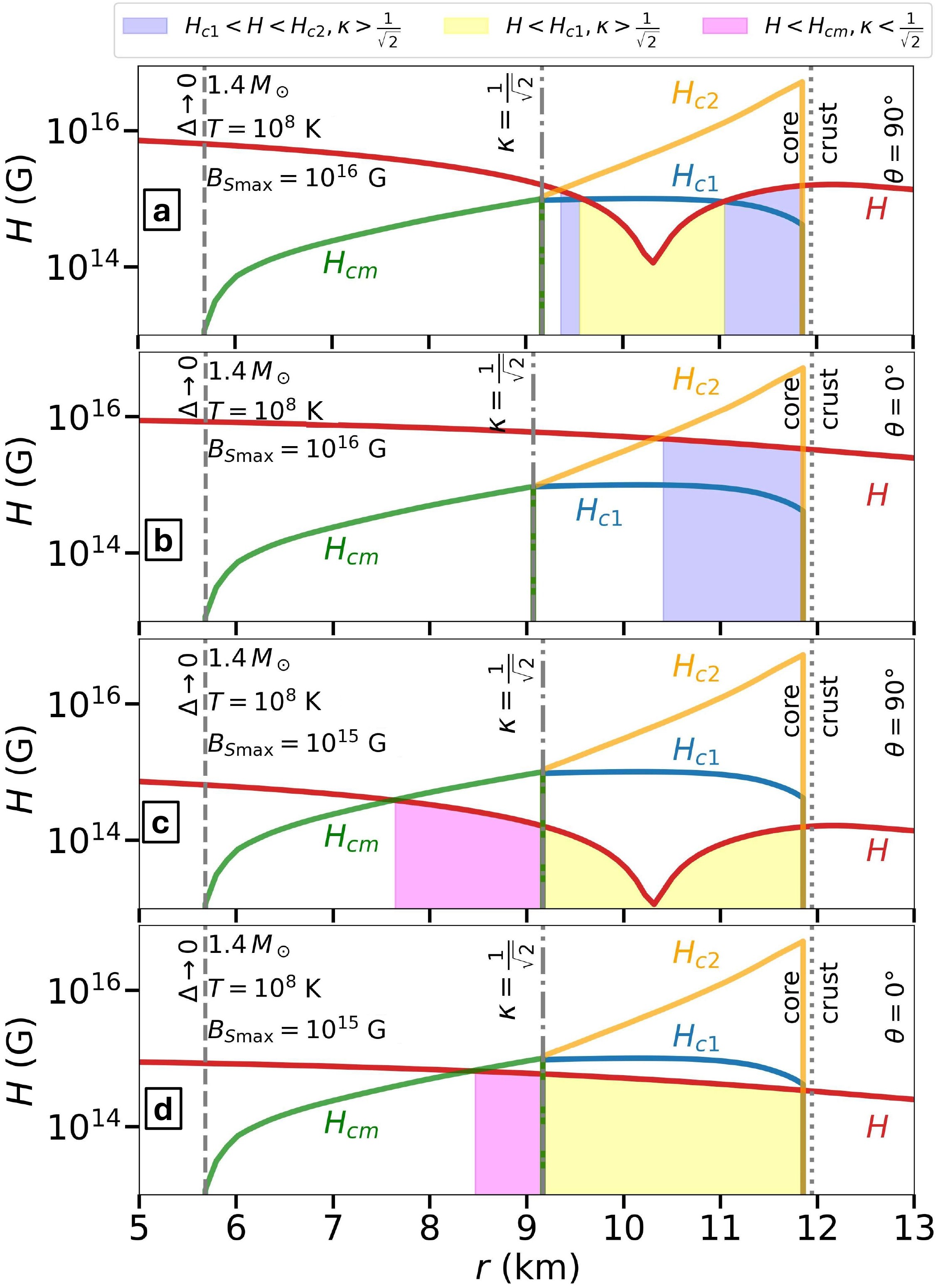}
    \caption{
      The radial profiles of various magnetic fields along 
      the equatorial ($\theta = 90^\circ$) and the polar directions ($\theta = 0^\circ$) and ranges occupied by various superconducting regions for an NS with $M = 1.4\,M_\odot$ and $T = 10^8$ K 
      constructed using DDME2 EoS. Panels  (a) and (b) correspond to the poloidal field with $B_{S\text{max}} = 10^{16}$~G, whereas panels (c) and (d) show the same but for $B_{S\text{max}} = 10^{15}$~G. The
      conventions are the same as Fig.~\ref{fig:4}.}
\label{fig:9}
\end{figure}

Next, we examine magnetar models with predominantly {\it poloidal magnetic fields}. For these configurations, the superconducting regions at the pole and equator (specified by $\theta$) are shown in one dimension in Fig.~\ref{fig:9} for NS models based on the DDME2 EoS with $M = 1.4\,M_\odot$, $R=13$ km, $T = 10^8$~K, and two values of the maximum stellar magnetic field, $B_{S\text{max}} = 10^{16}$~G and $10^{15}$~G. The corresponding 2D distributions are displayed in Fig.~\ref{fig:10}. Both figures consistently show that superconductivity is absent in a slightly prolate inner core region.

The weakening of the field strength near the equator in terms of the prominent dip at $r \sim 10$~km is apparent. For strong fields ($B_{S\text{max}} = 10^{16}$~G), this leads to a transition from the flux-tube phase to the type-II Meissner state (Fig.~\ref{fig:9}a). For weaker fields ($B_{S\text{max}} = 10^{15}$~G), the flux-tube phase is entirely replaced by the type-II Meissner phase, accompanied by the appearance of a type-I region at higher densities (Fig.~\ref{fig:9}c).

At the poles, a robust flux-tube phase is present adjacent to the crust–core boundary for strong fields (Fig.~\ref{fig:9}b), while for weaker fields it transforms into a combination of type-II Meissner and type-I phases at higher density. Overall, the differences of poloidal configuration from the toroidal configuration can be traced by visualizing the different $r$ and $\theta$ dependence of the corresponding magnetic fields (Fig.~\ref{fig:7} vs. Fig.~\ref{fig:9}).  

The 2D maps (Fig.~\ref{fig:10}) clearly illustrate that the depletion of the poloidal field strength near the equator triggers the transition from the flux-tube phase to the type-II Meissner state, where $H$ reaches a local minimum (see Fig.~\ref{fig:1}b) for strong fields. This behavior contrasts with the toroidal case, in which superconductivity is suppressed within a torus-shaped region with maximum magnetic field (Fig.~\ref{fig:1}a). Apart from this feature, the phase distribution remains nearly spherically symmetric with slight prolateness, reflecting the poloidal symmetry of the $H$-field.

\begin{figure}[t!]
\centering
    \includegraphics[width=\columnwidth]{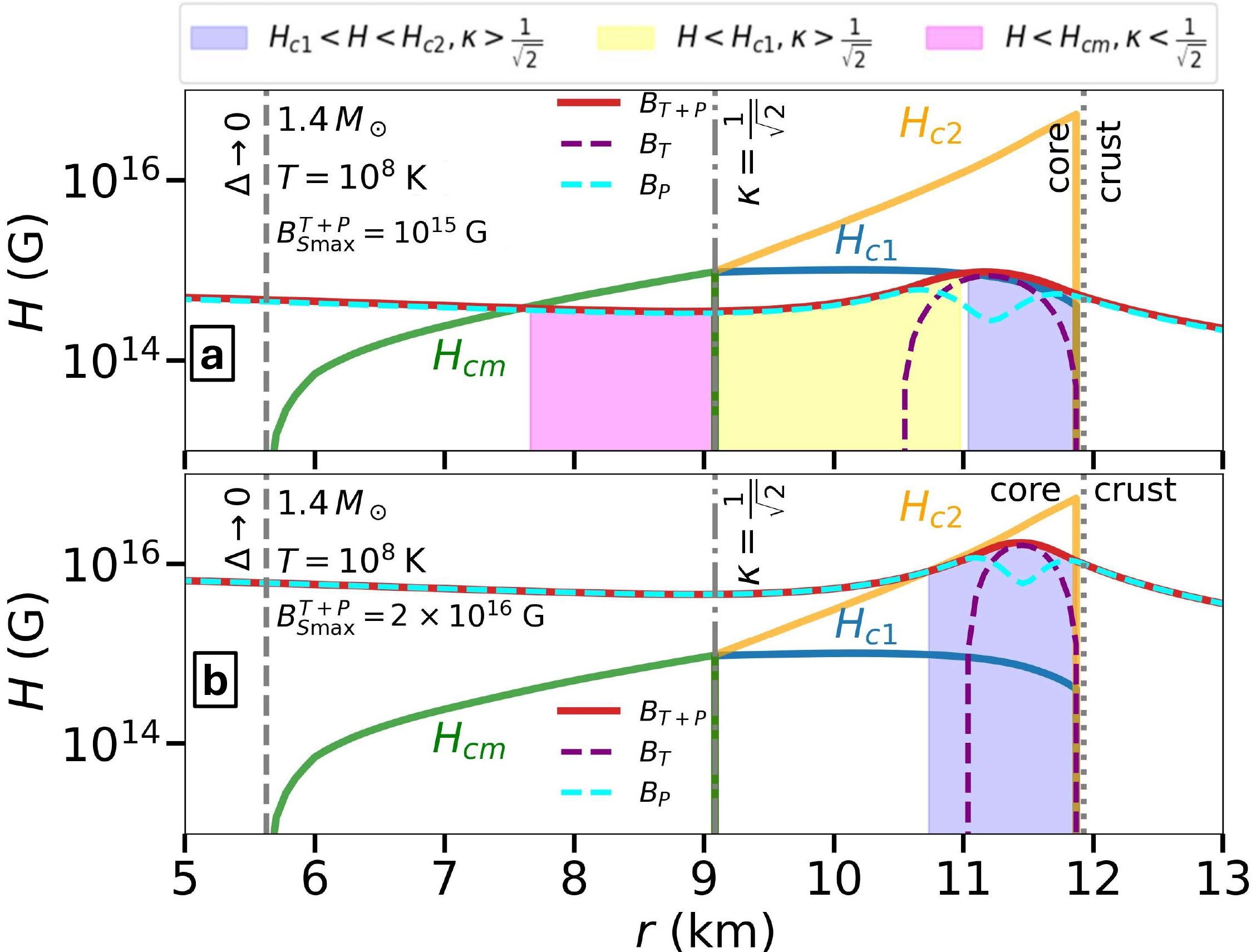}
    \caption{
      The radial profiles of various magnetic fields along the equatorial direction
      and ranges occupied by various superconducting regions  for an NSs with $M = 1.4\,M_\odot$ and $T = 10^8$ K, modeled using DDME2 EoS assuming twisted-torus configuration. Panel (a) corresponds to $B_{S\text{max}}^{T+P} = 10^{15}$~G, while panel (b) -- to  $2\times 10^{16}$~G. The
      conventions are the same as Fig.~\ref{fig:4}.}
\label{fig:tt1}
\end{figure}

Motivated by the stability analysis based on magnetic configurations given in the Appendix, in what follows we will primarily focus on toroidal magnetic field configurations to explore the GW emission from the NSs. Nevertheless, before doing so, it is instructive to briefly illustrate the structure of {\it twisted-torus} equilibria, which consist of a large-scale poloidal field accompanied by a localized toroidal component. For this purpose, we consider two cases, where we specify
the parameters used in the \texttt{XNS} solver as follows: 
\begin{description}
\item  (a) $B_{S\text{max}}^{T+P} = 10^{15}$, $B_{S\text{max}}^{T} =9.5\times 10^{14}$, $B_{S\text{max}}^{P} = 6\times10^{14}$~G;
\item  (b) $B_{S\text{max}}^{T+P} = 2.2\times10^{16}$, $B_{S\text{max}}^{T} = 2\times10^{16}$, $B_{S\text{max}}^{P} = 10^{16}$~G.
\end{description}
To estimate the magnetic energy contributions in comparison with the stability criteria introduced in Appendix, we express the toroidal and poloidal energies in terms of the  field strengths and volume–profile factors:
\begin{eqnarray}
E_{\rm tor} \propto f_{\rm tor}\, (B_{S\text{max}}^{T})^2, \quad
E_{\rm pol} \propto f_{\rm pol}\, (B_{S\text{max}}^{P})^2,
\end{eqnarray}
where $f_{\rm tor}$ and $f_{\rm pol}$ account for the distribution of magnetic energy throughout the stellar volume. The total magnetic energy is then $E_{\rm mag} = E_{\rm pol} + E_{\rm tor}$, with the poloidal energy fraction defined as $\Lambda_B \equiv E_{\rm pol}/E_{\rm mag}$, as in Eq.~\eqref{Elimit}.

For case (a), assuming a twisted-torus configuration with the toroidal field confined to a narrow internal region, the filling factor ratio is typically $f_{\rm tor}/f_{\rm pol} \simeq 0.015$. This yields
\begin{equation}
\frac{E_{\rm tor}}{E_{\rm pol}} = 0.038, \quad
\frac{E_{\rm tor}}{E_{\rm mag}} \approx 0.037, \quad
\Lambda_B \approx 0.95.
\end{equation}
Similarly, for case (b) with the same filling-factor ratio, we obtain
\begin{equation}
\frac{E_{\rm tor}}{E_{\rm pol}} = 0.04, \quad
\frac{E_{\rm tor}}{E_{\rm mag}} \approx 0.035, \quad
\Lambda_B \approx 0.96.
\end{equation}
These values lie near the upper limit of stability for the poloidal energy fraction $\Lambda_B$ found in stably stratified twisted–torus equilibria~\cite{MSM2015}.

Figures~\ref{fig:tt1} and \ref{fig:tt2} show the 1D superconducting profiles and 2D topologies for NS models, respectively, computed with the DDME2 EoS. The overall superconducting topology is similar to that of purely poloidal configurations, except in regions where the poloidal field exhibits a local dip. In these regions, the added toroidal component increases the local field strength, inducing a transition from type-II Meissner to flux-tube superconductivity in case (a), and from flux-tube to non-superconducting behavior in case (b), confined to the small toroidal ring (Fig.~\ref{fig:1}c). These trends are consistent with those observed for purely toroidal fields, given our parameter choices.

Strongly toroidally dominated configurations do not violate stability criteria, since analytic studies of mixed-field equilibria in stably stratified, non-barotropic stars~\cite{Akgun2013} demonstrate that a modest poloidal component can stabilize a much stronger toroidal field. Therefore, in the following, toroidally dominated configurations are considered in our estimates of GW limits from NSs.

\section{Imprint on Gravitational Waves}
\label{sec:msp}

A promising approach to indirectly probe superconductivity in neutron stars is to search for continuous gravitational waves emitted by isolated, magnetically deformed, obliquely rotating MSPs, in which the magnetic and rotational axes are misaligned~\cite{sold2021,das-mukhopadhyay}. In this work, we model selected MSPs using their observed rotation frequencies within the \texttt{XNS} code to construct the stellar structure and compute the resulting magnetic deformation, assuming the presence of a strong toroidal magnetic field buried beneath the stellar surface. The corresponding CGW amplitude is then evaluated under the assumption of a misalignment between the magnetic and rotational axes.
The conjecture that MSPs may host significantly stronger internal magnetic fields than those inferred from their observed dipole spin-down has been discussed in the context of magnetic-field burial during past accretion episodes~\cite{Haskell18}. For magnetic stability, the toroidal field component beneath the stellar surface is expected to exceed the surface poloidal field by at least an order of magnitude, as implied by Eq.~\eqref{Elimit} for toroidal-dominated stable configurations (since $E_{\rm tor/pol} \propto B_{\rm tor/pol}^2$). The core toroidal field may be further enhanced by three to four orders of magnitude (see Fig.~\ref{fig:1}a). Consequently, in scenarios where the surface field is substantially screened, the internal toroidal field $B_{S\text{max}}$ may reach values up to $5 \times 10^4$ times larger than the dipolar surface field $B_P^{\text{pole}}$. This internal magnetic-field configuration strongly influences the stellar deformation, commonly quantified by the ellipticity $\epsilon$, which in turn determines the amplitude of the resulting CGW signal.

\begin{figure}[t]
\begin{center}
\includegraphics[width=\columnwidth]{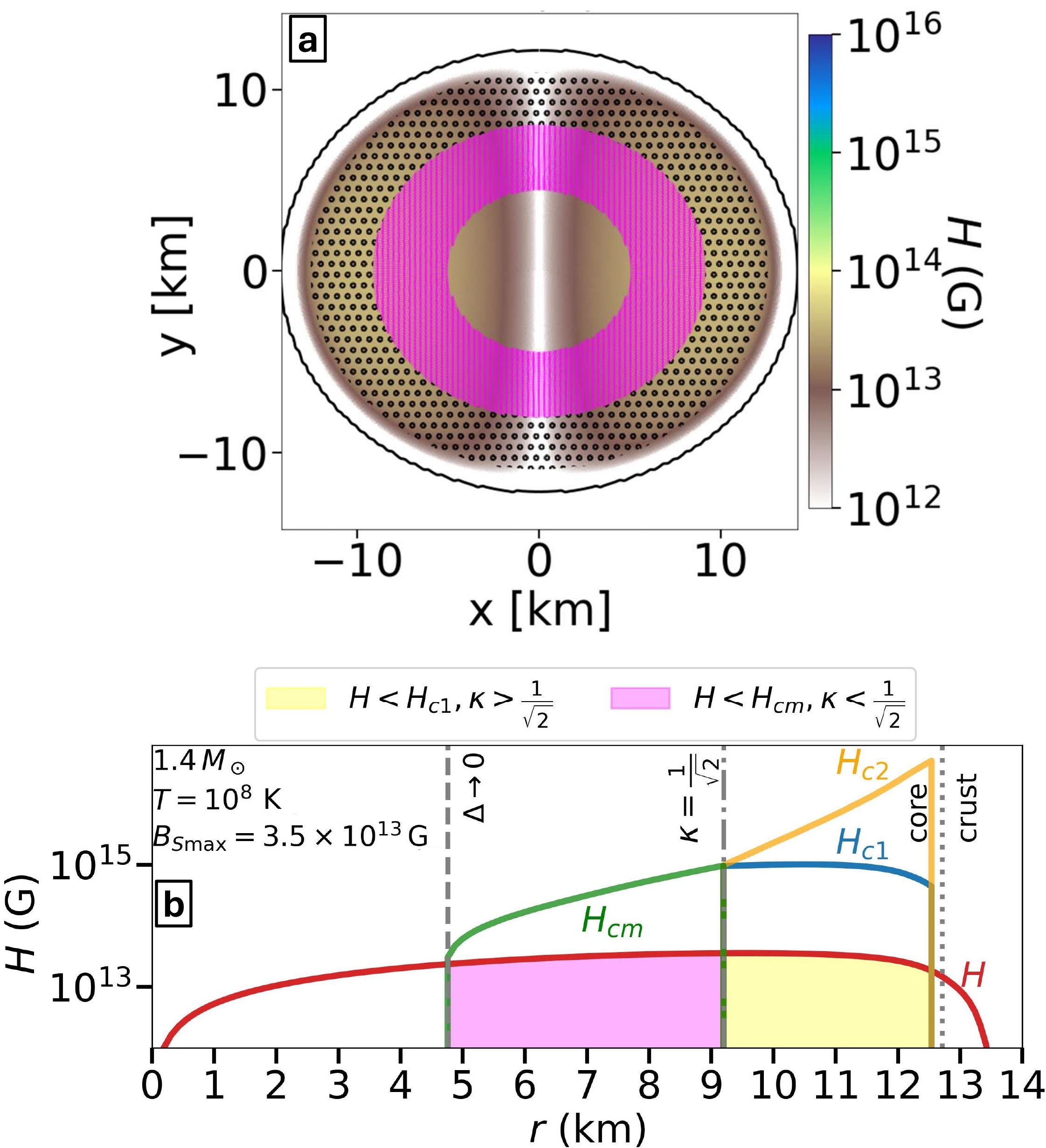}
\caption{(a) Color map of the superconducting regions 
of the MSP PSR J1939+2134 in 2D using the same conventions 
as in Fig.~\ref{fig:5}. (b) The 1D equatorial slice
showing the radial profiles of various magnetic fields and
    ranges occupied by various superconducting regions.  
 The conventions are the same as in Fig.~\ref{fig:4}.}
\label{fig:msp}
\end{center}
\end{figure}

We model several known MSPs listed in Table \ref{tab:msp_obs_sim} using \texttt{XNS}, based on their observed mass, rotational frequency $\nu$, and estimated internal toroidal field strength $B_{S\text{max}}$. The field profile is obtained by solving the Einstein-Maxwell equations by choosing a purely toroidal internal field and a stellar rotation at the observed frequency. The internal field strength is estimated from surface dipolar field inferred from observations as $B_{S\text{max}} \approx 5 \times 10^4 \times B_P^{\text{pole}}$. The masses estimated from observational data are reproduced by adjusting the central density using the DDME2 EoS. For each MSP, we compute the superconducting volume fraction $V_s$ (as a percentage of the total stellar volume), based on the magnetic field and temperature criteria for superconductivity described in Section \ref{sec:tc}, and report the values in Table \ref{tab:msp_obs_sim}. Even though MSPs are fast-spinning, the ratio of kinetic to gravitational energy  is at most 0.05 for those MSP models which satisfy $E_{\rm mag}/E_{\rm grav}\le 0.23$, so that the star should not go  through dynamical rotational instability~\cite{Baiotti2007}. For the well-known MSP PSR J1939+2134, one of the fastest-spinning NSs, we present a 2D visualization of the superconducting regions in Fig.~\ref{fig:msp}, which clearly shows that the star is almost entirely superconducting. This is due to the fact that the internal magnetic field remains below both $H_{c1}$ and $H_{cm}$. However, this conclusion may vary depending on the EoS; for example, in the case of a softer EoS, the superconducting region would be narrower, as discussed in Section \ref{sec:eosmicro}.

\begin{table*}
\centering
\caption{Observed properties and simulated characteristic gravitational-wave strain of selected MSPs. Columns 1 to 6 list the pulsar name, rotation frequency, distance, mass, poloidal magnetic-field strength, and the upper limit on the ellipticity. The subsequent columns provide simulated quantities: the assumed maximum internal magnetic field for the model, the percentage of the volume occupied by the superconducting region $V_s$, the ratio of ellipticities in the non-superconducting and superconducting cases, $\mathcal{F} = \epsilon_n / \epsilon_s$, the model values of ellipticities $\epsilon_n$ and $\epsilon_s$, and the corresponding characteristic strains $h_n^{\rm signal}$ and $h_s^{\rm signal}$. }
\small
\begin{tabular}{|l|c|c|c|c|c||c|c|c|c|c|c|c||c|}
\hline
\multicolumn{6}{|c||}{\textbf{Observed}} & \multicolumn{7}{c||}{\textbf{Simulated}} & \textbf{Ref.} \\
\hline
\begin{tabular}{c} {MSP PSR} \end{tabular} &
\begin{tabular}{c} $\nu$ \\ (Hz) \end{tabular} &
\begin{tabular}{c} $d$ \\ (kpc) \end{tabular} &
\begin{tabular}{c} $M$ \\ ($M_\odot$) \end{tabular} &
\begin{tabular}{c} $B_P^{\text{pole}}$ \\ (G) \end{tabular} &
\begin{tabular}{c} $\epsilon_{95\%}$ \\ $(\times 10^{-8})$\end{tabular} &
\begin{tabular}{c} $B_{\text{max}}$ \\ (G) \end{tabular} &
\begin{tabular}{c} $V_s$ \\ (\%) \end{tabular} &
\begin{tabular}{c} $\mathcal{F}$ \end{tabular} &
\begin{tabular}{c} $\epsilon_n$ \end{tabular} &
\begin{tabular}{c} $\epsilon_s $ \end{tabular} &
\begin{tabular}{c} $h_n^{\text{signal}}$ \end{tabular} &
\begin{tabular}{c} $h_s^{\text{signal}}$ \end{tabular} &
\begin{tabular}{c} \end{tabular} \\
\hline
J1810+1744    & 601.4  & 2.36 & 2.13  & \( 2 \times 10^8 \) &  \( 2.3  \) & \( 1 \times 10^{13} \) & 55 & 203 & \(5.5 \times 10^{-11}\) & \(1 \times 10^{-8}\) & \(4 \times 10^{-25}\) & \(8 \times 10^{-23}\) & \cite{romani2021} \\[1ex]
J0955$-$6150  & 500.2  & 2.17 & 1.71  & \( 2 \times 10^8 \)  & \( 3.6  \) & \( 1 \times 10^{13} \)  & 65 & 214 & \(5.4 \times 10^{-11}\) & \(1 \times 10^{-8}\) & \(3 \times 10^{-25}\) & \(5.5 \times 10^{-23}\) & \cite{ser2022} \\[1ex]
J1939+2134\footnote{This MSP model is visualized in Fig. \ref{fig:msp}.}   & 641.9  & 4.8 & $1.4$\footnote{Assumed mass; not directly measured observationally.} & \( 7 \times 10^8 \) & \( 6.1  \) & \( 3.5 \times 10^{13} \) & 71 & 74 & \(6.7 \times 10^{-10}\) & \(4.9 \times 10^{-8}\) & \(3 \times 10^{-24}\) & \(2.5 \times 10^{-22}\) & \cite{cordes2002} \\[1ex]
J1431$-$4715  & 497.0  & 1.55 & $1.4^b$ & \( 2 \times 10^8 \) & \( 2.3 \) & \( 1 \times 10^{13} \) & 71 & 258 & \(5.5 \times 10^{-11}\) & \(1 \times 10^{-8}\) & \(4 \times 10^{-25}\) & \(1 \times 10^{-22}\) & \cite{bates2015} \\[1ex]
J1513$-$2550  & 471.9  & 3.69 & $1.4^b$ & \( 4 \times 10^8 \) & \( 5.2  \) & \( 2 \times 10^{13} \) & 71 & 129 & \(2.2 \times 10^{-10}\) & \(2.8 \times 10^{-8}\) & \(5 \times 10^{-25}\) & \(7 \times 10^{-23}\) & \cite{sanpa2016} \\[1ex]
J0034$-$0534  & 532.7  & 1.35 & $1.4^b$         & \( 1 \times 10^8 \) & \( 1.2  \) & \( 5 \times 10^{12} \) & 71 & 520 & \(1.3 \times 10^{-11}\) & \(6.8 \times 10^{-9}\) & \(1 \times 10^{-25}\) & \(6 \times 10^{-23}\) & \cite{abdo2010} \\[1ex]
J1543$-$5149  & 486.2  & 1.15 & 1.35 & \( 3 \times 10^8 \) & \( 1.3  \) & \( 1.5 \times 10^{13} \) & 72 & 164 & \(1 \times 10^{-10}\) & \(1.6 \times 10^{-8}\) & \(8 \times 10^{-25}\) & \(1.2 \times 10^{-22}\) & \cite{chisabi2025} \\[1ex]
\hline
\end{tabular}
\label{tab:msp_obs_sim}
\end{table*}

The key parameter to interpret CGW emission from MSPs is $\epsilon$, defined as $\epsilon = |I_{yy} - I_{xx}| / I_{xx}$ for axisymmetric systems, where $I_{ij}$ are components of the moment of inertia tensor with $\hat{y}$ aligned along the magnetic axis. These components are extracted from \texttt{XNS} simulations. For non-superconducting models, the resulting ellipticity $\epsilon_n$, where $n$ refers to ellipticity in the absence of superconductivity,
is tabulated in Table~\ref{tab:msp_obs_sim}.
Note that \texttt{XNS} code solves the Einstein-Maxwell equations assuming normal matter
under a single-fluid approximation, i.e., it neglects the multifluid composition of neutron-proton mixtures in the presence of superconductivity. Importantly, this prescription also neglects the enhancement of magnetic stress due to the ordered flux tube lattice in type-II superconductors, which significantly alters stellar deformation and thus $\epsilon$. In this regime, when $B_S < H_{c1}$, the magnetic stress is enhanced by flux tubes by a factor of $H_{c1}/B_S$, leading to an amplification of stellar deformation \cite{easson1977,freibenrezz2012}. While deformation in normal matter scales as $\propto B_S^2$, in superconducting matter it scales as $\propto B_S H_{c1}$.

Although our primary focus has been on identifying where superconductivity sets in, we now estimate the ellipticity $\epsilon_s$ in the presence of type-II superconductivity. In this state, the magnetic field is frozen into flux tubes even when $B_S < H_{c1}$, due to extremely long field decay timescales arising from high electrical conductivity~\cite{Baym1969}. Thus, the enhanced magnetic stress in type-II regions is physically justified. However, the magnetic flux distribution in type-I regions is 
less certain and can be disordered.
Hence, we conservatively assume that the flux tube enhancement of magnetic stress in type-I zones is absent. Given that $\epsilon$ is proportional to the ratio of magnetic to gravitational binding energy \cite{freibenrezz2012}, the enhanced ellipticity $\epsilon_s$ in a superconducting star can be related to the normal-matter value $\epsilon_n$ through the ratio of total magnetic energies in the two cases as
\begin{align}
    \epsilon_s &= \epsilon_n \times \mathcal{F}, \nonumber \\
    \mathcal{F} &= \frac{\frac{1}{8\pi} \int_{\text{n}} B_S^2 \, dV + \frac{1}{8\pi} \int_{\text{s}} B_S H_{c1} \, dV}{\frac{1}{8\pi} \int_0^R B_S^2 \, dV}.
\end{align}
The integrals over the regions labeled `$n$' and `$s$' correspond to the normal and type-II superconducting domains of the star, respectively. For example, in the case of PSR J1939+2134, these regions span approximately $r \sim 0-9.2$~km (non-superconducting and type-I superconducting core) and $r \sim 9.2-12.5$~km (type-II superconducting shell), as shown in Fig.~\ref{fig:msp}. These contributions are computed grid-wise using the 2D magnetic field profile $B_S(r, \theta)$ and the critical field $H_{c 1}(r, \theta)$. Notably, throughout much of the star, the inequality $H_{c 1} \gg B_S$ holds, which implies that the type-II superconducting regions dominate the magnetic energy budget. As a result, these regions contribute significantly to the stellar deformation, leading to a substantially enhanced ellipticity. This amplification can reach up to two to three orders of magnitude, as evidenced by the numerical values presented in Table~\ref{tab:msp_obs_sim}.

We now compute the GW strain $h$ from the MSPs listed in Table \ref{tab:msp_obs_sim}, using stellar structure parameters extracted from our \texttt{XNS} models and observational data. The relevant quantities include $\epsilon$, $I_{xx}$ (e.g., $5 \times 10^{44}\,\mathrm{g\,cm^2}$ for PSR J1939+2134), the spin frequency $\nu$, and distance $d$ inferred from parallax measurements, where available, or dispersion measure estimates \cite{fermilat}.  The GW strain for non-superconducting and superconducting stars is given by \cite{sold2021,das-mukhopadhyay}
\begin{equation}
    h_{n/s} = f(\chi,i)\frac{2G}{c^4} \frac{(2\pi\nu)^2 \epsilon_{n/s} I_{xx}}{d},
    \label{eq:gwh}
  \end{equation}
  where $f(\chi, i)$ encodes the dependence on the magnetic obliquity angle $\chi$ (between the magnetic and spin axes) and the inclination angle $i$ (between the spin axis and the observer line of sight). Following \cite{DasMB2025}, it is given by:
\begin{align}
f(\chi,i) &= \left(2\cos^2\chi - \sin^2\chi\right)\frac{\sin\chi}{2\sqrt{2}} \notag\\
&\quad \times \left[ 
\cos^2 i \sin^2 i \cos^2 \chi + \sin^2 i \cos^2 \chi \right. \notag\\
&\quad \left. + (1 + \cos^2 i)^2 \sin^2 \chi + 4\cos^2 i \sin^2 \chi 
\right]^{1/2}.
\label{eq:h_avg_formula}
\end{align}
For PSR J1939+2134, both $\chi$ and $i$ have been estimated to be $\approx 80^\circ$ based on pulse profile fitting~\cite{guill2012}, yielding $f(80^\circ,80^\circ) \approx 0.17$. For other MSPs in the table, $\chi$ and $i$ are not observationally constrained; we therefore adopt the angular average $\langle f(\chi, i) \rangle \approx 0.147$ for them~\cite{DasMB2025}. The characteristic signal strength, relevant for CGW searches over an observational period $t \sim 1$ year, is given in Ref.~\cite{Moore2015}
\begin{equation}
    h_{n/s}^{\text{signal}} = h_{n/s} \sqrt{N_c},
    \label{eq:h_char}
  \end{equation}
where $N_c = \nu t$ is the total number of accumulated cycles coherently stacked by the detector. Since $\epsilon_s = \mathcal{F} \times \epsilon_n$, it follows that $h_s^{\text{signal}} = \mathcal{F} \times h_n^{\text{signal}}$. These values are reported in Table~\ref{tab:msp_obs_sim}.

\begin{figure}
\begin{center}
\includegraphics[width=1.05\columnwidth]{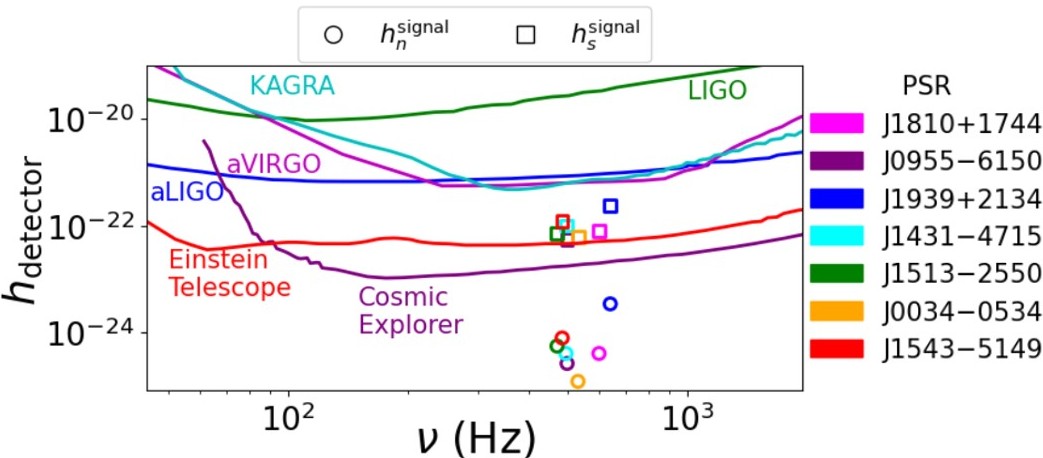}
\caption{
The estimated strain of the gravitational-wave signal as a function of frequency, emitted by MSPs listed in Table~\ref{tab:msp_obs_sim},  for the case of a non-superconducting interior (circles) and in the presence of superconductivity (squares). The sensitivity curves of current and future GW detectors are shown as labeled.}
\label{fig:mspsens}
\end{center}
\end{figure}

To evaluate detectability, we compare $h^{\text{signal}}$ with the detector sensitivity $h^{\text{noise}}$. A signal is considered detectable only if the signal-to-noise ratio (SNR),
\begin{equation}
    \text{SNR} = \frac{h^{\text{signal}}}{h^{\text{noise}}},
\end{equation}
exceeds the threshold $\text{SNR}_{\text{th}} \approx 11.4$, for more than 95\% detection efficiency \cite{Cieslar2021}. The corresponding detectability threshold of the detector is then
\begin{equation}
    h^{\text{detector}} = \text{SNR}_{\text{th}} \times h^{\text{noise}}.
\end{equation}
In Fig. \ref{fig:mspsens}, we compare $h^{\text{signal}}$ with $h^{\text{detector}}$ for the listed MSPs with the condition for detection being $h^{\text{signal}} > h^{\text{detector}}$.

All MSPs in Table \ref{tab:msp_obs_sim} have already been targeted in dedicated CGW searches by LIGO-Virgo collaboration~\cite{ligo2022}, with no detections so far. These null results provide upper limits on the ellipticity, denoted $\epsilon_{95\%}$, which are also listed in the table \cite{ligo2022}. From Fig.~\ref{fig:mspsens}, we see that even if these MSPs are superconducting, their predicted GW signal remains below the current LIGO detection threshold. However, our models predict that such signals could possibly be detectable by next-generation detectors such as the Einstein Telescope and Cosmic Explorer within a one-year integration time. In contrast, the same sources would remain undetectable if they were non-superconducting. Thus, the detection of a CGW signal from such an MSP would provide strong evidence for bulk superconductivity and for the presence of a buried internal magnetic field, which amplifies magnetic stresses through the superconducting response of the stellar matter. Conversely, continued non-detection of CGW may suggest either a weaker internal magnetic field or a superconducting region that is limited in size, or possibly absent altogether.  The size of the superconducting region depends on the combined miscophysics of EoS and pairing, which can be degenerate, because the softer the EoS or the stronger the three-body repulsion, the smaller is the size of the superconducting domain. Thus, a combination of a stiff EoS with strong three-body coupling would produce a result that is similar to that of a soft EoS and weak three-body force. Therefore, their individual contributions are difficult to disentangle when using the decoupling approximation; it should be noted, however, that a consistent many-body framework inherently incorporates both pairing correlations and three-body forces as integral components.

\section{Discussion and conclusion}
\label{sec:Conclusions}

Given the lack of direct experimental access to cold, dense nuclear matter, substantial uncertainties remain in our understanding of both the EoS and the pairing interaction responsible for superconductivity in NSs. Although indirect evidence -- such as the rapid cooling of Cas~A and pulsar glitches -- supports the presence of superfluid and superconducting components, direct observational constraints are still lacking. In this work, we have explored an additional avenue for probing the nature of superconductivity in magnetars, focusing on the distinction between type-I and type-II behavior. We have also highlighted the potential GW signatures from MSPs, which may become detectable with next-generation observatories such as the 
Einstein Telescope and Cosmic Explorer.

Among the EoS considered, the microscopic BBB model, while theoretically well-motivated, is too soft to support the observed mass of PSR~J0740+6620~\cite{Cromartie2019}. In contrast, the stiffer RMF EoS DDME2 produces a higher maximum mass consistent with this constraint. However, DDME2 predicts a tidal deformability $\Lambda_{1.4}$ for a $1.4\,M_\odot$ NS that exceeds the observational upper bound $\Lambda_{1.4} \lesssim 580$ inferred from GW170817~\cite{Abbot2018lambda}, as also noted in Ref.~\cite{zuraiq2024}. Tidal deformability, which quantifies the star’s susceptibility to deformation by an external gravitational perturbation, constrains the EoS stiffness: stiffer EoS yield larger radii and thus larger $\Lambda$. The softer BBB EoS used here predicts a relatively small $\Lambda_{1.4}$ that remains consistent with the lower observational limit $\Lambda_{1.4} \gtrsim 292$~\cite{Krastev2018}.

Taken together, the BBB and DDME2 EoS represent two limiting cases: the former is too soft to support the highest precisely measured NS mass, whereas the latter is too stiff to remain consistent with the tidal deformability inferred from GW170817. The true physical EoS likely lies between these extremes. Consequently, our results can be viewed as bounding estimates for the size of superconducting regions. A complementary analysis employing EoS that satisfy both mass and tidal-deformability constraints -- such as those listed in Ref.~\cite{Biswas2022} -- could provide more realistic predictions. 

We have systematically investigated the spatial size and topology of type-I and type-II superconducting regions in magnetized NSs using the \texttt{XNS} code. Considering both toroidal and poloidal magnetic field configurations, we have explored a broad parameter space encompassing stellar mass (and thus central density), temperature, magnetic field strength, and nuclear microphysics, including the effects of EoS stiffness and three-body nuclear forces on proton pairing.  Our results indicate that superconducting regions contract with increasing stellar mass. In magnetars with predominantly toroidal fields, the superconducting phase develops characteristic geometric structures, such as torus-shaped voids and polar Meissner layers, governed by the intersection of the magnetic field with the critical field thresholds $H_{c1}$ and $H_{c2}$. The two-dimensional topology evolves significantly across the toroidal field models, exhibiting a transition from fully closed, ellipsoidal type-II Meissner regions at low field strength to open-polar flux-tube configurations at higher fields. The type-I domains, featuring an outer prolate envelope and an inner toroidal core, extend prominently toward the poles. In contrast, poloidal field configurations favor a more uniformly prolate superconducting region, with weaker magnetic suppression in the outer-core mid-latitude zone, where the field strength reaches its minimum.

The impact of proton superconductivity on observable phenomena depends sensitively on both the EoS and the proton pairing interaction. In our study, the stiff DDME2 RMF EoS yields spatially more extended superconducting regions than the softer BBB EoS {\it for NSs of the same mass}. On the other hand, the inclusion of repulsive three-body forces reduces the pairing gap at high densities (as has been implemented in our estimate of GW emission), leading to more reduced (conservative) estimates of the superconducting volume compared to models including only two-body interactions. These two competing effects can partially offset each other, potentially leading to degenerate outcomes for the size of the superconducting region. Uncertainties in the true EoS stiffness may be reduced through future improved measurements of NS radii and GW observations of binary NS mergers, leaving the accurate treatment of the pairing as a key missing ingredient.

Our results indicate that the inner core of an NS often remains in a normal (non-superconducting) state unless the EoS is exceptionally stiff. This challenges the common assumption, often implicit in many studies, that superconductivity extends throughout the entire stellar interior (see, e.g., Ref.~\cite{Sinha2015PhRvC}). For canonical $1.4\,M_{\odot}$ NSs, even at asymptotically low temperatures and for magnetic fields below the upper critical value for superconductivity suppression, the proton-superconducting phase typically occupies at most $\sim70\%$ of the stellar volume. These findings highlight that the interplay between proton superconductivity and magnetic equilibria in NS cores remains an open and theoretically rich problem, warranting further investigation.

Finally, we have examined the implications for GW emission from MSPs. Should current or future detectors such as the Einstein Telescope or Cosmic Explorer detect CGWs from these sources, the inferred high ellipticities, which are enhanced by type-II superconductivity, could provide strong evidence for type-II superconductivity and reveal hidden internal magnetic fields. Conversely, non-detection would constrain either the internal field strength or the existence of superconductivity, offering a rare observational probe of NS microphysics. 

\section*{Acknowledgments}  MD thanks Bhaskar Biswas (Hamburg Observatory) for fruitful discussion about tidal deformability, and Suprovo Ghosh (U. Southampton) for suggestion regarding LIGO non-detection upper limits. MD and BM thank Tanmoy Das (IISc) for the fruitful discussion about interaction potentials. MD is especially grateful to Soumallya Mitra for his constant support and insightful discussions on computational implementation. She also acknowledges the Prime Minister’s Research Fellowship (PMRF) scheme, with Ref. No. TF/PMRF-22-5442.03. BM acknowledges a project funded by SERB, India, with Ref. No. CRG/2022/003460, for partial support towards this research. AS acknowledges support through Deutsche Forschungsgemeinschaft Grant No. SE 1836/6-1 and the Polish National Science  Centre (NCN)
Grant No. 2023/51/B/ST9/02798.

\appendix
\section{Review of stability of magnetic configurations}

Early theoretical work~\cite{Tayler1973, Markey1973, Wright1973} showed that purely poloidal (all field lines going from pole to pole) and purely toroidal (all field lines wrapped around azimuthally inside the star) magnetic configurations in a fluid star are generically unstable to ideal MHD instabilities on the Alfv\'en timescale (typically milliseconds to seconds for NSs). For poloidal fields, a small perturbation in the closed-field-line region (near the equator inside the star) can cause displacements that grow and tangle the field. For toroidal fields, the Tayler instability of an $m=1$ kink mode quickly distorts purely toroidal fields.  In newly born stars, before solidification of the crust, and inside the core, which remains fluid, these instabilities imply that neither pure component can survive long after birth. Our results in this work, particularly the stellar models under consideration, remain obviously valid if one of the components of the field (poloidal or toroidal) is dominating. Therefore, it is imperative to discuss the potential stable structures in magnetars. Note that local instabilities can lead to phase separation in the core and the crust of a magnetar - a phenomenon that is not included in the global stability analysis so far~\cite{Rau2021,Rau2023} and will be ignored below.

However, the understanding of stable magnetic equilibria in NSs and magnetars remains incomplete, particularly regarding the allowed ratio of poloidal to toroidal field strengths.  In {\it barotropic} (non-stratified) stars, the toroidal energy fraction is severely limited ($<10\%$)~\cite{Lander2009,Ciolfi2009}. However,
{\it non-barotropic}, stably stratified stars, where composition or entropy gradients provide buoyancy, can support a much wider range of stable poloidal-toroidal field
equilibria~\cite{Braithwaite2009,Akgun2013,Mastrano2011,Mastrano2015}. While the overall stability of axisymmetric equilibria is generally accepted, their full three-dimensional stability against non-axisymmetric perturbations remains debated.

\underline{\it The Twisted–Torus Configuration}.

A leading candidate for a long-lived internal magnetic structure is
the {\it twisted-torus} geometry. This conclusion is supported by semi-analytical studies~\cite{Mastrano2011,Mastrano2015} and both Newtonian and relativistic MHD simulations~\cite{Braithwaite2006,BraithwaiteNordlund2006,Braithwaite2009,Ciolfi2009,CR2013}.
It consists of: (a)  a {\it poloidal field} extending throughout the star and into the exterior; (b)
 a {\it toroidal field} confined within the closed-field-line region, encircling the magnetic axis.
Although fully evolved configurations are non-axisymmetric, axisymmetric approximations are often used in simulations. Such equilibria can survive for many Alfvén timescales ($>10^5$ years), consistent with magnetar lifetimes.
Observationally, only the poloidal surface field (from spin-down) is accessible, while a much stronger internal toroidal component may remain hidden -- storing magnetic energy that powers magnetar bursts and crustal stresses. The ratio of poloidal to toroidal energies thus strongly influences magneto-thermal evolution and crustal yielding.

\underline{\it Magnetic Energy Budget.}
The total magnetic energy is given by
\begin{eqnarray}
E_{\rm mag} = \frac{1}{8\pi}\int_V B^2\, dV = E_{\rm pol} + E_{\rm tor},
\end{eqnarray}
where $E_{\rm pol}$ and $E_{\rm tor}$ are the poloidal and toroidal components.

Numerical simulations~\cite{Braithwaite2009} found that long-lived equilibria can exist for a wide range of energy ratios:
\begin{eqnarray}
0.25 \lesssim \frac{E_{\rm tor}}{E_{\rm pol}} = \frac{1 - \Lambda_B}{\Lambda_B} \lesssim 10^2\text{–}10^3,
\quad \Lambda_B \equiv \frac{E_{\rm pol}}{E_{\rm mag}},
\label{Elimit}
\end{eqnarray}
for field strengths $B \sim 10^{15}\text{–}10^{16}\,\mathrm{G}$.

\underline{\it Stability Limits and Constraints.}
Stable, non-rotating, poloidally dominated configurations generally satisfy
\begin{eqnarray}
\frac{E_{\rm tor}}{E_{\rm mag}} \lesssim 0.1
\end{eqnarray}
in both Newtonian and relativistic models~\cite{Ciolfi2009,Mastrano2011,Akgun2013}.
However, the toroidal field may still exceed the poloidal one locally, since it occupies only a limited internal region.
Excessive toroidal energy triggers Tayler instability, while too little toroidal energy destabilizes the closed-field-line region of the poloidal field.
Ref.~\cite{CR2013} introduced an alternative current prescription allowing stronger toroidal components, reaching
\begin{eqnarray}
\frac{E_{\rm tor}}{E_{\rm mag}} \lesssim 0.9.
\end{eqnarray}
Another stability condition relates the magnetic and gravitational energies~\cite{Braithwaite2009,Akgun2013}:
\begin{eqnarray}
\frac{E_{\rm pol}}{E_{\rm tor}} \gtrsim 2a \frac{E_{\rm tor}}{E_{\rm grav}},
\end{eqnarray}
where $E_{\rm grav}$ is the gravitational binding energy and $a\sim 200-800$.
Since $E_{\rm pol} / E_{\rm grav} \ll 1$, this implies that the toroidal component can dominate the magnetic energy and serve as the main energy reservoir behind magnetar activity. Such strong internal fields can also induce stellar deformations and CGW emission.

\underline{\it Stability Ranges in the Twisted-Torus Model.}
Analyses of twisted-torus equilibria~\cite{Mastrano2011,MSM2015} found stability over a broad range of poloidal energy fractions: $
0.01 \lesssim \Lambda_B \lesssim 0.8.$ For barotropic stars, this range narrows to approximately $0.5 \lesssim \Lambda_B \lesssim 0.9$, whereas non-barotropic, stratified models can sustain much stronger toroidal dominance $(\Lambda_B \sim 0.01)$.

In summary, stable magnetic equilibria in NSs depend sensitively on stratification and on the balance between the poloidal and toroidal components. Barotropic stars permit only weak toroidal fields, whereas stratified interiors can host strong, long-lived toroidal regions, embedded within poloidal fields. In such cases, the so-called twisted-torus configuration arises. There are also other possible mixed-field configurations with a stronger toroidal component. Such configurations provide a natural explanation for magnetar phenomenology: a modest observable surface dipole field concealing a much stronger internal toroidal component that stores magnetic energy, drives bursts, and induces stellar deformation. Consequently, understanding the relative poloidal–toroidal energy balance is crucial for modeling magnetar evolution, magnetic stress on the crust, and potential CGW emission.

\bibliographystyle{apsrev}
\bibliography{references2,PPNP_mag}

@article{Jones1975,
  author       = {Jones, P.~B.},
  title        = {Pulsar Electrodynamics: Superconducting Protons and Magnetic Fields},
  journal      = {Astrophysical Journal},
  volume       = {201},
  pages        = {661--667},
  year         = {1975},
  doi          = {10.1086/153927}
}

@article{Cutler2002,
  author       = {Cutler, Curt},
  title        = {Gravitational waves from neutron stars with large toroidal B fields},
  journal      = {Physical Review D},
  volume       = {66},
  pages        = {084025},
  year         = {2002},
  doi          = {10.1103/PhysRevD.66.084025}
}

@article{Haskell2008,
  author       = {Haskell, Brynmor and Samuelsson, Lars and Glampedakis, Kostas and Andersson, Nils},
  title        = {Modelling magnetically deformed neutron stars},
  journal      = {MNRAS},
  volume       = {385},
  pages        = {531--542},
  year         = {2008},
  doi          = {10.1111/j.1365-2966.2008.12857.x}
}

@article{Akgun2018,
  author       = {{Akg{\"u}n}, Taner and {Cerd{\'a}-Dur{\'a}n}, Pablo and {Miralles}, Juan A. and {Pons}, Jos{\'e} A.},
  title        = {Magnetically deformed neutron stars: spin evolution and gravitational waves},
  journal      = {MNRAS},
  volume       = {481},
  pages        = {5331--5345},
  year         = {2018},
  doi          = {10.1093/mnras/sty2668}
}

@article{GugerCAlpar2014,
  author       = {G{\"u}gercinoglu, Erbil and Alpar, M. Ali},
  title        = {Toroidal flux lines in neutron star superconductors and vortex pinning},
  journal      = {Astrophysical Journal},
  volume       = {788},
  pages        = {L11},
  year         = {2014},
  doi          = {10.1088/2041-8205/788/1/L11}
}

@article{GugerCAlpar2016,
  author       = {G{\"u}gercinoglu, Erbil and Alpar, M. Ali},
  title        = {Neutron Star Superconductivity, Toroidal Flux Tubes, and Vortex Dynamics},
  journal      = {MNRAS},
  volume       = {462},
  pages        = {1453--1462},
  year         = {2016},
  doi          = {10.1093/mnras/stw1736}
}

@article{SedrakyanHayrapetyan2015,
  author       = {Sedrakyan, D.~M. and Hayrapetyan, M.~V.},
  title        = {Toroidal superconducting structures in neutron stars},
  journal      = {Astrophysics},
  volume       = {58},
  pages        = {120--129},
  year         = {2015},
  doi          = {10.1007/s10511-015-9369-1}
}

@article{Tamagaki1970,
  author  = {Tamagaki, Tamio},
  title   = {Superfluid State in Neutron Star Matter. I. Generalized Pairing of Neutrons},
  journal = {Progress of Theoretical Physics},
  volume  = {44},
  pages   = {905--928},
  year    = {1970},
  doi     = {10.1143/PTP.44.905}
}

@article{Hoffberg1970,
  author  = {Hoffberg, M. and Richardson, A. E. and Ruderman, M. and Glassgold, A. E.},
  title   = {Anisotropic Superfluidity in Nuclear Matter},
  journal = {Physical Review Letters},
  volume  = {24},
  pages   = {775--777},
  year    = {1970},
  doi     = {10.1103/PhysRevLett.24.775}
}

@article{Clark1972,
  author  = {Clark, J. W. and Yang, C. H.},
  title   = {Superfluidity in Nuclear Matter},
  journal = {Nuclear Physics A},
  volume  = {186},
  pages   = {1--12},
  year    = {1972},
  doi     = {10.1016/0375-9474(72)90661-3}
}

@article{Takatsuka1972,
  author  = {Takatsuka, Toshio},
  title   = {Proton Superfluidity in Dense Nuclear Matter},
  journal = {Progress of Theoretical Physics},
  volume  = {48},
  pages   = {1517--1529},
  year    = {1972},
  doi     = {10.1143/PTP.48.1517}
}

@article{Takatsuka1973,
  author  = {Takatsuka, Toshio},
  title   = {Pairing Correlations of Protons in Neutron Star Matter},
  journal = {Progress of Theoretical Physics},
  volume  = {50},
  pages   = {1754--1765},
  year    = {1973},
  doi     = {10.1143/PTP.50.1754}
}

@ARTICLE{Shternin2011,
       author = {{Shternin}, Peter S. and {Yakovlev}, Dmitry G. and {Heinke}, Craig O. and {Ho}, Wynn C.~G. and {Patnaude}, Daniel J.},
        title = "{Cooling neutron star in the Cassiopeia A supernova remnant: evidence for superfluidity in the core}",
      journal = {\mnras},
         year = 2011,
        month = mar,
       volume = {412},
       number = {1},
        pages = {L108-L112},
          doi = {10.1111/j.1745-3933.2011.01015.x},
archivePrefix = {arXiv},
       eprint = {1012.0045},
 primaryClass = {astro-ph.SR},
       adsurl = {https://ui.adsabs.harvard.edu/abs/2011MNRAS.412L.108S}
}

@ARTICLE{Alford2014ApJ,
       author = {{Alford}, Mark G. and {Schwenzer}, Kai},
        title = "{Gravitational Wave Emission and Spin-down of Young Pulsars}",
      journal = {\apj},
         year = 2014,
        month = jan,
       volume = {781},
       number = {1},
          eid = {26},
        pages = {26},
          doi = {10.1088/0004-637X/781/1/26},
archivePrefix = {arXiv},
       eprint = {1210.6091},
 primaryClass = {gr-qc},
       adsurl = {https://ui.adsabs.harvard.edu/abs/2014ApJ...781...26A}
}

@ARTICLE{Piccinni2022,
       author = {{Piccinni}, Ornella Juliana},
        title = "{Status and Perspectives of Continuous Gravitational Wave Searches}",
      journal = {Galaxies},
         year = 2022,
        month = may,
       volume = {10},
       number = {3},
          eid = {72},
        pages = {72},
          doi = {10.3390/galaxies10030072},
archivePrefix = {arXiv},
       eprint = {2202.01088},
 primaryClass = {gr-qc},
       adsurl = {https://ui.adsabs.harvard.edu/abs/2022Galax..10...72P}
}

@ARTICLE{Sinha2015PhRvC,
       author = {{Sinha}, Monika and {Sedrakian}, Armen},
        title = "{Magnetar superconductivity versus magnetism: Neutrino cooling processes}",
      journal = {\prc},
         year = 2015,
        month = mar,
       volume = {91},
       number = {3},
          eid = {035805},
        pages = {035805},
          doi = {10.1103/PhysRevC.91.035805},
archivePrefix = {arXiv},
       eprint = {1502.02979},
 primaryClass = {astro-ph.HE},
       adsurl = {https://ui.adsabs.harvard.edu/abs/2015PhRvC..91c5805S}
}

@ARTICLE{Raduta2019MNRAS,
       author = {{Raduta}, Adriana R. and {Li}, Jia Jie and {Sedrakian}, Armen and {Weber}, Fridolin},
        title = "{Cooling of hypernuclear compact stars: Hartree-Fock models and high-density pairing}",
      journal = {\mnras},
         year = 2019,
        month = aug,
       volume = {487},
       number = {2},
        pages = {2639-2652},
          doi = {10.1093/mnras/stz1459},
archivePrefix = {arXiv},
       eprint = {1903.01295},
 primaryClass = {nucl-th},
       adsurl = {https://ui.adsabs.harvard.edu/abs/2019MNRAS.487.2639R}
}

@ARTICLE{Alm1996NuPhA,
       author = {{Alm}, T. and {R{\"o}pke}, G. and {Sedrakian}, A. and {Weber}, F.},
        title = "{$^{3}$D $_{2}$ pairing in asymmetric nuclear matter}",
      journal = {\nphysa},
         year = 1996,
        month = feb,
       volume = {604},
       number = {4},
        pages = {491-504},
          doi = {10.1016/0375-9474(96)00153-4},
       adsurl = {https://ui.adsabs.harvard.edu/abs/1996NuPhA.604..491A}
}

@ARTICLE{Andersson2021Univ,
       author = {{Andersson}, Nils},
        title = "{A Superfluid Perspective on Neutron Star Dynamics}",
      journal = {Universe},
         year = 2021,
        month = jan,
       volume = {7},
       number = {1},
          eid = {17},
        pages = {17},
          doi = {10.3390/universe7010017},
archivePrefix = {arXiv},
       eprint = {2103.10218},
 primaryClass = {astro-ph.HE},
       adsurl = {https://ui.adsabs.harvard.edu/abs/2021Univ....7...17A}
}

@ARTICLE{Sedrakian2019EPJA,
       author = {{Sedrakian}, Armen and {Clark}, John W.},
        title = "{Superfluidity in nuclear systems and neutron stars}",
      journal = {European Physical Journal A},
         year = 2019,
        month = sep,
       volume = {55},
       number = {9},
          eid = {167},
        pages = {167},
          doi = {10.1140/epja/i2019-12863-6},
archivePrefix = {arXiv},
       eprint = {1802.00017},
 primaryClass = {nucl-th},
       adsurl = {https://ui.adsabs.harvard.edu/abs/2019EPJA...55..167S}
}

@book{Abrikosov:Fundamentals,
  author = 	 {A.A. Abrikosov},
  title={Fundamentals of the Theory of Metals},
  publisher = 	 {North-Holland},
  year = 	 1988,
  address = 	 {Amsterdam},
}

@ARTICLE{Sedrakian2005PhRvD,
       author = {{Sedrakian}, Armen},
        title = "{Type-I superconductivity and neutron star precession}",
      journal = {\prd},
         year = 2005,
        month = apr,
       volume = {71},
       number = {8},
          eid = {083003},
        pages = {083003},
          doi = {10.1103/PhysRevD.71.083003},
archivePrefix = {arXiv},
       eprint = {astro-ph/0408467},
 primaryClass = {astro-ph},
       adsurl = {https://ui.adsabs.harvard.edu/abs/2005PhRvD..71h3003S}
}

@ARTICLE{CompOSE2022EPJA,
       author = {{CompOSE Core Team} and {Typel}, S. and {Oertel}, M. and {Kl{\"a}hn}, T. and {Chatterjee}, D. and {Dexheimer}, V. and {Ishizuka}, C. and {Mancini}, M. and {Novak}, J. and {Pais}, H. and {Provid{\^e}ncia}, C. and {R. Raduta}, Ad. and {Servillat}, M. and {Tolos}, L.},
        title = "{CompOSE reference manual: Version 3.01, CompStar Online Supernov{\ae} Equations of State, ``harmonising the concert of nuclear physics and astrophysics'', https://compose.obspm.fr}",
      journal = {European Physical Journal A},
         year = 2022,
        month = nov,
       volume = {58},
       number = {11},
          eid = {221},
        pages = {221},
          doi = {10.1140/epja/s10050-022-00847-y},
archivePrefix = {arXiv},
       eprint = {2203.03209},
 primaryClass = {astro-ph.HE},
       adsurl = {https://ui.adsabs.harvard.edu/abs/2022EPJA...58..221C}
}

@ARTICLE{Muhlschlegel1959ZPhy,
       author = {{M{\"u}hlschlegel}, Bernhard},
        title = "{Die thermodynamischen Funktionen des Supraleiters}",
      journal = {Zeitschrift fur Physik},
         year = 1959,
        month = jun,
       volume = {155},
       number = {3},
        pages = {313-327},
          doi = {10.1007/BF01332932},
       adsurl = {https://ui.adsabs.harvard.edu/abs/1959ZPhy..155..313M}
}

@ARTICLE{Sedrakian1997MNRAS,
       author = {{Sedrakian}, D.~M. and {Sedrakian}, A.~D. and {Zharkov}, G.~F.},
        title = "{Type I superconductivity of protons in neutron stars}",
      journal = {\mnras},
         year = 1997,
        month = sep,
       volume = {290},
       number = {1},
        pages = {203-207},
          doi = {10.1093/mnras/290.1.203},
archivePrefix = {arXiv},
       eprint = {astro-ph/9710280},
 primaryClass = {astro-ph},
       adsurl = {https://ui.adsabs.harvard.edu/abs/1997MNRAS.290..203S}
}

@ARTICLE{Sedrakian1995ApJ,
       author = {{Sedrakian}, Armen D. and {Sedrakian}, David M.},
        title = "{Superfluid Core Rotation in Pulsars. I. Vortex Cluster Dynamics}",
      journal = {\apj},
         year = 1995,
        month = jul,
       volume = {447},
        pages = {305},
          doi = {10.1086/175876},
       adsurl = {https://ui.adsabs.harvard.edu/abs/1995ApJ...447..305S}
}

@Article{Dexheimer2022,
AUTHOR = {Dexheimer, Veronica and Mancini, Marco and Oertel, Micaela and Providência, Constança and Tolos, Laura and Typel, Stefan},
TITLE = {Quick Guides for Use of the CompOSE Data Base},
JOURNAL = {Particles},
VOLUME = {5},
YEAR = {2022},
NUMBER = {3},
PAGES = {346--360},
no-URL = {https://www.mdpi.com/2571-712X/5/3/28},
no-ISSN = {2571-712X},
DOI = {10.3390/particles5030028}
}

@ARTICLE{Haber2017PhRvD,
       author = {{Haber}, Alexander and {Schmitt}, Andreas},
        title = "{Critical magnetic fields in a superconductor coupled to a superfluid}",
      journal = {\prd},
         year = 2017,
        month = jun,
       volume = {95},
       number = {11},
          eid = {116016},
        pages = {116016},
          doi = {10.1103/PhysRevD.95.116016},
archivePrefix = {arXiv},
       eprint = {1704.01575},
 primaryClass = {hep-th},
       adsurl = {https://ui.adsabs.harvard.edu/abs/2017PhRvD..95k6016H}
}

@ARTICLE{Sinha2015PPN,
       author = {{Sinha}, M. and {Sedrakian}, A.},
        title = "{Upper critical field and (non)-superconductivity of magnetars}",
      journal = {Physics of Particles and Nuclei},
         year = 2015,
        month = sep,
       volume = {46},
       number = {5},
        pages = {826-829},
          doi = {10.1134/S1063779615050275},
archivePrefix = {arXiv},
       eprint = {1403.2829},
 primaryClass = {astro-ph.SR},
       adsurl = {https://ui.adsabs.harvard.edu/abs/2015PPN....46..826S}
}

@ARTICLE{Pons2009AA,
       author = {{Pons}, J.~A. and {Miralles}, J.~A. and {Geppert}, U.},
        title = "{Magneto-thermal evolution of neutron stars}",
      journal = {\aap},
         year = 2009,
        month = mar,
       volume = {496},
       number = {1},
        pages = {207-216},
          doi = {10.1051/0004-6361:200811229},
archivePrefix = {arXiv},
       eprint = {0812.3018},
 primaryClass = {astro-ph},
       adsurl = {https://ui.adsabs.harvard.edu/abs/2009A&A...496..207P}
}

@ARTICLE{Ascenzi2024MNRAS,
       author = {{Ascenzi}, Stefano and {Vigano}, Daniele and {Dehman}, Clara and {Pons}, Jos{\'e} A. and {Rea}, Nanda and {Perna}, Rosalba},
        title = "{3D code for MAgneto-Thermal evolution in Isolated Neutron Stars, MATINS: thermal evolution and lightcurves}",
      journal = {\mnras},
         year = 2024,
        month = jul,
          doi = {10.1093/mnras/stae1749},
archivePrefix = {arXiv},
       eprint = {2401.15711},
 primaryClass = {astro-ph.HE},
       adsurl = {https://ui.adsabs.harvard.edu/abs/2024MNRAS.tmp.1732A}
}

@ARTICLE{Rau2020,
       author = {{Rau}, Peter B. and {Wasserman}, Ira},
        title = "{Relativistic finite temperature multifluid hydrodynamics in a neutron star from a variational principle}",
      journal = {\prd},
         year = 2020,
        no-month = sep,
       volume = {102},
       number = {6},
          eid = {063011},
        pages = {063011},
          doi = {10.1103/PhysRevD.102.063011},
archivePrefix = {arXiv},
       eprint = {2004.07468},
 primaryClass = {astro-ph.HE},
       adsurl = {https://ui.adsabs.harvard.edu/abs/2020PhRvD.102f3011R}
}

@ARTICLE{Gusakov2016a,
       author = {{Gusakov}, M.~E. and {Dommes}, V.~A.},
        title = "{Relativistic dynamics of superfluid-superconducting mixtures in the presence of topological defects and an electromagnetic field with application to neutron stars}",
      journal = {\prd},
         year = 2016,
        no-month = oct,
       volume = {94},
       number = {8},
          eid = {083006},
        pages = {083006},
          doi = {10.1103/PhysRevD.94.083006},
archivePrefix = {arXiv},
       eprint = {1607.01629},
 primaryClass = {gr-qc},
       adsurl = {https://ui.adsabs.harvard.edu/abs/2016PhRvD..94h3006G}
}

@ARTICLE{Page2011PhRvL,
       author = {{Page}, Dany and {Prakash}, Madappa and {Lattimer}, James M. and {Steiner}, Andrew W.},
        title = "{Rapid Cooling of the Neutron Star in Cassiopeia A Triggered by Neutron Superfluidity in Dense Matter}",
      journal = {\prl},
         year = 2011,
        no-month = feb,
       volume = {106},
       number = {8},
          eid = {081101},
        pages = {081101},
          doi = {10.1103/PhysRevLett.106.081101},
archivePrefix = {arXiv},
       eprint = {1011.6142},
 primaryClass = {astro-ph.HE},
       adsurl = {https://ui.adsabs.harvard.edu/abs/2011PhRvL.106h1101P}
}

@ARTICLE{GlitchesZhou2022,
       author = {{Zhou}, Shiqi and {G{\"u}gercino{\u{g}}lu}, Erbil and {Yuan}, Jianping and {Ge}, Mingyu and {Yu}, Cong},
        title = "{Pulsar Glitches: A Review}",
      journal = {Universe},
         year = 2022,
        no-month = dec,
       volume = {8},
       number = {12},
          eid = {641},
        pages = {641},
          doi = {10.3390/universe8120641},
archivePrefix = {arXiv},
       eprint = {2211.13885},
 primaryClass = {astro-ph.HE},
       adsurl = {https://ui.adsabs.harvard.edu/abs/2022Univ....8..641Z}
}

@ARTICLE{Antonopoulou2022RPPh,
       author = {{Antonopoulou}, Danai and {Haskell}, Brynmor and {Espinoza}, Crist{\'o}bal M.},
        title = "{Pulsar glitches: observations and physical interpretation}",
      journal = {Reports on Progress in Physics},
         year = 2022,
        no-month = dec,
       volume = {85},
       number = {12},
          eid = {126901},
        pages = {126901},
          doi = {10.1088/1361-6633/ac9ced},
       adsurl = {https://ui.adsabs.harvard.edu/abs/2022RPPh...85l6901A}
}

@ARTICLE{Blaschke2013PhRvC,
       author = {{Blaschke}, D. and {Grigorian}, H. and {Voskresensky}, D.~N.},
        title = "{Nuclear medium cooling scenario in light of new Cas A cooling data and the 2M$_{{\ensuremath{\odot}}}$ pulsar mass measurements}",
      journal = {\prc},
         year = 2013,
        month = dec,
       volume = {88},
       number = {6},
          eid = {065805},
        pages = {065805},
          doi = {10.1103/PhysRevC.88.065805},
archivePrefix = {arXiv},
       eprint = {1308.4093},
 primaryClass = {nucl-th},
       adsurl = {https://ui.adsabs.harvard.edu/abs/2013PhRvC..88f5805B}
}

@ARTICLE{Baiotti2007,
       author = {{Baiotti}, Luca and {de Pietri}, Roberto and {Manca}, Gian Mario and {Rezzolla}, Luciano},
        title = "{Accurate simulations of the dynamical bar-mode instability in full general relativity}",
      journal = {\prd},
         year = 2007,
        month = feb,
       volume = {75},
       number = {4},
          eid = {044023},
        pages = {044023},
          doi = {10.1103/PhysRevD.75.044023},
archivePrefix = {arXiv},
       no-eprint = {astro-ph/0609473},
 primaryClass = {astro-ph},
       adsurl = {https://ui.adsabs.harvard.edu/abs/2007PhRvD..75d4023B}
}

@ARTICLE{Rau2021,
       author = {{Rau}, Peter B. and {Wasserman}, Ira},
        title = "{Magnetohydrodynamic stability of magnetars in the ultrastrong field regime I: the core}",
      journal = {\mnras},
         year = 2021,
        month = sep,
       volume = {506},
       number = {3},
        pages = {4632-4653},
          doi = {10.1093/mnras/stab1538},
archivePrefix = {arXiv},
       no-eprint = {2104.08563},
 primaryClass = {astro-ph.HE},
       adsurl = {https://ui.adsabs.harvard.edu/abs/2021MNRAS.506.4632R}
}

@ARTICLE{Rau2023,
       author = {{Rau}, Peter B. and {Wasserman}, Ira},
        title = "{Magnetohydrodynamic stability of magnetars in the ultrastrong field regime - II. The crust}",
      journal = {\mnras},
         year = 2023,
        month = mar,
       volume = {520},
       number = {1},
        pages = {1173-1192},
          doi = {10.1093/mnras/stad146},
archivePrefix = {arXiv},
       no-eprint = {2210.05774},
 primaryClass = {astro-ph.HE},
       adsurl = {https://ui.adsabs.harvard.edu/abs/2023MNRAS.520.1173R}
}

@ARTICLE{Sedrakian1999,
       author = {{Sedrakian}, Armen and {Cordes}, James M.},
        title = "{Vortex-interface interactions and generation of glitches in pulsars}",
      journal = {\mnras},
         year = 1999,
        month = aug,
       volume = {307},
       number = {2},
        pages = {365-375},
          doi = {10.1046/j.1365-8711.1999.02638.x},
archivePrefix = {arXiv},
       no-eprint = {astro-ph/9806042},
 primaryClass = {astro-ph},
       adsurl = {https://ui.adsabs.harvard.edu/abs/1999MNRAS.307..365S}
}

@ARTICLE{2014Goglu,
       author = {{G{\"u}gercino{\u{g}}lu}, Erbil and {Alpar}, M. Ali},
        title = "{Vortex Creep Against Toroidal Flux Lines, Crustal Entrainment, and Pulsar Glitches}",
      journal = {\apjl},
         year = 2014,
        month = jun,
       volume = {788},
       number = {1},
          eid = {L11},
        pages = {L11},
          doi = {10.1088/2041-8205/788/1/L11},
       adsurl = {https://ui.adsabs.harvard.edu/abs/2014ApJ...788L..11G}
}

@ARTICLE{Haskell2013,
       author = {{Haskell}, B. and {Pizzochero}, P.~M. and {Seveso}, S.},
        title = "{Investigating Superconductivity in Neutron Star Interiors with Glitch Models}",
      journal = {\apjl},
         year = 2013,
        month = feb,
       volume = {764},
       number = {2},
          eid = {L25},
        pages = {L25},
          doi = {10.1088/2041-8205/764/2/L25},
archivePrefix = {arXiv},
       no-eprint = {1209.6260},
 primaryClass = {astro-ph.SR},
       adsurl = {https://ui.adsabs.harvard.edu/abs/2013ApJ...764L..25H}
}

@ARTICLE{Baym1969Natur,
       author = {{Baym}, Gordon and {Pethick}, Christopher and {Pines}, David and {Ruderman}, Malvin},
        title = "{Spin Up in Neutron Stars : The Future of the Vela Pulsar}",
      journal = {\nat},
         year = 1969,
        month = nov,
       volume = {224},
       number = {5222},
        pages = {872-874},
          doi = {10.1038/224872a0},
       adsurl = {https://ui.adsabs.harvard.edu/abs/1969Natur.224..872B}
}

@ARTICLE{Andersson1975,
       author = {{Anderson}, P.~W. and {Itoh}, N.},
        title = "{Pulsar glitches and restlessness as a hard superfluidity phenomenon}",
      journal = {\nat},
         year = 1975,
        month = jul,
       volume = {256},
       number = {5512},
        pages = {25-27},
          doi = {10.1038/256025a0},
       adsurl = {https://ui.adsabs.harvard.edu/abs/1975Natur.256...25A}
}

@ARTICLE{Sedrakian_2025,
       author = {{Sedrakian}, Armen and {Rau}, Peter B.},
        title = "{Josephson currents in neutron stars}",
      journal = {\prd},
         year = 2025,
        month = jan,
       volume = {111},
       number = {2},
          eid = {023044},
        pages = {023044},
          doi = {10.1103/PhysRevD.111.023044},
archivePrefix = {arXiv},
       no-eprint = {2407.13686},
 primaryClass = {astro-ph.HE},
       adsurl = {https://ui.adsabs.harvard.edu/abs/2025PhRvD.111b3044S}
}

@article{V18,
  title = {Accurate nucleon-nucleon potential with charge-independence breaking},
  author = {Wiringa, R. B. and Stoks, V. G. J. and Schiavilla, R.},
  journal = {Phys. Rev. C},
  volume = {51},
  issue = {1},
  pages = {38--51},
  numpages = {0},
  year = {1995},
  month = {Jan},
  publisher = {American Physical Society},
  doi = {10.1103/PhysRevC.51.38}}

@article{UIX,
  title = {Quantum Monte Carlo Calculations of $\mathit{A}\ensuremath{\le}6$ Nuclei},
  author = {Pudliner, B. S. and Pandharipande, V. R. and Carlson, J. and Wiringa, R. B.},
  journal = {Phys. Rev. Lett.},
  volume = {74},
  issue = {22},
  pages = {4396--4399},
  numpages = {0},
  year = {1995},
  month = {May},
  publisher = {American Physical Society},
  doi = {10.1103/PhysRevLett.74.4396}}

@ARTICLE{Baldo2014,
       author = {{Baldo}, M. and {Burgio}, G.~F. and {Schulze}, H. -J. and {Taranto}, G.},
        title = "{Nucleon effective masses within the Brueckner-Hartree-Fock theory: Impact on stellar neutrino emission}",
      journal = {\prc},
         year = 2014,
        month = apr,
       volume = {89},
       number = {4},
          eid = {048801},
        pages = {048801},
          doi = {10.1103/PhysRevC.89.048801},
archivePrefix = {arXiv},
       no-eprint = {1404.7031},
 primaryClass = {nucl-th},
       adsurl = {https://ui.adsabs.harvard.edu/abs/2014PhRvC..89d8801B}
}

@ARTICLE{Sharma2015,
       author = {{Sharma}, B.~K. and {Centelles}, M. and {Vi{\~n}as}, X. and {Baldo}, M. and {Burgio}, G.~F.},
        title = "{Unified equation of state for neutron stars on a microscopic basis}",
      journal = {\aap},
         year = 2015,
        month = dec,
       volume = {584},
          eid = {A103},
        pages = {A103},
          doi = {10.1051/0004-6361/201526642},
archivePrefix = {arXiv},
       no-eprint = {1506.00375},
 primaryClass = {nucl-th},
       adsurl = {https://ui.adsabs.harvard.edu/abs/2015A&A...584A.103S}
}

@article{Abbott2018,
  author = {B. P. Abbott et al.},
  title = {GW170817: Measurements of neutron star radii and EOS},
  journal = {Phys. Rev. Lett.},
  volume = {121},
  pages = {161101},
  year = {2018}
}

@article{Brueckner1958,
  author = {K. A. Brueckner and J. L. Gammel},
  title = {Properties of nuclear matter},
  journal = {Phys. Rev.},
  volume = {109},
  pages = {1023},
  year = {1958}
}

@ARTICLE{Lander2009,
       author = {{Lander}, S.~K. and {Jones}, D.~I.},
        title = "{Magnetic fields in axisymmetric neutron stars}",
      journal = {\mnras},
         year = 2009,
        month = jun,
       volume = {395},
       number = {4},
        pages = {2162-2176},
          doi = {10.1111/j.1365-2966.2009.14667.x},
archivePrefix = {arXiv},
       no-eprint = {0903.0827},
 primaryClass = {astro-ph.SR},
       adsurl = {https://ui.adsabs.harvard.edu/abs/2009MNRAS.395.2162L}
}

@ARTICLE{Mastrano2015,
       author = {{Mastrano}, A. and {Suvorov}, A.~G. and {Melatos}, A.},
        title = "{Neutron star deformation due to poloidal-toroidal magnetic fields of arbitrary multipole order: a new analytic approach}",
      journal = {\mnras},
         year = 2015,
        month = mar,
       volume = {447},
       number = {4},
        pages = {3475-3485},
          doi = {10.1093/mnras/stu2671},
archivePrefix = {arXiv},
       no-eprint = {1501.01134},
 primaryClass = {astro-ph.HE},
       no-adsurl = {https://ui.adsabs.harvard.edu/abs/2015MNRAS.447.3475M}
}

@ARTICLE{Mastrano2011,
       author = {{Mastrano}, A. and {Melatos}, A. and {Reisenegger}, A. and {Akg{\"u}n}, T.},
        title = "{Gravitational wave emission from a magnetically deformed non-barotropic neutron star}",
      journal = {\mnras},
         year = 2011,
        month = nov,
       volume = {417},
       number = {3},
        pages = {2288-2299},
          doi = {10.1111/j.1365-2966.2011.19410.x},
archivePrefix = {arXiv},
       no-eprint = {1108.0219},
 primaryClass = {astro-ph.HE},
       adsurl = {https://ui.adsabs.harvard.edu/abs/2011MNRAS.417.2288M}
}

@ARTICLE{Akgun2013,
       author = {{Akg{\"u}n}, T. and {Reisenegger}, A. and {Mastrano}, A. and {Marchant}, P.},
        title = "{Stability of magnetic fields in non-barotropic stars: an analytic treatment}",
      journal = {\mnras},
         year = 2013,
        month = aug,
       volume = {433},
       number = {3},
        pages = {2445-2466},
          doi = {10.1093/mnras/stt913},
archivePrefix = {arXiv},
       no-eprint = {1302.0273},
 primaryClass = {astro-ph.SR},
       adsurl = {https://ui.adsabs.harvard.edu/abs/2013MNRAS.433.2445A}
}

@ARTICLE{Ciolfi2009,
       author = {{Ciolfi}, R. and {Ferrari}, V. and {Gualtieri}, L. and {Pons}, J.~A.},
        title = "{Relativistic models of magnetars: the twisted torus magnetic field configuration}",
      journal = {\mnras},
         year = 2009,
        month = aug,
       volume = {397},
       number = {2},
        pages = {913-924},
          doi = {10.1111/j.1365-2966.2009.14990.x},
archivePrefix = {arXiv},
       no-eprint = {0903.0556},
 primaryClass = {astro-ph.SR},
       adsurl = {https://ui.adsabs.harvard.edu/abs/2009MNRAS.397..913C}
}

@article{BraithwaiteNordlund2006,
  author       = {Braithwaite, J. and Nordlund, A.},
  title        = {Stable magnetic fields in stellar interiors},
  journal      = {Astronomy \& Astrophysics},
  year         = {2006},
  volume       = {450},
  pages        = {1077--1095},
  doi          = {10.1051/0004-6361:20053695},
  publisher    = {EDP Sciences}
}

@article{Cieslar2021,
  author = {Cieślar, M. and Bulik, T. and Curyło, M. and Kowalska-Leszczynska, I. and Tunstall, L. and Woźna, A. and Życki, P.},
  title = {Prospects for the detection of continuous gravitational waves from known pulsars with future ground-based detectors},
  journal = {Astronomy \& Astrophysics},
  volume = {649},
  pages = {A92},
  year = {2021},
  doi = {10.1051/0004-6361/202039998}
}

@article{zuraiq2024,
  title = {Massive neutron stars as mass gap candidates: Exploring equation of state and magnetic field},
  author = {Zuraiq, Zenia and Mukhopadhyay, Banibrata and Weber, Fridolin},
  journal = {Phys. Rev. D},
  volume = {109},
  issue = {2},
  pages = {023027},
  numpages = {17},
  year = {2024},
  month = {Jan},
  publisher = {American Physical Society},
  doi = {10.1103/PhysRevD.109.023027}
}

@ARTICLE{Biswas2022,
       author = {{Biswas}, Bhaskar},
        title = "{Bayesian Model Selection of Neutron Star Equations of State Using Multi-messenger Observations}",
      journal = {\apj},
         year = 2022,
        month = feb,
       volume = {926},
       number = {1},
          eid = {75},
        pages = {75},
          doi = {10.3847/1538-4357/ac447b},
archivePrefix = {arXiv},
       no-eprint = {2106.02644},
 primaryClass = {astro-ph.HE},
       adsurl = {https://ui.adsabs.harvard.edu/abs/2022ApJ...926...75B}
}

@article{Krastev2018,
  title={Imprints of the nuclear symmetry energy on the tidal deformability of neutron stars},
  author={Krastev, Plamen G. and Li, Bao-An},
  journal={Journal of Physics G: Nuclear and Particle Physics},
  volume={45},
  number={7},
  pages={074001},
  year={2018},
  doi={10.1088/1361-6471/aac6f1},
  archivePrefix={arXiv},
  no-eprint={1801.04620},
  primaryClass={nucl-th}
}

@ARTICLE{Abbot2018lambda,
       author = {{Abbott}, B.~P. and {Abbott}, R. and {Abbott}, T.~D. and {Acernese}, F. and {Ackley}, K. and {Adams}, C. and {Adams}, T. and {Addesso}, P. and {Adhikari}, R.~X. and {Adya}, V.~B. and {Affeldt}, C. and {Agarwal}, B. and {Agathos}, M. and {Agatsuma}, K. and {Aggarwal}, N. and {Aguiar}, O.~D. and {Aiello}, L. and {Ain}, A. and {Ajith}, P. and {Allen}, B. and {Allen}, G. and {Allocca}, A. and {Aloy}, M.~A. and {Altin}, P.~A. and {Amato}, A. and {Ananyeva}, A. and {Anderson}, S.~B. and {Anderson}, W.~G. and {Angelova}, S.~V. and {Antier}, S. and {Appert}, S. and {Arai}, K. and {Araya}, M.~C. and {Areeda}, J.~S. and {Ar{\`e}ne}, M. and {Arnaud}, N. and {Arun}, K.~G. and {Ascenzi}, S. and {Ashton}, G. and {Ast}, M. and {Aston}, S.~M. and {Astone}, P. and {Atallah}, D.~V. and {Aubin}, F. and {Aufmuth}, P. and {Aulbert}, C. and {AultONeal}, K. and {Austin}, C. and {Avila-Alvarez}, A. and {Babak}, S. and {Bacon}, P. and {Badaracco}, F. and {Bader}, M.~K.~M. and {Bae}, S. and {Baker}, P.~T. and {Baldaccini}, F. and {Ballardin}, G. and {Ballmer}, S.~W. and {Banagiri}, S. and {Barayoga}, J.~C. and {Barclay}, S.~E. and {Barish}, B.~C. and {Barker}, D. and {Barkett}, K. and {Barnum}, S. and {Barone}, F. and {Barr}, B. and {Barsotti}, L. and {Barsuglia}, M. and {Barta}, D. and {Bartlett}, J. and {Bartos}, I. and {Bassiri}, R. and {Basti}, A. and {Batch}, J.~C. and {Bawaj}, M. and {Bayley}, J.~C. and {Bazzan}, M. and {B{\'e}csy}, B. and {Beer}, C. and {Bejger}, M. and {Belahcene}, I. and {Bell}, A.~S. and {Beniwal}, D. and {Bensch}, M. and {Berger}, B.~K. and {Bergmann}, G. and {Bernuzzi}, S. and {Bero}, J.~J. and {Berry}, C.~P.~L. and {Bersanetti}, D. and {Bertolini}, A. and {Betzwieser}, J. and {Bhandare}, R. and {Bilenko}, I.~A. and {Bilgili}, S.~A. and {Billingsley}, G. and {Billman}, C.~R. and {Birch}, J. and {Birney}, R. and {Birnholtz}, O. and {Biscans}, S. and {Biscoveanu}, S. and {Bisht}, A. and {Bitossi}, M. and {Bizouard}, M.~A. and {Blackburn}, J.~K. and {Blackman}, J. and {Blair}, C.~D. and {Blair}, D.~G. and {Blair}, R.~M. and {Bloemen}, S. and {Bock}, O. and {Bode}, N. and {Boer}, M. and {Boetzel}, Y. and {Bogaert}, G. and {Bohe}, A. and {Bondu}, F. and {Bonilla}, E. and {Bonnand}, R. and {Booker}, P. and {Boom}, B.~A. and {Booth}, C.~D. and {Bork}, R. and {Boschi}, V. and {Bose}, S. and {Bossie}, K. and {Bossilkov}, V. and {Bosveld}, J. and {Bouffanais}, Y. and {Bozzi}, A. and {Bradaschia}, C. and {Brady}, P.~R. and {Bramley}, A. and {Branchesi}, M. and {Brau}, J.~E. and {Briant}, T. and {Brighenti}, F. and {Brillet}, A. and {Brinkmann}, M. and {Brisson}, V. and {Brockill}, P. and {Brooks}, A.~F. and {Brown}, D.~D. and {Brunett}, S. and {Buchanan}, C.~C. and {Buikema}, A. and {Bulik}, T. and {Bulten}, H.~J. and {Buonanno}, A. and {Buskulic}, D. and {Buy}, C. and {Byer}, R.~L. and {Cabero}, M. and {Cadonati}, L. and {Cagnoli}, G. and {Cahillane}, C. and {Calder{\'o}n Bustillo}, J. and {Callister}, T.~A. and {Calloni}, E. and {Camp}, J.~B. and {Canepa}, M. and {Canizares}, P. and {Cannon}, K.~C. and {Cao}, H. and {Cao}, J. and {Capano}, C.~D. and {Capocasa}, E. and {Carbognani}, F. and {Caride}, S. and {Carney}, M.~F. and {Carullo}, G. and {Casanueva Diaz}, J. and {Casentini}, C. and {Caudill}, S. and {Cavagli{\`a}}, M. and {Cavalier}, F. and {Cavalieri}, R. and {Cella}, G. and {Cepeda}, C.~B. and {Cerd{\'a}-Dur{\'a}n}, P. and {Cerretani}, G. and {Cesarini}, E. and {Chaibi}, O. and {Chamberlin}, S.~J. and {Chan}, M. and {Chao}, S. and {Charlton}, P. and {Chase}, E. and {Chassande-Mottin}, E. and {Chatterjee}, D. and {Chatziioannou}, K. and {Cheeseboro}, B.~D. and {Chen}, H.~Y. and {Chen}, X. and {Chen}, Y. and {Cheng}, H. -P. and {Chia}, H.~Y. and {Chincarini}, A.},
        title = "{GW170817: Measurements of Neutron Star Radii and Equation of State}",
      journal = {\prl},
         year = 2018,
        month = oct,
       volume = {121},
       number = {16},
          eid = {161101},
        pages = {161101},
          doi = {10.1103/PhysRevLett.121.161101},
archivePrefix = {arXiv},
       no-eprint = {1805.11581},
 primaryClass = {gr-qc},
       adsurl = {https://ui.adsabs.harvard.edu/abs/2018PhRvL.121p1101A}
}

@article{DasMB2025,
  author  = {Das, Mayusree and Mukhopadhyay, Banibrata and Bulik, Tomasz},
  title   = {Continuous Gravitational Waves from Magnetized White Dwarfs: Quantifying the Detection Plausibility by LISA},
  journal = {ApJ},
  year    = {2025},
  doi     = {10.3847/1538-4357/ae1732},
  note    = {Accepted for publication; in press.}
}

@article{Fortin2018,
  author    = {M. Fortin and G. Taranto and G. F. Burgio and P. Haensel and H.-J. Schulze and J. L. Zdunik},
  title     = {Neutron star radii and crusts: Uncertainties and unified equations of state},
  journal   = {\mnras},
  volume    = {475},
  number    = {4},
  pages     = {5010--5022},
  year      = {2018},
  doi       = {10.1093/mnras/sty181},
  publisher = {Oxford University Press}
}

@article{Cromartie2019,
  author       = {H. T. Cromartie and E. Fonseca and S. M. Ransom and P. B. Demorest and Z. Arzoumanian and H. Blumer and P. R. Brook and M. E. DeCesar and T. Dolch and J. A. Ellis and R. D. Ferdman and E. C. Ferrara and N. Garver-Daniels and P. A. Gentile and M. L. Jones and M. T. Lam and D. R. Lorimer and R. S. Lynch and A. N. Lommen and C. Ng and D. J. Nice and T. T. Pennucci and S. M. Ransom and L. S. Ray and R. M. Shannon and I. H. Stairs and D. R. Stinebring and K. Stovall and J. K. Swiggum and W. Zhu},
  title        = {Relativistic Shapiro delay measurements of an extremely massive millisecond pulsar},
  journal      = {Nature Astronomy},
  year         = {2019},
  volume       = {4},
  pages        = {72--76},
  doi          = {10.1038/s41550-019-0880-2}
}

@article{pili2014,
  author       = {A. G. Pili and N. Bucciantini and L. Del Zanna},
  title        = {Axisymmetric equilibrium models for magnetized neutron stars in general relativity under the Conformally Flat Condition},
  journal      = {\mnras},
  volume       = {439},
  number       = {4},
  pages        = {3541--3563},
  year         = {2014},
  doi          = {10.1093/mnras/stu217},
  no-eprint       = {1312.3943},
  archivePrefix= {arXiv},
  primaryClass = {astro-ph.HE}
}

@book{Tinkham1996,
  author    = {Michael Tinkham},
  title     = {Introduction to Superconductivity},
  edition   = {2nd},
  publisher = {McGraw-Hill},
  year      = {1996}
}

@article{Abrikosov1957a,
  author  = {A. A. Abrikosov},
  title   = {On the Magnetic Properties of Superconductors of the Second Group},
  journal = {Soviet Physics JETP},
  volume  = {5},
  pages   = {1174--1182},
  year    = {1957}
}

@article{Sedrakian2019,
  author       = {Armen Sedrakian and John W. Clark},
  title        = {Superfluidity in nuclear systems and neutron stars},
  journal      = {European Physical Journal A},
  year         = {2019},
  volume       = {55},
  pages        = {167},
  doi          = {10.1140/epja/i2019-12863-6}
}

@article{Taranto2013,
  author = {G. Taranto and M. Baldo and G. F. Burgio},
  title = {Density dependence of symmetry energy},
  journal = {Phys. Rev. C},
  volume = {87},
  pages = {045803},
  year = {2013}
}

@article{Baldo1992,
  author       = {M. Baldo and J. Cugnon and A. Lejeune and U. Lombardo},
  title        = {Proton and neutron superfluidity in neutron star matter},
  journal      = {Nuclear Physics A},
  volume       = {536},
  number       = {3},
  pages        = {349--365},
  year         = {1992},
  doi          = {10.1016/0375-9474(92)90904-B}
}

@article{Tayler1973,
    author = {Tayler, R. J.},
    title = "{ The Adiabatic Stability of Stars Containing Magnetic Fields–I: T OROIDAL F IELDS}",
    journal = {MNRAS},
    volume = {161},
    number = {4},
    pages = {365-380},
    year = {1973},
    month = {04},
    issn = {0035-8711},
    doi = {10.1093/mnras/161.4.365},
}

@ARTICLE{Wright1973,
       author = {{Wright}, G.~A.~E.},
        title = "{Pinch instabilities in magnetic stars}",
      journal = {\mnras},
         year = 1973,
        month = feb,
       volume = {162},
        pages = {339-358},
          doi = {10.1093/mnras/162.4.339},
       adsurl = {https://ui.adsabs.harvard.edu/abs/1973MNRAS.162..339W}
}

@ARTICLE{Braithwaite2006,
       author = {{Braithwaite}, J.},
        title = "{The stability of toroidal fields in stars}",
      journal = {\aap},
         year = 2006,
        month = jul,
       volume = {453},
       number = {2},
        pages = {687-698},
          doi = {10.1051/0004-6361:20041282},
archivePrefix = {arXiv},
       no-eprint = {astro-ph/0512182},
 primaryClass = {astro-ph},
       adsurl = {https://ui.adsabs.harvard.edu/abs/2006A&A...453..687B}
}

@ARTICLE{Markey1973,
       author = {{Markey}, P. and {Tayler}, R.~J.},
        title = "{The adiabatic stability of stars containing magnetic fields. II. Poloidal fields}",
      journal = {\mnras},
         year = 1973,
        month = mar,
       volume = {163},
        pages = {77-91},
          doi = {10.1093/mnras/163.1.77},
       adsurl = {https://ui.adsabs.harvard.edu/abs/1973MNRAS.163...77M}
}

@ARTICLE{CR2013,
       author = {{Ciolfi}, R. and {Rezzolla}, L.},
        title = "{Twisted-torus configurations with large toroidal magnetic fields in  relativistic stars.}",
      journal = {\mnras},
         year = 2013,
        month = aug,
       volume = {435},
        pages = {L43-L47},
          doi = {10.1093/mnrasl/slt092},
archivePrefix = {arXiv},
       no-eprint = {1306.2803},
 primaryClass = {astro-ph.SR},
       adsurl = {https://ui.adsabs.harvard.edu/abs/2013MNRAS.435L..43C}
}

@article{moore2015,
  author       = {Christopher J. Moore and Robert H. Cole and Christopher P. L. Berry},
  title        = {Gravitational-wave sensitivity curves},
  journal      = {Classical and Quantum Gravity},
  volume       = {32},
  number       = {1},
  pages        = {015014},
  year         = {2015}}

@ARTICLE{romani2021,
       author = {{Romani}, Roger W. and {Kandel}, D. and {Filippenko}, Alexei V. and {Brink}, Thomas G. and {Zheng}, WeiKang},
        title = "{PSR J1810+1744: Companion Darkening and a Precise High Neutron Star Mass}",
      journal = {\apjl},
         year = 2021,
        month = feb,
       volume = {908},
       number = {2},
          eid = {L46},
        pages = {L46},
          doi = {10.3847/2041-8213/abe2b4},
archivePrefix = {arXiv},
       no-eprint = {2101.09822},
 primaryClass = {astro-ph.HE},
       adsurl = {https://ui.adsabs.harvard.edu/abs/2021ApJ...908L..46R}
}

@PHDTHESIS{sanpa2016,
       author = {{Sanpa-Arsa}, Siraprapa},
        title = "{Searching for new millisecond pulsars with the GBT in Fermi unassociated sources}",
       school = {University of Virginia},
         year = 2016,
        month = jan,
       adsurl = {https://ui.adsabs.harvard.edu/abs/2016PhDT.......539S}
}

@ARTICLE{bates2015,
       author = {{Bates}, S.~D. and {Thornton}, D. and {Bailes}, M. and {Barr}, E. and {Bassa}, C.~G. and {Bhat}, N.~D.~R. and {Burgay}, M. and {Burke-Spolaor}, S. and {Champion}, D.~J. and {Flynn}, C.~M.~L. and {Jameson}, A. and {Johnston}, S. and {Keith}, M.~J. and {Kramer}, M. and {Levin}, L. and {Lyne}, A. and {Milia}, S. and {Ng}, C. and {Petroff}, E. and {Possenti}, A. and {Stappers}, B.~W. and {van Straten}, W. and {Tiburzi}, C.},
        title = "{The High Time Resolution Universe survey - XI. Discovery of five recycled pulsars and the optical detectability of survey white dwarf companions}",
      journal = {\mnras},
         year = 2015,
        month = feb,
       volume = {446},
       number = {4},
        pages = {4019-4028},
          doi = {10.1093/mnras/stu2350},
archivePrefix = {arXiv},
       no-eprint = {1411.1288},
 primaryClass = {astro-ph.SR},
       adsurl = {https://ui.adsabs.harvard.edu/abs/2015MNRAS.446.4019B}
}

@ARTICLE{abdo2010,
       author = {{Abdo}, A.~A. and {Ackermann}, M. and {Ajello}, M. and {Allafort}, A. and {Baldini}, L. and {Ballet}, J. and {Barbiellini}, G. and {Bastieri}, D. and {Bechtol}, K. and {Bellazzini}, R. and {Berenji}, B. and {Blandford}, R.~D. and {Bloom}, E.~D. and {Bonamente}, E. and {Borgland}, A.~W. and {Bouvier}, A. and {Bregeon}, J. and {Brez}, A. and {Brigida}, M. and {Bruel}, P. and {Burnett}, T.~H. and {Buson}, S. and {Caliandro}, G.~A. and {Cameron}, R.~A. and {Camilo}, F. and {Caraveo}, P.~A. and {Carrigan}, S. and {Casandjian}, J.~M. and {Cecchi}, C. and {{\c{C}}elik}, {\"O}. and {Chekhtman}, A. and {Cheung}, C.~C. and {Chiang}, J. and {Ciprini}, S. and {Claus}, R. and {Cognard}, I. and {Cohen-Tanugi}, J. and {Conrad}, J. and {Corbet}, R. and {DeCesar}, M.~E. and {Dermer}, C.~D. and {Desvignes}, G. and {de Angelis}, A. and {de Palma}, F. and {Digel}, S.~W. and {Dormody}, M. and {Silva}, E. do Couto e. and {Drell}, P.~S. and {Dubois}, R. and {Dumora}, D. and {Espinoza}, C. and {Farnier}, C. and {Favuzzi}, C. and {Fegan}, S.~J. and {Focke}, W.~B. and {Frailis}, M. and {Freire}, P.~C.~C. and {Fukazawa}, Y. and {Funk}, S. and {Fusco}, P. and {Gargano}, F. and {Gasparrini}, D. and {Gehrels}, N. and {Germani}, S. and {Giavitto}, G. and {Giglietto}, N. and {Giordano}, F. and {Glanzman}, T. and {Godfrey}, G. and {Grenier}, I.~A. and {Grondin}, M. -H. and {Grove}, J.~E. and {Guillemot}, L. and {Guiriec}, S. and {Hadasch}, D. and {Harding}, A.~K. and {Hays}, E. and {Hobbs}, G. and {Horan}, D. and {Hughes}, R.~E. and {J{\'o}hannesson}, G. and {Johnson}, A.~S. and {Johnson}, T.~J. and {Johnson}, W.~N. and {Johnston}, S. and {Kamae}, T. and {Katagiri}, H. and {Kataoka}, J. and {Kawai}, N. and {Kerr}, M. and {Kn{\"o}dlseder}, J. and {Kramer}, M. and {Kuss}, M. and {Lande}, J. and {Latronico}, L. and {Lemoine-Goumard}, M. and {Llena Garde}, M. and {Longo}, F. and {Loparco}, F. and {Lott}, B. and {Lovellette}, M.~N. and {Lubrano}, P. and {Lyne}, A.~G. and {Makeev}, A. and {Manchester}, R.~N. and {Marelli}, M. and {Mazziotta}, M.~N. and {McConville}, W. and {McEnery}, J.~E. and {McGlynn}, S. and {Meurer}, C. and {Michelson}, P.~F. and {Mitthumsiri}, W. and {Mizuno}, T. and {Moiseev}, A.~A. and {Monte}, C. and {Monzani}, M.~E. and {Morselli}, A. and {Moskalenko}, I.~V. and {Murgia}, S. and {Nolan}, P.~L. and {Norris}, J.~P. and {Noutsos}, A. and {Nuss}, E. and {Ohsugi}, T. and {Omodei}, N. and {Orlando}, E. and {Ormes}, J.~F. and {Ozaki}, M. and {Paneque}, D. and {Panetta}, J.~H. and {Parent}, D. and {Pelassa}, V. and {Pepe}, M. and {Pesce-Rollins}, M. and {Pierbattista}, M. and {Piron}, F. and {Porter}, T.~A. and {Rain{\`o}}, S. and {Rando}, R. and {Ransom}, S.~M. and {Razzano}, M. and {Reimer}, A. and {Reimer}, O. and {Reposeur}, T. and {Ripken}, J. and {Ritz}, S. and {Rochester}, L.~S. and {Rodriguez}, A.~Y. and {Romani}, R.~W. and {Roth}, M. and {Ryde}, F. and {Sadrozinski}, H.~F. -W. and {Sander}, A. and {Saz Parkinson}, P.~M. and {Scargle}, J.~D. and {Sgr{\`o}}, C. and {Siskind}, E.~J. and {Smith}, D.~A. and {Smith}, P.~D. and {Spandre}, G. and {Spinelli}, P. and {Stappers}, B.~W. and {Starck}, J. -L. and {Strickman}, M.~S. and {Suson}, D.~J. and {Takahashi}, H. and {Tanaka}, T. and {Thayer}, J.~B. and {Thayer}, J.~G. and {Theureau}, G. and {Thompson}, D.~J. and {Thorsett}, S.~E. and {Tibaldo}, L. and {Torres}, D.~F. and {Tosti}, G. and {Tramacere}, A. and {Usher}, T.~L. and {Van Etten}, A. and {Vasileiou}, V. and {Venter}, C. and {Vilchez}, N. and {Vitale}, V. and {Waite}, A.~P. and {Wallace}, E. and {Wang}, P. and {Weltevrede}, P. and {Winer}, B.~L. and {Wood}, K.~S. and {Ylinen}, T. and {Ziegler}, M.},
        title = "{Discovery of Pulsed {\ensuremath{\gamma}}-Rays from PSR J0034-0534 with the Fermi Large Area Telescope: A Case for Co-Located Radio and {\ensuremath{\gamma}}-Ray Emission Regions}",
      journal = {\apj},
         year = 2010,
        month = apr,
       volume = {712},
       number = {2},
        pages = {957-963},
          doi = {10.1088/0004-637X/712/2/957},
archivePrefix = {arXiv},
       no-eprint = {1002.2607},
 primaryClass = {astro-ph.HE},
       adsurl = {https://ui.adsabs.harvard.edu/abs/2010ApJ...712..957A}
}

@ARTICLE{ligo2022,
       author = {{Abbott}, R. and {Abe}, H. and {Acernese}, F. and {Ackley}, K. and {Adhikari}, N. and {Adhikari}, R.~X. and {Adkins}, V.~K. and {Adya}, V.~B. and {Affeldt}, C. and {Agarwal}, D. and {Agathos}, M. and {Agatsuma}, K. and {Aggarwal}, N. and {Aguiar}, O.~D. and {Aiello}, L. and {Ain}, A. and {Ajith}, P. and {Akutsu}, T. and {Albanesi}, S. and {Alfaidi}, R.~A. and {Allocca}, A. and {Altin}, P.~A. and {Amato}, A. and {Anand}, C. and {Anand}, S. and {Ananyeva}, A. and {Anderson}, S.~B. and {Anderson}, W.~G. and {Ando}, M. and {Andrade}, T. and {Andres}, N. and {Andr{\'e}s-Carcasona}, M. and {Andri{\'c}}, T. and {Angelova}, S.~V. and {Ansoldi}, S. and {Antelis}, J.~M. and {Antier}, S. and {Apostolatos}, T. and {Appavuravther}, E.~Z. and {Appert}, S. and {Apple}, S.~K. and {Arai}, K. and {Araya}, A. and {Araya}, M.~C. and {Areeda}, J.~S. and {Ar{\`e}ne}, M. and {Aritomi}, N. and {Arnaud}, N. and {Arogeti}, M. and {Aronson}, S.~M. and {Arun}, K.~G. and {Asada}, H. and {Asali}, Y. and {Ashton}, G. and {Aso}, Y. and {Assiduo}, M. and {de Souza Melo}, S. Assis and {Aston}, S.~M. and {Astone}, P. and {Aubin}, F. and {Aultoneal}, K. and {Austin}, C. and {Babak}, S. and {Badaracco}, F. and {Bader}, M.~K.~M. and {Badger}, C. and {Bae}, S. and {Bae}, Y. and {Baer}, A.~M. and {Bagnasco}, S. and {Bai}, Y. and {Bailes}, M. and {Baird}, J. and {Bajpai}, R. and {Baka}, T. and {Ball}, M. and {Ballardin}, G. and {Ballmer}, S.~W. and {Balsamo}, A. and {Baltus}, G. and {Banagiri}, S. and {Banerjee}, B. and {Bankar}, D. and {Barayoga}, J.~C. and {Barbieri}, C. and {Barish}, B.~C. and {Barker}, D. and {Barneo}, P. and {Barone}, F. and {Barr}, B. and {Barsotti}, L. and {Barsuglia}, M. and {Barta}, D. and {Bartlett}, J. and {Barton}, M.~A. and {Bartos}, I. and {Basak}, S. and {Bassiri}, R. and {Basti}, A. and {Bawaj}, M. and {Bayley}, J.~C. and {Bazzan}, M. and {Becher}, B.~R. and {B{\'e}csy}, B. and {Bedakihale}, V.~M. and {Beirnaert}, F. and {Bejger}, M. and {Belahcene}, I. and {Benedetto}, V. and {Beniwal}, D. and {Benjamin}, M.~G. and {Bennett}, T.~F. and {Bentley}, J.~D. and {Benyaala}, M. and {Bera}, S. and {Berbel}, M. and {Bergamin}, F. and {Berger}, B.~K. and {Bernuzzi}, S. and {Bersanetti}, D. and {Bertolini}, A. and {Betzwieser}, J. and {Beveridge}, D. and {Bhandare}, R. and {Bhandari}, A.~V. and {Bhardwaj}, U. and {Bhatt}, R. and {Bhattacharjee}, D. and {Bhaumik}, S. and {Bianchi}, A. and {Bilenko}, I.~A. and {Billingsley}, G. and {Bini}, S. and {Birney}, I.~A. and {Birnholtz}, O. and {Biscans}, S. and {Bischi}, M. and {Biscoveanu}, S. and {Bisht}, A. and {Biswas}, B. and {Bitossi}, M. and {Bizouard}, M. -A. and {Blackburn}, J.~K. and {Blair}, C.~D. and {Blair}, D.~G. and {Blair}, R.~M. and {Bobba}, F. and {Bode}, N. and {Bo{\"e}r}, M. and {Bogaert}, G. and {Boldrini}, M. and {Bolingbroke}, G.~N. and {Bonavena}, L.~D. and {Bondu}, F. and {Bonilla}, E. and {Bonnand}, R. and {Booker}, P. and {Boom}, B.~A. and {Bork}, R. and {Boschi}, V. and {Bose}, N. and {Bose}, S. and {Bossilkov}, V. and {Boudart}, V. and {Bouffanais}, Y. and {Bozzi}, A. and {Bradaschia}, C. and {Brady}, P.~R. and {Bramley}, A. and {Branch}, A. and {Branchesi}, M. and {Brau}, J.~E. and {Breschi}, M. and {Briant}, T. and {Briggs}, J.~H. and {Brillet}, A. and {Brinkmann}, M. and {Brockill}, P. and {Brooks}, A.~F. and {Brooks}, J. and {Brown}, D.~D. and {Brunett}, S. and {Bruno}, G. and {Bruntz}, R. and {Bryant}, J. and {Bucci}, F. and {Bulik}, T. and {Bulten}, H.~J. and {Buonanno}, A. and {Burtnyk}, K. and {Buscicchio}, R. and {Buskulic}, D. and {Buy}, C. and {Byer}, R.~L. and {Davies}, G.~S. Cabourn and {Cabras}, G. and {Cabrita}, R. and {Cadonati}, L. and {Caesar}, M. and {Cagnoli}, G.},
        title = "{Searches for Gravitational Waves from Known Pulsars at Two Harmonics in the Second and Third LIGO-Virgo Observing Runs}",
      journal = {\apj},
         year = 2022,
        month = aug,
       volume = {935},
       number = {1},
          eid = {1},
        pages = {1},
          doi = {10.3847/1538-4357/ac6acf},
archivePrefix = {arXiv},
       no-eprint = {2111.13106},
 primaryClass = {astro-ph.HE},
       adsurl = {https://ui.adsabs.harvard.edu/abs/2022ApJ...935....1A}
}

@ARTICLE{guill2012,
       author = {{Guillemot}, L. and {Johnson}, T.~J. and {Venter}, C. and {Kerr}, M. and {Pancrazi}, B. and {Livingstone}, M. and {Janssen}, G.~H. and {Jaroenjittichai}, P. and {Kramer}, M. and {Cognard}, I. and {Stappers}, B.~W. and {Harding}, A.~K. and {Camilo}, F. and {Espinoza}, C.~M. and {Freire}, P.~C.~C. and {Gargano}, F. and {Grove}, J.~E. and {Johnston}, S. and {Michelson}, P.~F. and {Noutsos}, A. and {Parent}, D. and {Ransom}, S.~M. and {Ray}, P.~S. and {Shannon}, R. and {Smith}, D.~A. and {Theureau}, G. and {Thorsett}, S.~E. and {Webb}, N.},
        title = "{Pulsed Gamma Rays from the Original Millisecond and Black Widow Pulsars: A Case for Caustic Radio Emission?}",
      journal = {\apj},
         year = 2012,
        month = jan,
       volume = {744},
       number = {1},
          eid = {33},
        pages = {33},
          doi = {10.1088/0004-637X/744/1/33},
archivePrefix = {arXiv},
       no-eprint = {1110.1271},
 primaryClass = {astro-ph.HE},
       adsurl = {https://ui.adsabs.harvard.edu/abs/2012ApJ...744...33G}
}

@ARTICLE{cordes2002,
       author = {{Cordes}, J.~M. and {Lazio}, T.~J.~W.},
        title = "{NE2001.I. A New Model for the Galactic Distribution of Free Electrons and its Fluctuations}",
      journal = {arXiv e-prints},
         year = 2002,
        month = jul,
          eid = {astro-ph/0207156},
        pages = {astro-ph/0207156},
          doi = {10.48550/arXiv.astro-ph/0207156},
archivePrefix = {arXiv},
       no-eprint = {astro-ph/0207156},
 primaryClass = {astro-ph},
       adsurl = {https://ui.adsabs.harvard.edu/abs/2002astro.ph..7156C}
}

@ARTICLE{chisabi2025,
       author = {{Chisabi}, M. and {Andrianomena}, S. and {Enwelum}, U. and {Gasennelwe}, E.~G. and {Idris}, A. and {Idogbe}, E.~A. and {Shilunga}, S. and {Geyer}, M. and {Reardon}, D.~J. and {Okany}, C.~F. and {Shamohammadi}, M. and {Shannon}, R.~M. and {Krishnan}, V. Venkatraman and {Abbate}, F. and {Kramer}, M.},
        title = "{Timing and noise analysis of five millisecond pulsars observed with MeerKAT}",
      journal = {\mnras},
         year = 2025,
        month = mar,
       volume = {537},
       number = {3},
        pages = {2462-2470},
          doi = {10.1093/mnras/staf100},
archivePrefix = {arXiv},
       no-eprint = {2501.07728},
 primaryClass = {astro-ph.HE},
       adsurl = {https://ui.adsabs.harvard.edu/abs/2025MNRAS.537.2462C}
}

@ARTICLE{ser2022,
       author = {{Serylak}, M. and {Venkatraman Krishnan}, V. and {Freire}, P.~C.~C. and {Tauris}, T.~M. and {Kramer}, M. and {Geyer}, M. and {Parthasarathy}, A. and {Bailes}, M. and {Bernadich}, M.~C. i. and {Buchner}, S. and {Burgay}, M. and {Camilo}, F. and {Karastergiou}, A. and {Lower}, M.~E. and {Possenti}, A. and {Reardon}, D.~J. and {Shannon}, R.~M. and {Spiewak}, R. and {Stairs}, I.~H. and {van Straten}, W.},
        title = "{The eccentric millisecond pulsar, PSR J0955{\ensuremath{-}}6150. I. Pulse profile analysis, mass measurements, and constraints on binary evolution}",
      journal = {\aap},
         year = 2022,
        month = sep,
       volume = {665},
          eid = {A53},
        pages = {A53},
          doi = {10.1051/0004-6361/202142670},
archivePrefix = {arXiv},
       no-eprint = {2203.00607},
 primaryClass = {astro-ph.HE},
       adsurl = {https://ui.adsabs.harvard.edu/abs/2022A&A...665A..53S}
}

@book{landau1980statistical,
  title        = {Statistical Physics, Part 2},
  author       = {Landau, L. D. and Lifshitz, E. M.},
  series       = {Course of Theoretical Physics},
  volume       = {9},
  edition      = {1st},
  year         = {1980},
  publisher    = {Pergamon Press},
  address      = {Oxford},
  isbn         = {9780080230393}
}

@ARTICLE{Buckley2004,
       author = {{Buckley}, Kirk B. and {Metlitski}, Max A. and {Zhitnitsky}, Ariel R.},
        title = "{Neutron Stars as Type-I Superconductors}",
      journal = {\prl},
         year = 2004,
        month = apr,
       volume = {92},
       number = {15},
          eid = {151102},
        pages = {151102},
          doi = {10.1103/PhysRevLett.92.151102},
archivePrefix = {arXiv},
       no-eprint = {astro-ph/0308148},
 primaryClass = {astro-ph},
       adsurl = {https://ui.adsabs.harvard.edu/abs/2004PhRvL..92o1102B}
}

@ARTICLE{Suvorov2025,
       author = {{Suvorov}, Arthur G. and {Stefanou}, Petros and {Pons}, Jos{\'e} A.},
        title = "{Universality of gravitational radiation from magnetar magnetospheres}",
      journal = {arXiv e-prints},
         year = 2025,
        month = jul,
          eid = {arXiv:2507.14634},
        pages = {arXiv:2507.14634},
          doi = {10.48550/arXiv.2507.14634},
archivePrefix = {arXiv},
       no-eprint = {2507.14634},
 primaryClass = {gr-qc},
       adsurl = {https://ui.adsabs.harvard.edu/abs/2025arXiv250714634S}
}

@ARTICLE{Larson1999,
       author = {{Larson}, Michelle B. and {Link}, Bennett},
        title = "{Superfluid Friction and Late-Time Thermal Evolution of Neutron Stars}",
      journal = {\apj},
         year = 1999,
        month = aug,
       volume = {521},
       number = {1},
        pages = {271-280},
          doi = {10.1086/307532},
archivePrefix = {arXiv},
       no-eprint = {astro-ph/9810441},
 primaryClass = {astro-ph},
       adsurl = {https://ui.adsabs.harvard.edu/abs/1999ApJ...521..271L}
}

@ARTICLE{Fujiwara2024,
       author = {{Fujiwara}, Motoko and {Hamaguchi}, Koichi and {Nagata}, Natsumi and {Ramirez-Quezada}, Maura E.},
        title = "{Vortex creep heating in neutron stars}",
      journal = {\jcap},
         year = 2024,
        month = mar,
       volume = {2024},
       number = {3},
          eid = {051},
        pages = {051},
          doi = {10.1088/1475-7516/2024/03/051},
archivePrefix = {arXiv},
       no-eprint = {2308.16066},
 primaryClass = {astro-ph.HE},
       adsurl = {https://ui.adsabs.harvard.edu/abs/2024JCAP...03..051F}
}

@ARTICLE{Alford2005,
       author = {{Alford}, Mark and {Good}, Gerald and {Reddy}, Sanjay},
        title = "{Isospin asymmetry and type-I superconductivity in neutron star matter}",
      journal = {\prc},
         year = 2005,
        month = nov,
       volume = {72},
       number = {5},
          eid = {055801},
        pages = {055801},
          doi = {10.1103/PhysRevC.72.055801},
archivePrefix = {arXiv},
       no-eprint = {nucl-th/0505025},
 primaryClass = {nucl-th},
       adsurl = {https://ui.adsabs.harvard.edu/abs/2005PhRvC..72e5801A}
}

@ARTICLE{Brandt2003,
       author = {{Brandt}, Ernst Helmut},
        title = "{Properties of the ideal Ginzburg-Landau vortex lattice}",
      journal = {\prb},
         year = 2003,
        month = aug,
       volume = {68},
       number = {5},
          eid = {054506},
        pages = {054506},
          doi = {10.1103/PhysRevB.68.054506},
archivePrefix = {arXiv},
       no-eprint = {cond-mat/0304237},
 primaryClass = {cond-mat.supr-con},
       adsurl = {https://ui.adsabs.harvard.edu/abs/2003PhRvB..68e4506B}
}

@ARTICLE{Das2025,
       author = {{Das}, Mayusree and {Sedrakian}, Armen and {Mukhopadhyay}, Banibrata},
        title = "{Superconductivity in magnetars: Exploring type-I and type-II states in toroidal magnetic fields}",
      journal = {\prd},
         year = 2025,
        month = apr,
       volume = {111},
       number = {8},
          eid = {L081307},
        pages = {L081307},
          doi = {10.1103/PhysRevD.111.L081307},
archivePrefix = {arXiv},
       no-eprint = {2503.14594},
 primaryClass = {astro-ph.SR},
       adsurl = {https://ui.adsabs.harvard.edu/abs/2025PhRvD.111h1307D}
}

@ARTICLE{Baldo1997,
       author = {{Baldo}, M. and {Bombaci}, I. and {Burgio}, G.~F.},
        title = "{Microscopic nuclear equation of state with three-body forces and neutron star structure}",
      journal = {\aap},
         year = 1997,
        month = dec,
       volume = {328},
        pages = {274-282},
          doi = {10.48550/arXiv.astro-ph/9707277},
archivePrefix = {arXiv},
       no-eprint = {astro-ph/9707277},
 primaryClass = {astro-ph},
       adsurl = {https://ui.adsabs.harvard.edu/abs/1997A&A...328..274B}
}

@ARTICLE{Baldo2007,
       author = {{Baldo}, M. and {Schulze}, H. -J.},
        title = "{Proton pairing in neutron stars}",
      journal = {\prc},
         year = 2007,
        month = feb,
       volume = {75},
       number = {2},
          eid = {025802},
        pages = {025802},
          doi = {10.1103/PhysRevC.75.025802},
       adsurl = {https://ui.adsabs.harvard.edu/abs/2007PhRvC..75b5802B}
}

@ARTICLE{AW2008MNRAS,
       author = {{Akg{\"u}n}, T. and {Wasserman}, I.},
        title = "{Toroidal magnetic fields in type II superconducting neutron stars}",
      journal = {\mnras},
         year = 2008,
        month = feb,
       volume = {383},
       number = {4},
        pages = {1551-1580},
          doi = {10.1111/j.1365-2966.2007.12660.x},
archivePrefix = {arXiv},
       no-eprint = {0705.2195},
 primaryClass = {astro-ph},
       adsurl = {https://ui.adsabs.harvard.edu/abs/2008MNRAS.383.1551A}
}

@ARTICLE{Hu1972,
       author = {{Hu}, Chia-Ren},
        title = "{Numerical Constants for Isolated Vortices in Superconductors}",
      journal = {\prb},
         year = 1972,
        month = sep,
       volume = {6},
       number = {5},
        pages = {1756-1760},
          doi = {10.1103/PhysRevB.6.1756},
       adsurl = {https://ui.adsabs.harvard.edu/abs/1972PhRvB...6.1756H}
}

@ARTICLE{das-mukhopadhyay,
       author = {{Das}, Mayusree and {Mukhopadhyay}, Banibrata},
        title = "{Detection Possibility of Continuous Gravitational Waves from Rotating Magnetized Neutron Stars}",
      journal = {\apj},
         year = 2023,
        month = sep,
       volume = {955},
       number = {1},
          eid = {19},
        pages = {19},
          doi = {10.3847/1538-4357/aceb63},
archivePrefix = {arXiv},
       no-eprint = {2302.03706},
 primaryClass = {astro-ph.HE},
       adsurl = {https://ui.adsabs.harvard.edu/abs/2023ApJ...955...19D}
}

@ARTICLE{sold2021main,
       author = {{Soldateschi}, J. and {Bucciantini}, N. and {Del Zanna}, L.},
        title = "{Quasi-universality of the magnetic deformation of neutron stars in general relativity and beyond}",
      journal = {\aap},
         year = 2021,
        month = oct,
       volume = {654},
          eid = {A162},
        pages = {A162},
          doi = {10.1051/0004-6361/202141448},
archivePrefix = {arXiv},
       no-eprint = {2106.00603},
 primaryClass = {astro-ph.HE},
       adsurl = {https://ui.adsabs.harvard.edu/abs/2021A&A...654A.162S}
}

@ARTICLE{sold2021,
       author = {{Soldateschi}, Jacopo and {Bucciantini}, Niccol{\`o}},
        title = "{Detectability of Continuous Gravitational Waves from Magnetically Deformed Neutron Stars}",
      journal = {Galaxies},
         year = 2021,
        month = nov,
       volume = {9},
       number = {4},
          eid = {101},
        pages = {101},
          doi = {10.3390/galaxies9040101},
archivePrefix = {arXiv},
       no-eprint = {2110.06039},
 primaryClass = {astro-ph.HE},
       adsurl = {https://ui.adsabs.harvard.edu/abs/2021Galax...9..101S}
}

@ARTICLE{BPP1969,
       author = {{Baym}, Gordon and {Pethick}, Christopher and {Pines}, David},
        title = "{Superfluidity in Neutron Stars}",
      journal = {\nat},
         year = 1969,
        month = nov,
       volume = {224},
       number = {5220},
        pages = {673-674},
          doi = {10.1038/224673a0},
       adsurl = {https://ui.adsabs.harvard.edu/abs/1969Natur.224..673B}
}

@article{MSM2015,
    author = {Mastrano, A. and Suvorov, A. G. and Melatos, A.},
    title = "{Neutron star deformation due to poloidal–toroidal magnetic fields of arbitrary multipole order: a new analytic approach}",
    journal = {MNRAS},
    volume = {447},
    number = {4},
    pages = {3475-3485},
    year = {2015},
    month = {01},
    issn = {0035-8711},
    doi = {10.1093/mnras/stu2671},
}

@ARTICLE{Braithwaite2009,
       author = {{Braithwaite}, Jonathan},
        title = "{Axisymmetric magnetic fields in stars: relative strengths of poloidal and toroidal components}",
      journal = {\mnras},
         year = 2009,
        month = aug,
       volume = {397},
       number = {2},
        pages = {763-774},
          doi = {10.1111/j.1365-2966.2008.14034.x},
archivePrefix = {arXiv},
       no-eprint = {0810.1049},
 primaryClass = {astro-ph},
       adsurl = {https://ui.adsabs.harvard.edu/abs/2009MNRAS.397..763B}
}

@ARTICLE{sinha15,
       author = {{Sinha}, Monika and {Sedrakian}, Armen},
        title = "{Magnetar superconductivity versus magnetism: Neutrino cooling processes}",
      journal = {\prc},
         year = 2015,
        month = mar,
       volume = {91},
       number = {3},
          eid = {035805},
        pages = {035805},
          doi = {10.1103/PhysRevC.91.035805},
archivePrefix = {arXiv},
       no-eprint = {1502.02979},
 primaryClass = {astro-ph.HE},
       adsurl = {https://ui.adsabs.harvard.edu/abs/2015PhRvC..91c5805S}
}

@INPROCEEDINGS{Haskell18,
       author = {{Haskell}, Brynmor and {Sedrakian}, Armen},
        title = "{Superfluidity and Superconductivity in Neutron Stars}",
    booktitle = {ASSL},
         year = 2018,
       volume = {457},
        month = jan,
        pages = {401},
          doi = {10.1007/978-3-319-97616-7_8},
archivePrefix = {arXiv},
       no-eprint = {1709.10340},
 primaryClass = {astro-ph.HE},
       adsurl = {https://ui.adsabs.harvard.edu/abs/2018ASSL..457..401H}
}

@ARTICLE{gppva,
       author = {{Grill}, Fabrizio and {Pais}, Helena and {Provid{\^e}ncia}, Constan{\c{c}}a and {Vida{\~n}a}, Isaac and {Avancini}, Sidney S.},
        title = "{Equation of state and thickness of the inner crust of neutron stars}",
      journal = {\prc},
         year = 2014,
        month = oct,
       volume = {90},
       number = {4},
          eid = {045803},
        pages = {045803},
          doi = {10.1103/PhysRevC.90.045803},
archivePrefix = {arXiv},
       no-eprint = {1404.2753},
 primaryClass = {nucl-th},
       adsurl = {https://ui.adsabs.harvard.edu/abs/2014PhRvC..90d5803G}
}

@ARTICLE{fermilat,
       author = {{Smith}, D.~A. and {Abdollahi}, S. and {Ajello}, M. and {Bailes}, M. and {Baldini}, L. and {Ballet}, J. and {Baring}, M.~G. and {Bassa}, C. and {Gonzalez}, J. Becerra and {Bellazzini}, R. and {Berretta}, A. and {Bhattacharyya}, B. and {Bissaldi}, E. and {Bonino}, R. and {Bottacini}, E. and {Bregeon}, J. and {Bruel}, P. and {Burgay}, M. and {Burnett}, T.~H. and {Cameron}, R.~A. and {Camilo}, F. and {Caputo}, R. and {Caraveo}, P.~A. and {Cavazzuti}, E. and {Chiaro}, G. and {Ciprini}, S. and {Clark}, C.~J. and {Cognard}, I. and {Corongiu}, A. and {Orestano}, P. Cristarella and {Crnogorcevic}, M. and {Cuoco}, A. and {Cutini}, S. and {D'Ammando}, F. and {de Angelis}, A. and {DeCesar}, M.~E. and {De Gaetano}, S. and {de Menezes}, R. and {Deneva}, J. and {de Palma}, F. and {Di Lalla}, N. and {Dirirsa}, F. and {Di Venere}, L. and {Dom{\'\i}nguez}, A. and {Dumora}, D. and {Fegan}, S.~J. and {Ferrara}, E.~C. and {Fiori}, A. and {Fleischhack}, H. and {Flynn}, C. and {Franckowiak}, A. and {Freire}, P.~C.~C. and {Fukazawa}, Y. and {Fusco}, P. and {Galanti}, G. and {Gammaldi}, V. and {Gargano}, F. and {Gasparrini}, D. and {Giacchino}, F. and {Giglietto}, N. and {Giordano}, F. and {Giroletti}, M. and {Green}, D. and {Grenier}, I.~A. and {Guillemot}, L. and {Guiriec}, S. and {Gustafsson}, M. and {Harding}, A.~K. and {Hays}, E. and {Hewitt}, J.~W. and {Horan}, D. and {Hou}, X. and {Jankowski}, F. and {Johnson}, R.~P. and {Johnson}, T.~J. and {Johnston}, S. and {Kataoka}, J. and {Keith}, M.~J. and {Kerr}, M. and {Kramer}, M. and {Kuss}, M. and {Latronico}, L. and {Lee}, S. -H. and {Li}, D. and {Li}, J. and {Limyansky}, B. and {Longo}, F. and {Loparco}, F. and {Lorusso}, L. and {Lovellette}, M.~N. and {Lower}, M. and {Lubrano}, P. and {Lyne}, A.~G. and {Maan}, Y. and {Maldera}, S. and {Manchester}, R.~N. and {Manfreda}, A. and {Marelli}, M. and {Mart{\'\i}-Devesa}, G. and {Mazziotta}, M.~N. and {McEnery}, J.~E. and {Mereu}, I. and {Michelson}, P.~F. and {Mickaliger}, M. and {Mitthumsiri}, W. and {Mizuno}, T. and {Moiseev}, A.~A. and {Monzani}, M.~E. and {Morselli}, A. and {Negro}, M. and {Nemmen}, R. and {Nieder}, L. and {Nuss}, E. and {Omodei}, N. and {Orienti}, M. and {Orlando}, E. and {Ormes}, J.~F. and {Palatiello}, M. and {Paneque}, D. and {Panzarini}, G. and {Parthasarathy}, A. and {Persic}, M. and {Pesce-Rollins}, M. and {Pillera}, R. and {Poon}, H. and {Porter}, T.~A. and {Possenti}, A. and {Principe}, G. and {Rain{\`o}}, S. and {Rando}, R. and {Ransom}, S.~M. and {Ray}, P.~S. and {Razzano}, M. and {Razzaque}, S. and {Reimer}, A. and {Reimer}, O. and {Renault-Tinacci}, N. and {Romani}, R.~W. and {S{\'a}nchez-Conde}, M. and {Parkinson}, P.~M. Saz and {Scotton}, L. and {Serini}, D. and {Sgr{\`o}}, C. and {Shannon}, R. and {Sharma}, V. and {Shen}, Z. and {Siskind}, E.~J. and {Spandre}, G. and {Spinelli}, P. and {Stappers}, B.~W. and {Stephens}, T.~E. and {Suson}, D.~J. and {Tabassum}, S. and {Tajima}, H. and {Tak}, D. and {Theureau}, G. and {Thompson}, D.~J. and {Tibolla}, O. and {Torres}, D.~F. and {Valverde}, J. and {Venter}, C. and {Wadiasingh}, Z. and {Wang}, N. and {Wang}, N. and {Wang}, P. and {Weltevrede}, P. and {Wood}, K. and {Yan}, J. and {Zaharijas}, G. and {Zhang}, C. and {Zhu}, W.},
        title = "{The Third Fermi Large Area Telescope Catalog of Gamma-Ray Pulsars}",
      journal = {\apj},
         year = 2023,
        month = dec,
       volume = {958},
       number = {2},
          eid = {191},
        pages = {191},
          doi = {10.3847/1538-4357/acee67},
archivePrefix = {arXiv},
       no-eprint = {2307.11132},
 primaryClass = {astro-ph.HE},
       adsurl = {https://ui.adsabs.harvard.edu/abs/2023ApJ...958..191S}
}

@ARTICLE{bcs,
       author = {{Bardeen}, J. and {Cooper}, L.~N. and {Schrieffer}, J.~R.},
        title = "{Theory of Superconductivity}",
      journal = {Phys. Rev.},
         year = 1957,
        month = dec,
       volume = {108},
       number = {5},
        pages = {1175-1204},
          doi = {10.1103/PhysRev.108.1175},
       adsurl = {https://ui.adsabs.harvard.edu/abs/1957PhRv..108.1175B}
}

@ARTICLE{Chamel2017,
       author = {{Chamel}, N.},
        title = "{Superfluidity and Superconductivity in Neutron Stars}",
      journal = {\aap},
         year = 2017,
        month = sep,
       volume = {38},
       number = {3},
          eid = {43},
        pages = {43},
          doi = {10.1007/s12036-017-9470-9},
archivePrefix = {arXiv},
       no-eprint = {1709.07288},
 primaryClass = {astro-ph.HE},
       adsurl = {https://ui.adsabs.harvard.edu/abs/2017JApA...38...43C}
}

@ARTICLE{easson1977,
       author = {{Easson}, Ian and {Pethick}, C.~J.},
        title = "{Stress tensor of cosmic and laboratory type-II superconductors}",
      journal = {\prd},
         year = 1977,
        month = jul,
       volume = {16},
       number = {2},
        pages = {275-280},
          doi = {10.1103/PhysRevD.16.275},
       adsurl = {https://ui.adsabs.harvard.edu/abs/1977PhRvD..16..275E}
}

@ARTICLE{freibenrezz2012,
       author = {{Frieben}, J. and {Rezzolla}, L.},
        title = "{Equilibrium models of relativistic stars with a toroidal magnetic field}",
      journal = {\mnras},
         year = 2012,
        month = dec,
       volume = {427},
       number = {4},
        pages = {3406-3426},
          doi = {10.1111/j.1365-2966.2012.22027.x},
archivePrefix = {arXiv},
       no-eprint = {1207.4035},
 primaryClass = {gr-qc},
       adsurl = {https://ui.adsabs.harvard.edu/abs/2012MNRAS.427.3406F}
}

@ARTICLE{lander12,
       author = {{Lander}, S.~K. and {Andersson}, N. and {Glampedakis}, K.},
        title = "{Magnetic neutron star equilibria with stratification and type II superconductivity}",
      journal = {\mnras},
         year = 2012,
        month = jan,
       volume = {419},
       number = {1},
        pages = {732-747},
          doi = {10.1111/j.1365-2966.2011.19720.x},
archivePrefix = {arXiv},
       no-eprint = {1106.6322},
 primaryClass = {astro-ph.SR},
       adsurl = {https://ui.adsabs.harvard.edu/abs/2012MNRAS.419..732L}
}

@ARTICLE{Lala2005,
       author = {{Lalazissis}, G.~A. and {Nik{\v{s}}i{\'c}}, T. and {Vretenar}, D. and {Ring}, P.},
        title = "{New relativistic mean-field interaction with density-dependent meson-nucleon couplings}",
      journal = {\prc},
         year = 2005,
        month = feb,
       volume = {71},
       number = {2},
          eid = {024312},
        pages = {024312},
          doi = {10.1103/PhysRevC.71.024312},
       adsurl = {https://ui.adsabs.harvard.edu/abs/2005PhRvC..71b4312L}
}

@ARTICLE{Baym1969,
       author = {{Baym}, Gordon and {Pethick}, Christopher and {Pines}, David},
        title = "{Superfluidity in Neutron Stars}",
      journal = {\nat},
         year = 1969,
        month = nov,
       volume = {224},
       number = {5220},
        pages = {673-674},
          doi = {10.1038/224673a0},
       adsurl = {https://ui.adsabs.harvard.edu/abs/1969Natur.224..673B}
}

@ARTICLE{Page06,
       author = {{Page}, Dany and {Geppert}, Ulrich and {Weber}, Fridolin},
        title = "{The cooling of compact stars}",
      journal = {\nphysa},
         year = 2006,
        month = oct,
       volume = {777},
        pages = {497-530},
          doi = {10.1016/j.nuclphysa.2005.09.019},
archivePrefix = {arXiv},
       no-eprint = {astro-ph/0508056},
 primaryClass = {astro-ph},
       adsurl = {https://ui.adsabs.harvard.edu/abs/2006NuPhA.777..497P}
}
\end{document}